\documentclass[twocolumn,english,aps,prb]{revtex4-1}
\usepackage{CJK}

\usepackage{amsfonts}
\usepackage{amsmath}
\usepackage{ dsfont }
\usepackage{hyperref}
\usepackage{amssymb}
\usepackage{array,multirow}
\usepackage{longtable}
\usepackage{ragged2e}
\usepackage{xspace}
\usepackage{xcolor}
\usepackage{relsize}%\mathlarger command 
\usepackage[bottom]{footmisc}
\usepackage{color}
\usepackage{graphicx}
\usepackage[space]{grffile}
\usepackage{verbatim}
\usepackage{amssymb}
\usepackage{mathrsfs}
\usepackage{wasysym}
\usepackage[caption=false]{subfig}
\usepackage{url}
\usepackage{bbold}
\usepackage{slashed}
\usepackage{epstopdf}
\usepackage{braket}
\usepackage{float}
\usepackage[percent]{overpic}

\usepackage{tikz}
\usetikzlibrary{calc} 
\newcommand{\be}{\begin{equation}}
\newcommand{\ee}{\end{equation}}
\newcommand{\bea}{\begin{equation} \begin{aligned}}
\newcommand{\eea}{\end{aligned} \end{equation} }
\newcommand{\bi}{\begin{itemize}}
\newcommand{\ei}{\end{itemize}}

\newcolumntype{C}[1]{>{\centering\arraybackslash}p{#1}}

\usepackage{xspace}

\newcommand{\la}{\lambda}
\renewcommand{\be}{\beta}
\newcommand{\al}{\alpha}

\renewcommand{\th}{\theta}
\newcommand{\lp}{\left(}
\newcommand{\rp}{\right)}

\newcommand{\lag}{\mathcal{L}}

\newcommand{\mbf}[1]{\mathbf{#1}}

\newcommand{\bpm}{\begin{pmatrix}}
\newcommand{\epm}{\end{pmatrix}}
\renewcommand\arraystretch{1.3}

\newcommand{\ovl}[1]{\overline{#1}}

\DeclareRobustCommand{\App}[1]{App.~ \ref{#1}}

\DeclareRobustCommand{\Tab}[1]{Table~ \ref{#1}}
\DeclareRobustCommand{\Tabs}[2]{Tables~ \ref{#1} and  \ref{#2}}
\DeclareRobustCommand{\Fig}[1]{Fig.~ \ref{#1}}

\DeclareRobustCommand{\Eq}[1]{Eq.~ \ref{#1}}
\DeclareRobustCommand{\Eqs}[2]{Eqs.~\ref{#1} and  \ref{#2}}

\DeclareMathAlphabet\mathbfcal{OMS}{cmsy}{b}{n}
\RequirePackage[normalem]{ulem} %DIF PREAMBLE
\RequirePackage{color}\definecolor{RED}{rgb}{1,0,0}\definecolor{BLUE}{rgb}{0,0,1} %DIF PREAMBLE
\begin{document}

\title{Interacting Topological Quantum Chemistry in 2D with Many-body Real Space Invariants}
%\title{Many-Body Topological Quantum Chemistry Indices in 2D}
\author{Jonah Herzog-Arbeitman$^{1}$}
\author{B. Andrei Bernevig$^{1,2,3}$}
\author{Zhi-Da Song$^{1,4}$}

\affiliation{$^1$Department of Physics, Princeton University, Princeton, NJ 08544}
\affiliation{$^2$Donostia International Physics Center, P. Manuel de Lardizabal 4, 20018 Donostia-San Sebastian, Spain}
\affiliation{$^3$IKERBASQUE, Basque Foundation for Science, Bilbao, Spain}
\affiliation{$^4$International Center for Quantum Materials, School of Physics, Peking University, Beijing 100871, China}
\affiliation{$^5$Hefei National Laboratory, Hefei 230088, China}
\affiliation{$^6$Collaborative Innovation Center of Quantum Matter, Beijing 100871, China}

\begin{abstract}
The topological phases of non-interacting fermions have been classified by their symmetries, culminating in a modern electronic band theory where wavefunction topology can be obtained from momentum space. Recently, Real Space Invariants (RSIs) have provided a spatially local description of the global momentum space indices. The present work generalizes this real space classification to interacting 2D states. We construct many-body local RSIs as the quantum numbers of a set of symmetry operators on open boundaries, but which are independent of the choice of boundary. Using the $U(1)$ particle number, they yield many-body fragile topological indices, which we use to identify which single-particle fragile states are many-body topological or trivial at weak coupling. To this end, we construct an exactly solvable Hamiltonian with single-particle fragile topology that is adiabatically connected to a trivial state through strong coupling. We then define global many-body RSIs on periodic boundary conditions. They reduce to Chern numbers in the band theory limit, but also identify strongly correlated stable topological phases with no single-particle counterpart. Finally, we show that the many-body local RSIs appear as quantized coefficients of Wen-Zee terms in the topological quantum field theory describing the phase. 
\end{abstract}

\maketitle

\section{Introduction}

The symmetries of a Hamiltonian are essential to the classification of topological phases in crystals. For instance, the Ten-Fold Way \cite{2010NJPh...12f5010R,2009AIPC.1134...22K}, Topological Quantum Chemistry (TQC)\cite{2017Natur.547..298B,2020arXiv201000598E}, and symmetry indicators \cite{Aroyo:xo5013,2019arXiv190503262S,2017NatCo...8...50P,2017PhRvX...7d1069K} have redefined our understanding of non-interacting electronic states of matter with the symmetry group of the Hamiltonian taking center stage. The success of this program motivates us to extend its reach to interacting Hamiltonians where many-body effects accompany band topology in the groundstate \cite{PhysRevB.99.125122,PhysRevX.8.011040,2021arXiv210302624L,2021arXiv210104135I,2022PhRvR...4b3177M,2022PhRvB.105w5143B,2022PhRvL.128h7002H,2021arXiv210710837C,2015PhRvB..92h1304I,2022arXiv220900007H,2022PhRvB.105d5112H,PhysRevB.86.125119,2015NatCo...6.8944P,PhysRevB.103.205415,2022PhRvL.129g6401H,2020PhRvL.124p7002X,2016RvMP...88c5005C,2017PhRvB..96l5105H,2020PhRvB.101t5101K,PhysRevLett.120.096601,2018PhRvB..98c5151S,2021PhRvL.127x6403W,2021PhRvR...3a3040M,2021arXiv211215533W,2021arXiv210800008M,2020PhRvB.102h1110S,2022arXiv220811710F,2018RPPh...81k6501R,2022arXiv221107924K,2022arXiv220910556S,2022arXiv220910556S}. This work focuses on 2D systems with space group $G$ and $U(1)$ charge conservation at an integer filling per unit cell.

The classifications of single-particle topology originally relied on momentum space calculations such as the Wilson loop \cite{2018arXiv180409719B,2019PhRvX...9b1013A,2018arXiv180710676S,Alexandradinata:2012sp,2011PhRvB..84g5119Y,2018AdPhX...314631Y,Benalcazar_2017,andreibook} and band structure irreps \cite{2017NatCo...8...50P,2020arXiv200604890C,2018SciA....4.8685W,2017PhRvX...7d1069K,2018PhRvX...8c1070K,2018NatCo...9.3530S}. Physically, however, nontrivial topology is completely diagnosed in real space where a topological index serves as an obstruction to the atomic limit, defined by a representation of the groundstate with localized, symmetric Wannier states \cite{2012RvMP...84.1419M,2011PhRvB..83c5108S,PhysRevLett.98.046402}. Recently, Ref. \cite{rsis} extended this idea to symmetry-protected phases by developing Real Space Invariants (RSIs) --- local indices which can be considered as Noether charges associated to discrete symmetries --- which may be calculated from the irreps formed by the Wannier states in the unit cell, even in fragile phases \cite{PhysRevLett.121.126402,PhysRevLett.120.266401}. These RSIs are gauge-invariant and classify all 2D and 3D symmetry eigenvalue-indicated single-particle topology \cite{2020JPCM...32z3001P,Aroyo:xo5013, 2019arXiv190503262S,2021arXiv211102433X,2022arXiv220910559H}, but may not detect some topological states which are classified by cohomology or not protected by symmetry \cite{2011PhRvB..83h5426G,2011PhRvB..84l5132E,PhysRevResearch.4.033081,PhysRevB.105.155112,2010PhRvB..81m4509F}. Our paper extends this technique by defining {many-body} RSIs (henceforth referred to as RSIs for brevity) of two types. 

First we construct {local} RSIs on open boundary conditions (OBCs) in all 2D point groups. These RSIs classify adiabatically distinct many-body atomic states\cite{PhysRevB.99.125122,2017PhRvL.119x6402S,2020PhRvL.124x7001S,2020arXiv200613242B,Benalcazar_2017,2017PhRvB..96x5115B}.
We then define many-body fragile topological states \cite{PhysRevB.99.125122,2018arXiv180802482P,2019arXiv190503262S,2020arXiv200802288P,2019arXiv191006868P,2018arXiv180710676S,2018arXiv181002373W} by an obstruction to all many-body atomic limit states and derive their topological invariants in terms of inequalities between the RSIs and the $U(1)$ particle number. We also present an exactly solvable model verifying that interactions trivialize certain fragile non-interacting states identified by our classification. To study many-body stable topological states  \cite{PhysRevLett.61.2015,2006Sci...314.1757B,2005PhRvL..95n6802K,2017Natur.547..298B}, we introduce {global} RSIs defined on periodic boundary conditions and show that they are many-body stable topological invariants. Finally, we show that the local RSIs appear as quantized coefficients in the topological response theory generalizing the Chern-Simons action of a Chern insulator. 

Our theory provides an elementary classification of symmetry-protected Chern and fragile topological phases, provides an explicit connection between many-body and single-particle topological indices substantiated by exactly solvable model Hamiltonians, and proposes a fundamental relation between these topological indices defined on the lattice and the topological quantum field theory that describes their universal behavior. 

\section{Many-Body Local RSIs} 
Axiomatically, we define a many-body topological state by an obstruction to adiabatically deforming it into a many-body atomic (trivial) state while respecting the symmetries of the space group $G$. The results of this paper rely on the following definition. A many-body atomic state is any state which is adiabatically connected to a trivial many-body atomic {limit} which is $\mbf{(1)}$ non-degenerate, $\mathbf{(2)}$ spatially decoupled, and $\mbf{(3)}$ endowed with a many-body gap. This limit allows for arbitrarily strong interactions in the Hamiltonian as long as they are strictly local. A many-body atomic limit is the groundstate of the tight-binding Hamiltonian
\bea
\label{eq:HMBAL}
H_{AL} &= \sum_{\mbf{R},\mbf{r}} H_{\mbf{R},\mbf{r}}, \qquad  H_{\mbf{R},\mbf{r}} = T_\mbf{R} H_{\mbf{r}} T_\mbf{R}^\dag 
\eea
where $\mbf{R}$ are the lattice vectors, $T_\mbf{R} \in G$ are the translation operators, $\mbf{r}$ are the locations of orbitals in the unit cell, and $H_{\mbf{R},\mbf{r}}$ is supported only on the orbitals at $\mbf{R}+\mbf{r}$ (there are no hoppings). This ensures $[H_{\mbf{R},\mbf{r}} ,H_{\mbf{R}',\mbf{r}'} ] = 0 $ and thus the groundstate of $H$ can be written as $\ket{GS} = \prod_{\mbf{R},\mbf{r}} \mathcal{O}_{\mbf{R},\mbf{r}}\ket{0}$ and $\mathcal{O}_{\mbf{R},\mbf{r}}$ creates the (possibly correlated) groundstate of $H_{\mbf{R},\mbf{r}}$. In the thermodynamic limit, the filling is $\nu = N_{occ} / N_{orb}$ where $N_{orb}$ is the number of orbitals per unit cell and $N_{occ}$ is the filling, e.g. $\prod_{\mbf{r}} \mathcal{O}_{\mbf{R},\mbf{r}}$ creates $N_{occ}$ electrons. \Eq{eq:HMBAL}  ensures $\mbf{(2)}$ holds, and we {require} that $\mbf{(1)}$ and $\mathbf{(3)}$ are satisfied, as is natural in an insulator (see \App{app:RSIs}).

We now define local RSIs in many-body atomic limits at a Wyckoff position $\mbf{x}$ protected only by the symmetries of the point group $G_\mbf{x} \in G$. To do so, we identify a set of discrete symmetry operators whose eigenvalues are the local RSIs. This ensures the local RSIs are adiabatic invariants since they are discrete quantum numbers. To ensure locality, we define the local RSI on OBCs by imposing a spatial cutoff around $\mbf{x}$ {and} requiring invariance under the particular choice of cutoff. 

\begin{figure}
 \centering
  \includegraphics[width=.49\linewidth]{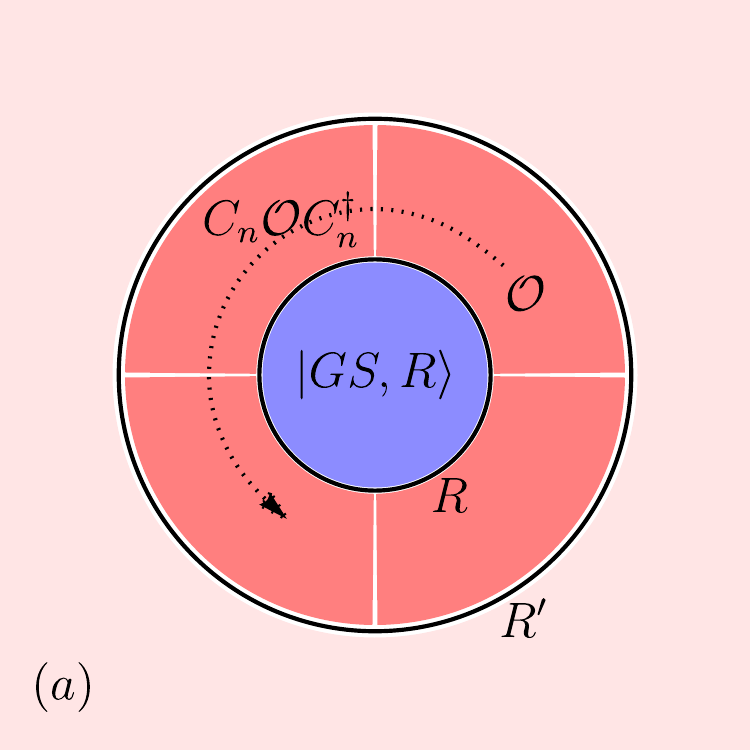}
  \includegraphics[width=.49\linewidth]{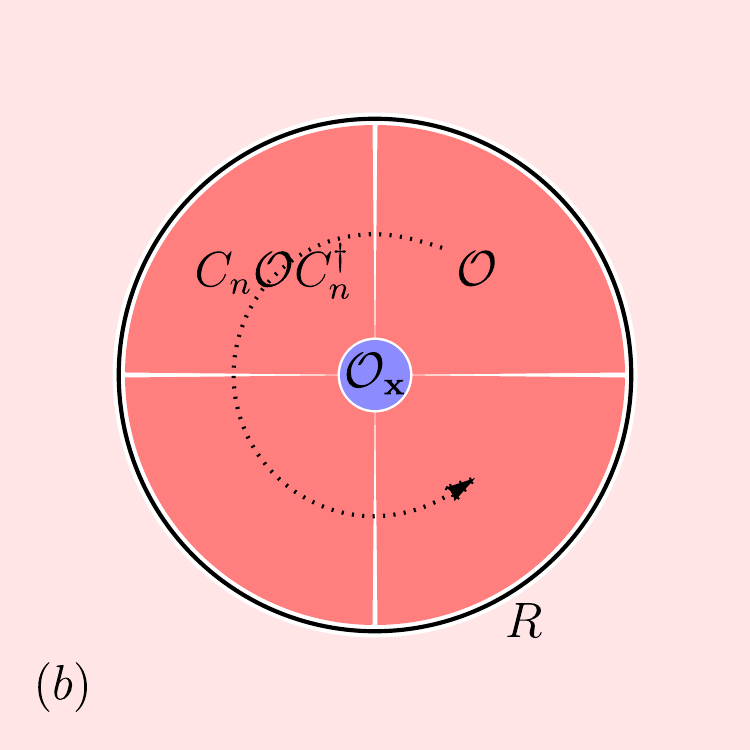}
\caption{Local Quantum Numbers. (a) We depict the groundstate $\ket{GS,R}$ on OBCs and the additional symmetry-related operators which are included upon expanding the cutoff to $R'$. The many-body local RSIs are invariant under the expansion. (b) Given a fixed cutoff $R$, all operators inside of $R$ but not at the $C_n$-invariant point $\mbf{x}$ are symmetry-related and so do not contribute to the many-body local RSI. Only the operator $O_\mbf{x}$ at $\mbf{x}$ transforms locally under $\mbf{x}$. Its quantum numbers determine the many-body local RSI. }
\label{fig:Ocalmain}
\end{figure}

Let $\ket{GS,R}$ be the groundstate of $H_{AL}$ but restricted to OBCs by including only sites $|\mbf{R}+\mbf{r} - \mbf{x}| \leq R$ in \Eq{eq:HMBAL} (see \Fig{fig:Ocalmain}a). The cutoff breaks translational symmetry but preserves the point group symmetries $g \in G_\mbf{x} \subset G$. The quantum numbers of $\ket{GS,R}$ are
\bea
\hat{N} \ket{GS,R} = N  \ket{GS,R} , \ g \ket{GS,R} = e^{i \la[g]} \ket{GS,R} 
\eea
where $\hat{N}$ is the number operator and $e^{i \la[g]}$ is a 1D irrep of $G_\mbf{x}$. Note that $\ket{GS,R}$ is a non-degenerate trivial many-body atomic limit and must transform in a 1D irrep $(\mbf{1})$. However, the quantum numbers $N$ and $\la[g]$ are not independent of the cutoff, as we now show, so they cannot be local RSIs. Consider the spinless rotation groups $G_\mbf{x} = n$ generated by the operator $C_n$. Because all terms in $H_{AL}$ are strictly local $\mbf{(2)}$, we can write the groundstate at a larger cutoff $R'>R$ as
\bea
\label{eq:GSR'}
\ket{GS,R'} = \prod_{i=0}^{n-1}  \mathcal{O}_i \, \ket{GS,R}, \quad \mathcal{O}_i =  \prod_{\mbf{R}+\mbf{r} \in \mathcal{D}} C^{i}_n  \mathcal{O}_{\mbf{R},\mbf{r}} C^{\dag i}_n  \\
\eea
where $\mathcal{D}$ is the annulus sector between $R$ and $R'$ of angle $2\pi/n$ shown in \Fig{fig:Ocalmain}a and $C_n \in G_\mbf{x}, C_n^n = +1$ is an $n$-fold rotation. Define the total charge $N_{\mathcal{O}} $ by $[\hat{N}, \mathcal{O}_i] = N_{\mathcal{O}} \mathcal{O}_i$ so that $\mathcal{O}_i \mathcal{O}_j = (-1)^{N_\mathcal{O}}\mathcal{O}_j \mathcal{O}_i$. Since the operators $\mathcal{O}_i$ commute/anti-commute if $N_\mathcal{O}$ is even/odd, \Eq{eq:GSR'} gives 
\bea
\label{eq:CnonR'}
C_n \ket{GS,R'} = e^{i \la[C_n]} (-1)^{N_\mathcal{O}} \ket{GS,R'} , \quad (n \text{ even}) \ .
\eea
Thus the $C_n$ eigenvalue $\la[C_n]$ is {not} invariant (for even $n$) under expanding the cutoff because $N_\mathcal{O}$ is arbitrary. Similarly, $\hat{N} \ket{GS,R'} = (N + n N_\mathcal{O})\ket{GS,R'}$ so $N$ is clearly not invariant. However, we can easily produce symmetry operators which are invariant under an arbitrary expansion of the cutoff. Indeed, $e^{i \frac{\pi}{n} \hat{N}}C_n$ is invariant because
\begin{align}
&\braket{GS,R'|e^{i \frac{\pi}{n} \hat{N}}C_n |GS,R'} = e^{i \frac{\pi}{n} (N+n N_\mathcal{O})} (-1)^{N_\mathcal{O}} e^{i \la[C_n]} \nonumber \\
&\qquad = e^{i\frac{\pi}{n} N} e^{i \la[C_n]}  = \braket{GS,R|e^{i \frac{\pi}{n} \hat{N}}C_n |GS,R} \ ,
\end{align}
and from \Eq{eq:CnonR'}, we see immediately that $C_n^2$ is also invariant (for $n$ even).  Hence we have found elementary symmetry operators whose eigenvalues only depend on the local properties of the groundstate near $\mbf{x}$ but are invariant under the imposed cutoff. Their eigenvalues are the local RSIs defined by
\bea
\label{eq:RSIsneven}
e^{i \frac{\pi}{n} \hat{N}}C_n \ket{GS} &= e^{i \frac{\pi}{n} \Delta_1}\ket{GS}, \quad \Delta_1 \in \mathds{Z}_{2n}  \\
C^2_n \ket{GS} &= e^{i \frac{2\pi}{n/2} \Delta_2}\ket{GS}, \quad \Delta_2 \in \mathds{Z}_{n/2}  \\
\eea
using $(e^{i \frac{\pi}{n} \hat{N}}C_n)^{2n} = (C_n^2)^{n/2} = +1$ which quantizes $\Delta_1,\Delta_2$. Although we derived \Eq{eq:RSIsneven} in many-body atomic {limits}, we now prove the local RSIs remain well-defined in general many-body atomic states. First, observe that the operators in \Eq{eq:RSIsneven} remain symmetries as hoppings and off-site interactions are added to $H_{AL}$ and their eigenvalues (the local RSIs) remain well-defined. Then because we assume a many-body gap $\mbf{(3)}$ and the local RSIs are quantized, they cannot change as $H_{AL}$ is adiabatically deformed out of the strict atomic limit. 

We have explicitly constructed local RSIs at a single Wyckoff position $\mbf{x}$ on OBCs. But on infinite boundary conditions, the unit cell contains multiple Wyckoff positions and it should be possible to define local RSIs $\Delta_{\mbf{x},1},\Delta_{\mbf{x},2}$ at each $\mbf{x}$. For instance in the wallpaper group $G=p2$ which is generated by  $C_2$ and translations, there are four Wyckoff positions 1a $= (0,0)$, 1b $= (1/2,0)$, 1c $= (0,1/2)$, and 1d $= (1/2,1/2)$ whose point groups are generated by $C_2, T_1 C_2, T_2 C_2$ and $T_1 T_2 C_2$ respectively. Even though imposing OBCs at one Wyckoff position breaks the symmetries of the others, we argue that the local RSIs are still well-defined on infinite boundary conditions. This is because the local RSIs are independent of the cutoff, so sending the cutoff to infinity recovers the full space group by restoring translations.

In \App{app:RSIs}, we extend our results to construct local RSIs in all 2D spinless and spinful point groups (see \Tabs{tab:RSInoSOC}{tab:RSIswithsoc}). In the spinless groups, we find that mirrors and time-reversal restrict the $C_n$ eigenvalue on the groundstate to be real, reducing the $\mathds{Z}_{2n} \times \mathds{Z}_{n/2}$ classification of \Eq{eq:RSIsneven} to $\mathds{Z}_{2n}$ for even $n$. For odd $n$, we find a $\mathds{Z}_n \times \mathds{Z}_n$ classification which is reduced to $\mathds{Z}_n$. For even $n$ in the spinful groups, mirrors and time-reversal also force $C_n = \pm 1$ to be real on any non-degenerate state, which yields a $\mathds{Z}_2$ factor in the local RSI classification. We check that $C_n = +1$ holds on all product states (Slater determinants). The $-1$ eigenvalue is only possible with strong interactions, and can be obtained in trivial atomic Mott insulators' \cite{2010PhRvL.105p6402Y,2022arXiv220910556S}. In all cases, the classifying groups are abelian, so the local RSIs are additive: the local RSIs of a tensor product of states is the sum of their individual local RSIs. This is crucial for defining many-body fragile topology. 

\section{Many-body Fragile Topology} 
We now consider a state on infinite boundary conditions with charge density $\nu = N_{occ}/N_{orb}$. Recall that single-particle fragile topology is characterized by an obstruction to adiabatic deformation into an atomic state, but this obstruction is removed if additional (trivial) orbitals in a specific representation are added \cite{2019arXiv190503262S}. The topological indices for the single-particle fragile states are inequalities and mod equations relating $N_{occ}$ and the RSIs in the unit cell \cite{rsis}.

This structure extends to the many-body case. We define a topological state with $N_{occ}$ particles per unit cell as many-body fragile iff it can be adiabatically connected to a many-body trivial atomic state with $N_{occ} + \tilde{N}$ particles per unit cell by the addition of $\tilde{N} > 0$ many-body atomic states\cite{PhysRevB.99.125122,2021PhRvB.103d5128L}. Let $\Delta_{N_{occ} +\tilde{N}}$ (resp. $\Delta_{\tilde{N}}$) denote the set of RSIs of the trivial state with $N_{occ} + \tilde{N}$ particles (resp. the trivial state of the additional $\tilde{N}$ orbitals) at all Wyckoff positions in the unit cell. The local RSIs of the $N_{occ}$-particle many-body fragile state are defined by $\Delta_{frag} = \Delta_{N_{occ} +\tilde{N}} - \Delta_{\tilde{N}}$ (see \App{app:fragile} for details). Crucially, $\Delta_{N_{occ} +\tilde{N}}$ is well-defined because the $N_{occ} + \tilde{N}$-particle state is trivial atomic. This underlies the essentially difference between fragile and stable topological many-body states (to be defined shortly), where the latter {cannot} be connected to any many-body atomic state via the addition of any many-body atomic states\cite{rsis}. 

To assess whether a state is topological given a set of local RSIs, we can enumerate all possible atomic limits in $G$ formed from $N_{occ}$ orbitals --- note that only a finite number are possible. If the local RSIs of the groundstate do {not} appear in this set, then the state must be fragile topological by definition. In practice, we find a simpler method by deriving inequality constraints that relate the local RSIs and the $U(1)$ electron density. To illustrate this, we again consider $G= p2$ with Wyckoff positions $\mbf{x} = $1a,1b,1c,1d. There are four RSIs given by the eigenvalues $e^{i \frac{\pi}{2} \Delta_{1,\mbf{x}}}$ of $e^{i \frac{\pi}{2} \hat{N}}C_{2, \mbf{x}}$ where $C_{2,\mbf{x}}$ is a rotation centered at $\mbf{x}$. It is convenient to define $\Delta_{1,\mbf{x}} \in \{-1,0,1,2\}$. For many-body atomic states on arbitrary OBCs respecting $G_\mbf{x}$, it is easy to prove (see \App{app:fragile}) that $N_\mbf{x} \geq |\Delta_{1,\mbf{x}}|$ where $N_\mbf{x}$ is the total number of particles. Note that $\Delta_{1\mbf{x}}$ is does not depend on the OBC cutoff, whereas
 $N_\mbf{x}$ obviously does. In a many-body atomic {limit} where we can take $N_\mbf{x} = N_\mathcal{O}$ (see \Eq{eq:GSR'}) by choosing a cutoff surrounding $\mbf{x}$ only (see \Fig{fig:Ocalmain}b), we can bound the total density by summing over the number of states at the high-symmetry Wyckoff positions in a single unit cell:
\bea
\label{eq:p2fragile}
N_{occ} &\geq \sum_{\mbf{x} = 1a,1b,1c,1d} N_{\mbf{x}} \geq \sum_{\mbf{x} = 1a,1b,1c,1d} |\Delta_{1, \mbf{x}} | 
\eea
which is a lower bound because only the high-symmetry Wyckoff positions are counted. The bound in \Eq{eq:p2fragile} is ultimately written in terms of RSIs and the charge density which are well-defined quantum numbers  in any many-body atomic state. \Eq{eq:p2fragile} holds in all many-body atomic states. Hence if \Eq{eq:p2fragile} is {violated}, then the RSIs impose an obstruction to deformation into a many-body atomic state, proving many-body fragile topology. 

We can now prove that certain single-particle fragile states remain fragile topological as interactions are added. As a first step, we determine a formula for the local RSI when acting on product states (which are groundstates without interactions). With $C_2$, the irreps are $A$ and $B$ which are even and odd under $C_{2}$ respectively. Noting that $e^{i \frac{\pi}{2} \hat{N}} C_{2} \ket{GS} = e^{i \frac{\pi}{2} (m(A) + m(B)) + i \pi m(B)} \ket{GS}$,  \Eq{eq:RSIsneven} yields 
\bea
\Delta_{1} = m(A) - m(B) \mod 4 
\eea
 where $m(\rho)$ is the multiplicity of the $\rho$ irrep in the product state. In fact, this expression can be understood perturbatively. The single-particle RSI with $C_2$ is $\delta_1 = m(B) - m(A) \in \mathds{Z}$ \cite{rsis}. With interactions, a state with $m(A) = 2$ can be scattered into a state with $m(B) = 2$ since both have even parity. Thus states with $\delta_1 = \pm 2$ are identified with interactions \cite{2019PhRvX...9c1003L}, and only $\delta_1 \mod 4 = \Delta_1$ is invariant. Let us now consider the single-particle $N_{occ} = 2$ fragile state $2 \Gamma_1 \oplus 2 X_1 \oplus 2 Y_1 \oplus 2 M_2 = (A_{1a} \oplus A_{1b}\oplus A_{1c} \ominus A_{1d}) \uparrow G $ which can be thought as stacking a Chern $+1$ state with a Chern $-1$ state. (Here $\uparrow$ denotes the Frobenius induction\cite{rsis,2020arXiv200604890C} of the irreps $\rho_\mbf{x} \in G_\mbf{x}$ to the full wallpaper group $G$ containing all Wyckoff positions $\mbf{x}$ as high symmetry points.) Although their total Chern number vanishes, there is still fragile topology protected by $C_2$. We compute the RSIs to be $\Delta_{1a} = \Delta_{1b}= \Delta_{1c} =1,  \Delta_{1d} = -1$. Evaluating the topological obstruction $\sum_\mbf{x} |\Delta_{\mbf{x}}| = 4$ in \Eq{eq:p2fragile}, we find that this single-particle state violates the inequality since $N_{occ} = 2$ and is {many-body} fragile. Adiabatically adding interactions cannot trivialize the state. 

We generalize the inequality criterion of \Eq{eq:p2fragile} to all wallpaper groups in \App{app:fragile}, obtaining topological invariants of many-body fragile phases. In \Tabs{tab:rsiredint}{tab:mpgrsi}, we give expressions for the local RSIs on product states in terms of irrep multiplicities and also single-particle RSIs which are readily computed from momentum space irreps \cite{rsis}. Our results determine the stability of any single-particle fragile phase when interactions are added. 

\section{Trivializing Single-Particle Fragile Topology.}
\begin{figure}
 \centering
\begin{overpic}[height=0.18\textwidth,tics=10]{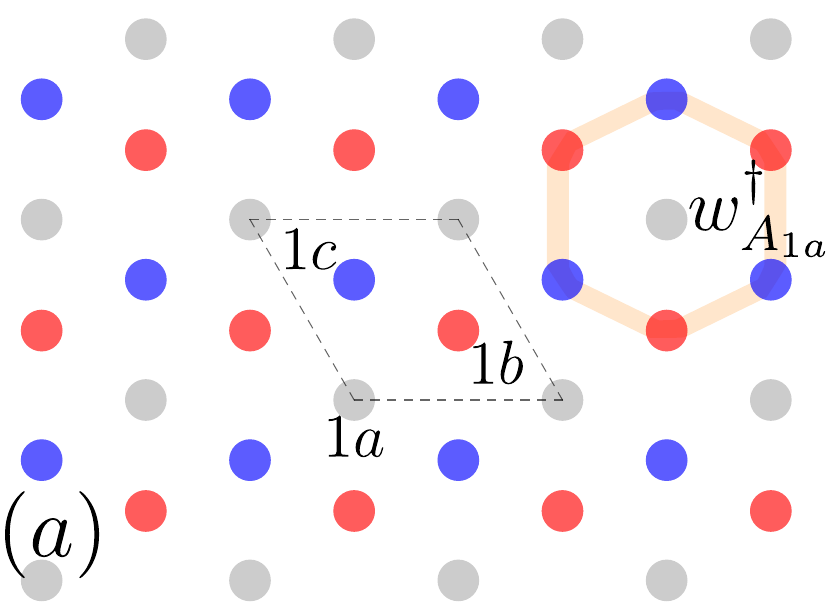}
\end{overpic} \qquad
\begin{overpic}[height=0.18\textwidth,tics=10]{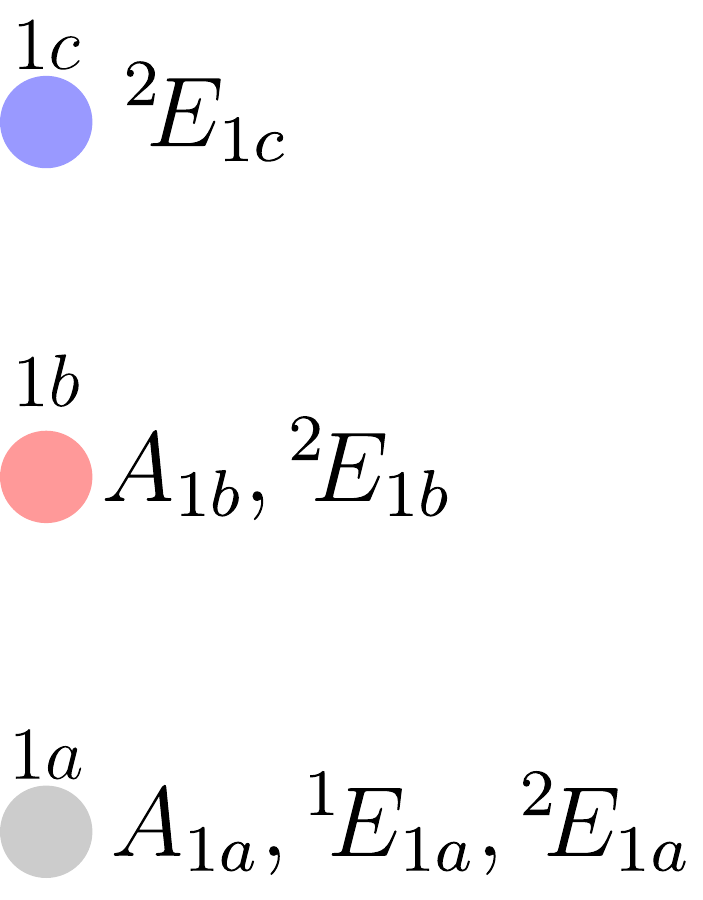}
\end{overpic} \\
\begin{overpic}[height=0.17\textwidth,tics=10]{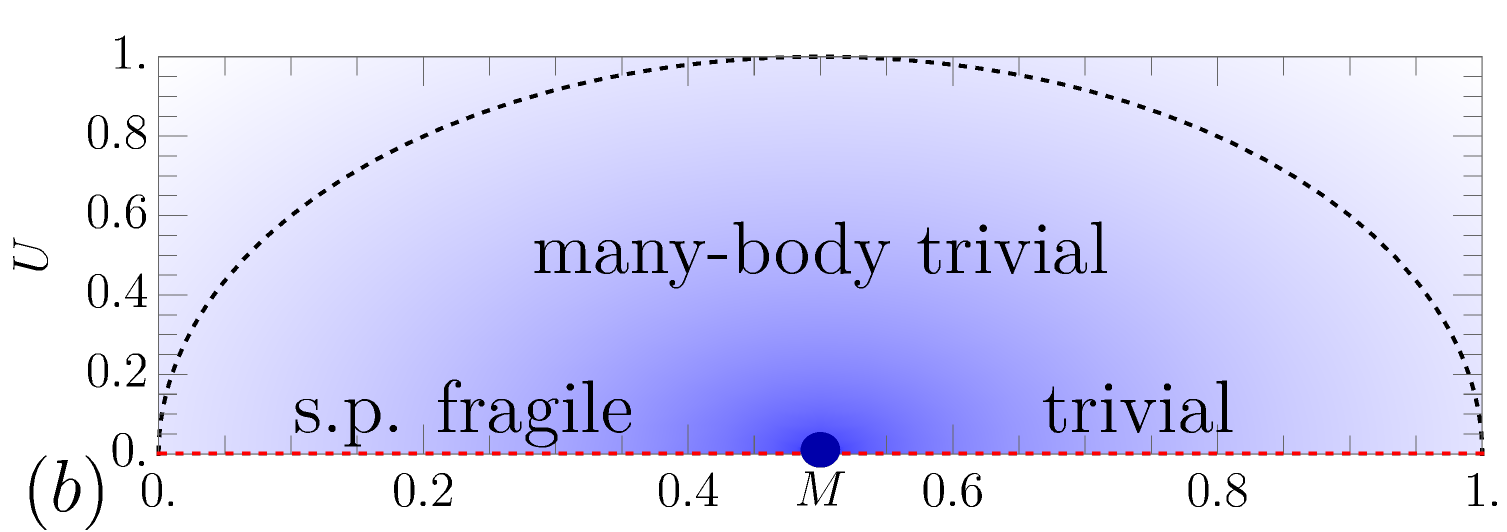}
\end{overpic} 
\caption{Trivializing fragile topology. $(a)$ Orbitals and Wyckoff positions with $w^\dag_{A_{1a}}$ shown in orange. $(b)$ Phase diagram of $H_0 + H_I$ where shading denotes the many-body gap. The only gapless point (blue) occurs at $M= 1/2, U=0$ separating the single-particle fragile and trivial phases at $U=0$. Both phases have the same (trivial) {many-body} RSIs. Along the dashed line, the many-body gap is equal to 1 (see \App{app:modelfragile}). 
}
\label{fig:model}
\end{figure}
If the local RSIs are compatible with a many-body atomic state, our method indicates there is no obstruction to trivialization even if the single-particle RSIs are nontrivial. We now present an exactly solvable model where we adiabatically deform a single-particle fragile state into a single-particle trivial state through a strong coupling region \cite{2010PhRvB..81m4509F}. 

Our strategy is to build a non-interacting Hamiltonian with fragile valence bands and obstructed atomic conduction bands. Importantly, we choose the conduction bands to have Wannier functions which are nonzero on a {finite} number of orbitals \cite{2021arXiv210713556S}. We choose $G=p3$ which has three Wyckoff positions shown in \Fig{fig:model}a. Each has spinless PG 3, whose irreps we denote $A, {}^1\!E, {}^2\!E$ carrying $C_3$ eigenvalues $1, e^{i 2\pi/3}, e^{-i 2\pi/3}$ respectively. We create the atomic orbitals $A_{1b}, {}^2\!E_{1b}, {}^2\!E_{1c}$ and form the state
\bea
w^\dag_{0,A_{1a}} \ket{0}= \frac{1}{3} \sum_{j=0}^2 C_3^j (c^\dag_{0, A_{1b}} + c^\dag_{0, {}^2\!E_{1b}}+ c^\dag_{0, {}^2\!E_{1c}} ) \ket{0}
\eea
in the $\mbf{R} = 0$ unit cell and $w^\dag_{\mbf{R},A_{1a}} = T_\mbf{R} w^\dag_{0,A_{1a}}  T_\mbf{R}^\dag$. $w^\dag_{0,A_{1a}}$ creates an $A$ irrep at 1a (see \Fig{fig:model}a). The complementary two bands on the $A_{1b}, {}^2\!E_{1b}, {}^2\!E_{1c}$ orbitals are fragile and cannot be induced from local orbitals (see \App{app:modelfragile}). We also add the atomic orbitals ${}^1E_{1a}$ and  ${}^2E_{1a}$ to the valence band (which do not {not} trivialize the fragile topology) and a $A_{1a}$ orbital to the conduction band. Note that $A_{1a}$ would trivialize the valence band. To construct the non-interacting Hamiltonian $H_0$, we choose all bands to be perfectly flat so $H_0$ is local in the Wannier basis. As discussed in \App{app:modelfragile}, we set 
\bea
H_{0} &= \sum_{\mbf{R}} (1-M) w^\dag_{\mbf{R},A_{1a}} w_{\mbf{R},A_{1a}} \\
&\qquad + M (n_{\mbf{R},{}^1\!E_{1a}} +n_{\mbf{R},{}^2\!E_{1a}} ) + (1-M) n_{\mbf{R},A_{1a}} 
\eea
where $n_{\mbf{R},\rho} = c^\dag_{\mbf{R},\rho} c_{\mbf{R},\rho}$. All terms in $H_0$ are strictly local because $w_{\mbf{R},A_{1a}}$ is finitely supported. At filling $\nu = N_{occ}/N_{orb} = 4/6$,  $M$ tunes between a fragile phase:
\bea
\label{eq:fraggs}
{}^1\!E_{1a} \oplus {}^2\!E_{1a} \oplus [A_{1b} \oplus {}^2\!E_{1b} \oplus {}^2\!E_{1c}  \ominus A_{1a} ] \ \uparrow G
\eea
for $M\in (0,1/2)$ and a trivial phase $A_{1a} \oplus A_{1b} \oplus {}^2\!E_{1b} \oplus {}^2\!E_{1c} $ for  $M\in (1/2,1)$. The two fragile bands in \Eq{eq:fraggs} are in brackets. A gap closing at $M=1/2$ separates the two phases. However \Tab{tab:mpgrsi} shows the RSIs are the same in both phases: $\Delta_{1a} = (1,0), \ \Delta_{1b} = (2,1), \ \Delta_{1c} = (1,1)$ and indicate a many-body atomic limit (as can be checked using the topological indices in \Tab{tab:fragilenoSOC}). Accordingly, the single-particle fragile phase can be connected to the trivial phase without a gap closing by adding interactions. We now add the symmetry-preserving term
\bea
H_I &= U \sum_{\mbf{R}} w^\dag_{\mbf{R}, A_{1a}} c^\dag_{\mbf{R}, A_{1a}} c_{\mbf{R},{}^1\!E_{1a}}c_{\mbf{R},{}^2\!E_{1a}} + h.c.
\eea
which is strictly local. Physically, $H_I$ implements the interaction-allowed conversion\cite{PhysRevB.99.125122} $ {}^1\!E_{1a} \oplus {}^2\!E_{1a} \to A_{1a} \oplus A_{1a}$ and removes the fragile obstruction symbolized as $\ominus A_{1a}$ in \Eq{eq:fraggs} (but note that $H_I$ annihilates the fragile bands). Because $H_0 + H_I$ acts independently on the Wannier states in each unit cell, the Hamiltonian is entirely decoupled and is trivial to solve \cite{2021arXiv210802414M}. We show the phase diagram in \Fig{fig:model}b and see that the gap closing at $U=0$ separating the single-particle phases can be opened with interactions, adiabatically connecting the phases. 

\section{Many-body Stable Topology}
 Non-interacting stable topological states (such as Chern and quantum spin Hall insulators where many-body invariants are known \cite{PhysRevLett.49.405,2006PhRvB..74s5312F,2021PhRvB.103g5102D,PhysRevX.2.031008,2008PhRvL.100r6807L}) cannot be trivialized by coupling to any local orbitals -- unlike fragile topology. This is reflected in the single-particle RSIs, which take on fractional values in stable states \cite{rsis}. For instance with $C_2$, the real-space derivation of the single-particle RSI $\delta_1 = m(A) - m(B) \in \mathds{Z}$ relies on the existence of Wannier functions such that $m(A),m(B)$ are well-defined integers. It is only by generalizing the definition of $\delta_1$ to momentum space (on periodic boundary conditions) that the possibility of fractional values emerges. In analogy to the non-interacting case, we define a many-body stable topological phase to be robust against coupling to all many-body atomic states. It is impossible to compute local (many-body) RSIs on OBCs in this case because the edge states, a signature of stable topology, prevent our assumption $\mbf{(1)}$ of non-degeneracy and cannot be removed by coupling to any atomic states.

We now propose a definition of {global} RSIs $\Delta^G_{\mbf{x},i}$ in many-body stable topological phases at Wyckoff position $\mbf{x}$ in the unit cell. Their definition is identical to \Eq{eq:RSIsneven} but evaluated on a spatial torus, i.e. periodic boundary conditions (PBCs). Explicitly, with $C_n \in G_\mbf{x}$ for $n$ even, 
\begin{align}
\label{eq:globalrsi}
e^{i \frac{\pi}{n} \hat{N}} C_n  \ket{GS,PBC}  &= e^{i \frac{\pi}{n} \Delta^G_{\mbf{x},1}} \ket{GS,PBC},\nonumber \\
 C_n^2 \ket{GS,PBC}  &= e^{i \frac{2\pi}{n/2} \Delta^G_{\mbf{x},2}} \ket{GS,PBC} 
\end{align}
defines the global RSIs. The PBCs are essential for $\ket{GS,PBC} $ to be non-degenerate so $\Delta^G_{\mbf{x},i}$ are quantum numbers. We claim that if $\Delta^G_{\mbf{x},i} \neq 0$, $\ket{GS,PBC}$ is many-body stable topological. In other words, the global RSIs are topological invariants. 

To support this claim, we prove two properties of $\Delta^G_{\mbf{x},i}$: $\mbf{(1)}$ All 2D many-body atomic phases have $\Delta^G_{\mbf{x},i} = 0$, from which it follows that all many-body fragile topological phases also have $\Delta^G_{\mbf{x},i} = 0$ and $\mbf{(2)}$:  $\Delta^G_{\mbf{x},i}$ is determined by the Chern number $C$ in non-interacting Slater determinant states. This demonstrates the well-known fact that Chern insulators are robust to weak interactions. 

\begin{figure}
 \centering
\begin{overpic}[width=0.3\textwidth,tics=10]{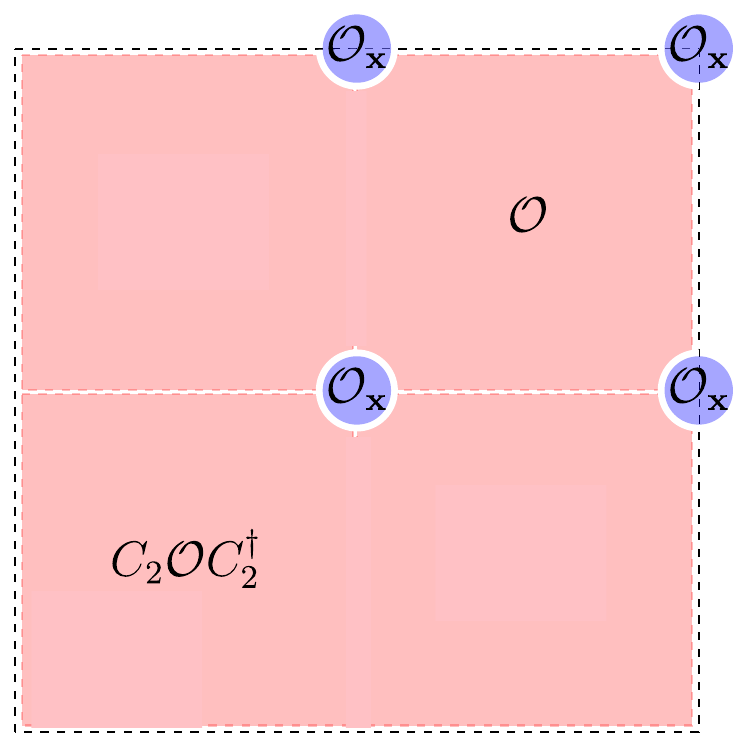}
\end{overpic} 
\caption{Local Operators on Periodic Boundary Conditions. We depict the partitioning of a many-body atomic limit state on a spatial torus (PBCs) into rotation-related operators $\mathcal{O}, C_2 \mathcal{O} C_2^\dag$ and the locally-transforming operators $\mathcal{O}_\mbf{x}$ operators at fixed points of the rotation $\mbf{x} \in \mbf{x}^G$.}
\label{fig:PGglobalmain}
\end{figure}

We will first prove $\mbf{(1)}$. Consider a many-body atomic limit $\prod_{\mbf{R},\mbf{r}} \mathcal{O}_{\mbf{R},\mbf{r}} \ket{0}$ with $G=p2$ generated by $C_2, T_\mbf{R}$ where $C_2 \in G_\mbf{x}$ is a rotation around the point $\mbf{x} = (0,0)$. Now consider placing the state on $L_1\times L_2$ PBCs with $L_1,L_2$ even. Observe that there are four points invariant under $C_2$ denoted $\mbf{x}^G = \{(0,0), (L_1/2,0),(0,L_2/2), (L_1/2,L_2/2) \}$. Since the $C_2$ operator is a symmetry of each point, it protects a local RSI $\Delta_{\mbf{x},1}$ given by $e^{i \frac{\pi}{2} \Delta_{1,\mbf{x}}}\mathcal{O}_\mbf{x}  =  (e^{i \frac{\pi}{2} \hat{N}}C_2) \mathcal{O}_\mbf{x} ( e^{i \frac{\pi}{2} \hat{N}}C_2)^\dag$ for each $\mbf{x} \in \mbf{x}^G$. In fact, the $\Delta_{1,\mbf{x}}$ is the same at each $\mbf{x} \in \mbf{x}^G$ because of translations: $\mathcal{O}_{\mbf{x}} = T_{\mbf{x}} \mathcal{O}_{(0,0)} T_{\mbf{x}}^\dag$ and $C_2 T_{\mbf{x}} C_2^\dag = T_{\mbf{x}}$ since $\mbf{x} = - \mbf{x}$ for $\mbf{x} \in \mbf{x}^G$ on PBCs. 

Now we compute the global RSI of the many-body atomic limit. Using the $C_n$ symmetry, the atomic limit groundstate can generically be written as (see \Fig{fig:PGglobalmain})
\bea
\ket{GS,PBC} = \prod_{\mbf{x} \in \mbf{x}^G}\! \mathcal{O}_\mbf{x} \ \prod_{i=1}^2 C^i_2 \mathcal{O} C_2^{i \dag} \ket{0}  \\
\eea
for some $\mathcal{O}$ which creates the correlated but strictly local groundstates in one half of the spatial torus shown in \Fig{fig:PGglobalmain}. The operators $\mathcal{O}_\mbf{x}$ create the states at the corners of the torus which are locally $C_2$ symmetric. Following \Eq{eq:CnonR'}, we compute
\bea
e^{i \frac{\pi}{2} \hat{N}} C_2 \ket{GS,PBC} =  \prod_{\mbf{x} \in \mbf{x}^G} e^{i \frac{\pi}{2} \Delta_{\mbf{x},1}} \ \ket{GS,PBC} 
\eea
which is intuitive because operators $\mathcal{O}$ off the $C_2$ centers contribute trivially. Using \Eq{eq:globalrsi}, the global RSI of the many-body atomic state is
\bea
\label{eq:gfourL}
\Delta_{\mbf{x},1}^G = \sum_{\mbf{x} \in \mbf{x}^G} \Delta_{\mbf{x},1} = 4 \Delta_{\mbf{x},1} = 0 \mod 4
\eea
since we proved that $\Delta_{\mbf{x},1}$ are all equal. The cancelation shown here is due to unexpected coincidence of the global RSI being defined mod 4, and the 4 $C_2$ symmetric points on PBCs each contributing an equal (integer) local RSI to the global RSIs. We extend this proof to all point groups in \App{app:strong}.

Next to prove $\mbf{(2)}$, we relate $\Delta^G_{\mbf{x},1}$ to the Chern number $C$ in product states. We need two existing results. With $C_2$, $(-1)^C$ is equal to the product of inversion eigenvalues at the high-symmetry points in the Brillouin zone (BZ)\cite{PhysRevB.83.245132, 2020JPCM...32z3001P,2017NatCo...8...50P,2020arXiv200604890C,PhysRevLett.120.096601}, and secondly $C_2 \ket{GS,PBC} = (-1)^C (-1)^{N/2}  \ket{GS,PBC}$ where $N$ is the number of states in the BZ which must be even since $L_1,L_2$ are even \cite{2013PhRvB..87c5119F}. Evaluating the global RSI with \Eq{eq:globalrsi} yields $\Delta_{1a,1}^G = 2C \mod 4$. We can also compute $\Delta^G_{\mbf{x},1}$ at other Wyckoff positions taking e.g. $C_2 \to T_1 C_2$. Then because $\ket{GS,PBC}$ has zero total many-body momentum in 2D, we find
\bea
\label{eq:delta4}
\Delta^G_{1a,1} = \Delta^G_{1b,1} = \Delta^G_{1c,1} = \Delta^G_{1d,1} = 2C \mod 4 \ .
\eea
This result is reminiscent of the half-integer valued single-particle RSIs in Chern insulators: \Eq{eq:gfourL} gives $\Delta^G_{\mbf{x},1} = 4 \Delta_{\mbf{x},1}$ when local RSIs are well-defined, so \Eq{eq:delta4} is suggestive of a half-integer local RSI in odd Chern states. 

We compute the global RSIs of all Slater determinants in \App{app:strong}. Our classification reveals the possibility of many stable topological phases which cannot exist in band theory but are enabled by strong interactions, agreeing with and extending earlier results \cite{2019PhRvX...9c1003L,PhysRevLett.118.216402,PhysRevB.95.205139}. We give a three illustrative examples. In $p2$, a state $\ket{\psi}$ with $\Delta_{1a,1}^G = \Delta_{1d,1}^G = 2$, $\Delta_{1b,1}^G = \Delta_{1c,1}^G = 0$ obeys $e^{i \frac{\pi}{2} \hat{N}} C_n \ket{\psi} = -\ket{\psi}$ like in a Chern state, but with nonzero total momentum $T_1 \ket{\psi} = T_2 \ket{\psi} = - \ket{\psi}$. Such a state carries a many-body Chern number \cite{PhysRevLett.120.096601}, but is adiabatically disconnected from any single-particle Chern state. In $p2mm$, mirrors ensure $C=-C = 0$ without interactions. But the global RSI $\Delta^G_1$ retains a $\mathds{Z}_4$ classification after adding mirrors, allowing an  interaction-enabled state like \Eq{eq:delta4} following the heuristic $2C = - 2C \mod 4$ \cite{2014arXiv1409.1234L}. Finally, \Eq{eq:delta4} shows that $\Delta^G_{\mbf{x},1}$ must be even in gapped Slater determinants. Since an $\Delta^G_{\mbf{x},1}$ odd requires odd particle number but $L_1,L_2$ are even, this suggests a gapless state. This is evidence that odd $\Delta^G_{\mbf{x},1}$, defined by \Eq{eq:globalrsi}, is a many-body semi-metal invariant \cite{PhysRevX.8.031069,2022arXiv221103802H}.

%Finally, \Eq{eq:delta4} also shows that $\Delta^G_{\mbf{x},1}$ must be even in Slater determinants, so $\Delta^G_{\mbf{x},1}$ odd is interaction-enabled. By \Eq{eq:globalrsi}, odd $ \Delta^G_{\mbf{x},1}$ requires an odd number of particles in the groundstate and indicates a gapless state (a correlated topological semimetal) since $L_1,L_2$ are even\cite{PhysRevX.8.031069,2022arXiv221103802H}.

\section{Topological Response Theory} 
RSIs are quantized, symmetry-protected invariants beyond the Chern number. Then, since a nonzero Chern number is encoded in the continuum topological response theory as a Chern-Simons term\cite{PhysRevB.78.195424}, it might be expected that RSIs appear as well. In the presence of crystalline symmetries, the response theory includes Wen-Zee-type terms\cite{PhysRevLett.69.953,PhysRevB.87.165107,2021PhRvR...3a3040M,2022PhRvB.105w5143B,PhysRevD.88.045013,2022PhRvB.105d5112H,PhysRevB.90.014435,2022arXiv220405320Z,2017arXiv170504394M,PhysRevB.91.165306,PhysRevB.91.125303,2021PhRvR...3a3040M,2022PhRvB.105w5143B,barrynew}:
\bea
\label{eq:lag}
\lag &= \frac{C}{4\pi} A \, dA + \frac{s}{2\pi} A \, d \omega + \frac{\ell}{4\pi} \omega \, d \omega 
\eea
where $A = A_\mu dx^\mu$ is the $U(1)$ gauge field and $\omega = \omega_\mu dx^\mu$ is the rotational gauge field, or spin connection \cite{PhysRevB.90.014435,PhysRevX.8.011040,2021PhRvR...3a3040M,2021PhRvX..11d1051E}. Physically, $\int \!d\!A$ and $\int \!d\omega$ are the total flux and total disclination angle. \Eq{eq:lag} neglects translational gauge fields\cite{PhysRevB.87.165107,2022arXiv221109127Z} and hence ignores the unit cell structure, so $\lag$ describes an expansion around a fixed Wyckoff position $\mbf{x}$. We will show that $s$ and $\ell$ are the local RSIs at $\mbf{x}$. 

The coefficients $s$ and $\ell$ can be understood from the equation of motion for the charge density $\rho$ and angular momentum density $L$:
\bea
\label{eq:eom}
\rho &= \frac{\delta \lag}{\delta A_0} = \frac{C}{2\pi}dA  + \frac{s}{2\pi} d\omega, \ L = \frac{\delta \lag}{\delta \omega_0} = \frac{s}{2\pi} dA + \frac{\ell}{2\pi} d\omega \ .
\eea
Let us first consider $\rho$. If $d\omega = 0$,  \Eq{eq:eom} reduces to the Streda formula\cite{1982JPhC...15L.717S}. If $dA = 0$,  $s$ describes the charge bound to a disclination center. A partial disclination with $s\neq 0$ reveals the fractional charge at $\mbf{x}$\cite{2020PhRvB.101k5115L,2021arXiv210800008M,2022arXiv220405320Z,2022arXiv220406268J}. As such, taking $\int d\omega = 2\pi$ to be a complete disclination (in analogy to inserting a full flux), \Eq{eq:eom} shows $s = \int \rho $ is the total charge at $\mbf{x}$. Since the total charge is related to the local RSIs, \Eq{eq:RSIsneven} gives (for $n$ even)
\bea
\label{eq:shifttoRSI}
e^{i \frac{2\pi}{n} s} = e^{i \frac{2\pi}{n} \hat{N}} = (e^{i \frac{\pi}{n} \hat{N}}C_n)^2 (C_n^2)^\dag = e^{i \frac{2\pi}{n}(\Delta_{1,\mbf{x}} - 2 \Delta_{2,\mbf{x}})}
\eea 
acting on $\ket{GS}$ with OBCs. Hence $s$ is the local charge
\bea
\label{eq:sRSIs}
s = \Delta_{1,\mbf{x}} - 2\Delta_{2,\mbf{x}} \mod n \quad \text{($n$ even)}  \ .
\eea
We remark that the physical charge bound to the disclination core is a well-defined local observable and can take any (rational) value. However, \Eq{eq:sRSIs} shows that its value mod $n$ is determined solely by the many-body RSIs of the defect-less groundstate and is universal. This can be understood from the Lagrangian in \Eq{eq:lag}: although $\lag$ is well-defined for all $s$, adiabatic deformations of the groundstate leave only $s \mod n$ constant. In many-body fragile or atomic states where $\Delta_{i,\mbf{x}}$ are integers, $s \in \mathds{Z}_n$. \Eq{eq:sRSIs} suggests a straightforward generalization to Chern states. Without interactions, the single-particle local RSIs are well-defined but fractional, and a formula for the charge $s$ at $\mbf{x}$ is known \cite{rsis,2022arXiv220910559H}. Proving the many-body extension of this result will be the subject of forthcoming work. For now in the $C_2$ case, note that \Eq{eq:gfourL} (proved only at $C=0$) shows $4 \Delta_{\mbf{x},1} = \Delta^G_{\mbf{x},1}$, while \Eq{eq:delta4} shows $\Delta^G_{\mbf{x},1} = 2 C \mod 4$. Then at least heuristically, the local RSI $\Delta_{\mbf{x},1} = C/2 \mod 4$ can be half-integer in agreement with Ref. \cite{2022arXiv220405320Z}.  

We now consider $L$ in \Eq{eq:eom}. Setting $d\omega =0$ shows that $s$ shifts the angular momentum after inserting a full flux. Indeed, single-particle RSIs can enforce irrep flow due to angular momentum pumping in flux\cite{2022arXiv220910559H,firstpaper}. Setting $dA = 0$ shows that $\ell$ describes the angular momentum bound to a disclination center and should be identified with the eigenvalue of the rotation operator $e^{i \frac{\pi}{n} \hat{N}} C_n$. Hence we propose $\ell = \Delta_{1,\mbf{x}}\in  \mathds{Z}_{2n}$ for $n$ even, matching the classification of Ref. \cite{2022arXiv220405320Z}. 
%For $n$ odd, the local RSIs $e^{i \frac{2\pi}{n} \Delta_{i,\mbf{x}}}$ are the eigenvalues of $C_n, e^{i \frac{2\pi}{n} \hat{N}}$ respectively, giving $s = \Delta_{2,\mbf{x}}$ and $\ell = \Delta_{1,\mbf{x}}$. 

\section{Discussion} 

TQC is a unifying theory of non-interacting materials. Given atomic orbitals, their symmetries, and the number of electrons, the topological invariants of TQC classify possible gapped, degenerate, and gapless phases. The present work achieves the first case in interacting Hamiltonians by defining local and global RSIs, many-body topological indices, and the effective field theory that governs them. In so doing, we revealed which single-particle fragile phases survive interactions and identified undiscovered stable topological states with no single-particle counterpart. We anticipate that many features of single-particle topology may be generalized to the many-body case using this formalism, for instance regarding bounds on quantum geometry \cite{2023PhRvB.107l5116K,2022PhRvL.128h7002H, PhysRevB.108.L140503,2021PNAS..11806744V,2023PNAS..12017816M}. Another perspective is offered by Ref. \cite{PhysRevLett.120.096601}, which shows that a generalization of the single-particle Brillouin zone is the flux torus obtained from twisted boundary conditions (on which the gapped, many-body groundstate can smoothly be defined), since in both cases the Berry curvature and symmetry eigenvalues can be defined. This connection may facilitate the computation of many-body RSIs without open boundary conditions and reveal connections between the momentum space and real space theories. We leave the study of global and magnetic symmetries, spontaneous symmetry breaking, the numerical investigation of interaction-enabled many-body stable topology, and the study of boundary signatures to future work.

\section{Acknowledgements.} We thank Adrian Po for influential comments in the early stages of this work and Abhinav Prem for his insight and careful explanation.
Additionally we thank Frank Schindler for useful guidance, Jiabin Yu, and Biao Lian for enlightening consultations, and Titus Neupert, Glenn Wagner, and Martina Soldini for sharing their unpublished paper on many-body atomic limits in the final states of this work. JHA thanks the Donostia International Physics Center for their hospitality during the completion of this manuscript. 
Funding: Z.-D. S. were supported by National Key Research and Development Program of China (No. 2021YFA1401903), National Natural Science Foundation of China (General Program No. 12274005), and Innovation Program for Quantum Science and Technology (No. 2021ZD0302403). B.A.B. and Z-D.S. were supported by the European Research Council (ERC) under the European Union's Horizon 2020 research and innovation programme (grant agreement No. 101020833), the ONR Grant No. N00014-20-1-2303, the Schmidt Fund for Innovative Research, Simons Investigator Grant No. 404513, the Packard Foundation, the Gordon and Betty Moore Foundation through the EPiQS Initiative, Grant GBMF11070 and Grant No. GBMF8685 towards the Princeton theory program. Further support was provided by the NSF-MRSEC Grant No. DMR-2011750, BSF Israel US foundation Grant No. 2018226, and the Princeton Global Network Funds. JHA is supported by a Hertz Fellowship and ONR Grant No. N00014-20-1-2303, as well as a Marshall Scholarship  during the early stages of this project.

\section{Inclusion and Ethics.} A portion of this work was completed at Princeton University, which occupies part of the ancient homeland and traditional territory of the Lenape people. We pay respect to Lenape peoples past, present, and future and their continuing presence in the homeland and throughout the Lenape diaspora.

%%%%%%%%%%%%%%%%%%%%%%%%%%%%%%%%%%
\let\oldaddcontentsline\addcontentsline% Store \addcontentsline
\renewcommand{\addcontentsline}[3]{}% Make \addcontentsline a no-op
%%%
\bibliography{finalbib}
\bibliographystyle{aipnum4-1}
\bibliographystyle{unsrtnat}
%%%
\let\addcontentsline\oldaddcontentsline
%%%%%%%%%%%%%%%%%%%%%%%%%%%%%%%%%%
\setcitestyle{numbers,square}

\onecolumngrid
\appendix

\newpage

\tableofcontents

\section{Symmetries in Interacting Hamiltonians}
\label{app:definitions}

In this Appendix, we lay out our notations for the Hamiltonians, symmetries, and irreducible representations (irreps) defined in this work. \Tabs{tab:2D-char}{tab:2D-charTRS} list the character tables in all point groups (PGs) with and without time-reversal symmetry.  

Let $\mbf{a}_1,\mbf{a}_2$ be the 2D Bravais lattice vectors, which we normalize for convenience by requiring $\mbf{a}_1\times \mbf{a}_2 = +1$, fixing the area of the unit cell to 1.  Greek indices $\al =1, \dots, N_{orb}$ denote the orbitals of the unit cell (i.e. the local Hilbert space dimension). We find it convenient to work in an index convention where $c^\dag_{\mbf{R},\al}$ is the creation operator in the $\mbf{R} = r_1 \mbf{a}_1 + r_2 \mbf{a}_2, r_i \in \mathds{Z}$ unit cell of orbital $\al$, which is located at $\mbf{R} + \pmb{\delta}_{\al}$. We raise and lower indices on the electron operators to save space in some expressions.

We now discuss crystalline symmetries which form the space group (also called a wallpaper group in two spatial dimensions) $G$. Comprehensive details of the space group may be found in Ref. \cite{2020arXiv200604890C, 2017Natur.547..298B}. In brief, $G$ contains an infinite subgroup composed of the translations $T_i$ and a finite set of generators of rotations and in-plane mirrors which yield the point groups (PGs). In this work, we restrict ourselves to these symmorphic (PG) symmetries. $G$ may also contain time-reversal $\mathcal{T}$, but we do not consider magnetic space groups explicitly here. However, one can extend our classification to this case.

A space group symmetry $g\in G$ is defined by its action on the election operators which reads
\bea
\label{eq:defg}
g^\dag c_{\mbf{R},\al} g &= \sum_\be D[g]^\al_{\be} c_{g(\mbf{R} + \pmb{\delta}_\al) - \pmb{\delta}_{\be},\be} \ .
\eea
where $D[g]^\al_{\be}$, the $N_{orb} \times N_{orb}$ representation matrix of $g$ on the orbitals, is nonzero only when $g\pmb{\delta}_\al = \pmb{\delta}_\be \mod \mbf{a}_i$. Hence $g(\mbf{R} + \pmb{\delta}_\al) - \pmb{\delta}_{\be}$ is a lattice vector and \Eq{eq:defg} is well-defined. Spinless representations describe systems without spin-orbit coupling (SOC) and have $D[C_n]^n = D[M]^2 = +1$ and spinful systems with SOC satisfy $D[C_n]^n = D[M]^2 = -1$, where $C_n$ is an $n$-fold rotation and $M$ is a mirror. In groups with time-reversal $\mathcal{T}$, spinless particles have $\mathcal{T}^2 = +1$ and spinful particles have $\mathcal{T}^2 = -1$. The representation of $\mathcal{T}$ on the orbitals is denoted $D[\mathcal{T}]K$ where $K$ is the complex conjugation operator, and hence $D[\mathcal{T}]D^*[\mathcal{T}] = \pm1$ for spinless/spinful particles. The irreps in all 2D PGs with and without SOC, with and without time-reversal are shown in \Tabs{tab:2D-char}{tab:2D-charTRS}. They can also be found, with full group theory data, on the Bilbao Crystallographic Server \cite{Aroyo:firstpaper,Aroyo:xo5013} as the $k_z = 0$ slice of the 3D layer groups. 

We can now write down a Hamiltonian $H$ with arbitrary interactions that preserve the space group symmetries. We decompose $H = \sum_q H_q$ where $H_q$ is a ``$q$-body'' term. We call $H$ interacting if it contains $2$-body or higher terms. Explicitly, the $q$-body term is written 
\bea
\label{eq:HamHq}
H_q &= \sum_{\mbf{R}'_1 \dots \mbf{R}'_q,\mbf{R}_1\dots \mbf{R}_1 }  t^{\al_1 \dots \al_q}_{\be_1 \dots \be_q}(\{\mbf{R}',\mbf{R} \}) c^\dag_{\mbf{R}'_1,\al_1} \dots c^\dag_{\mbf{R}_q',\al_q} c_{\mbf{R}_1,\be_1} \dots  c_{\mbf{R}_q,\be_q} \\
\eea
where we have summed over the repeated greek indices. The tensor $t$ obeys
\bea
\text{fermion anti-symmetry}: \ & t^{\al_2 \al_1  \dots \al_q}_{\be_1 \dots \be_q}(\mbf{R}'_2 \mbf{R}_1', \dots) = - t^{\al_1 \al_2  \dots \al_q}_{\be_1 \dots \be_q}(\mbf{R}_1' \mbf{R}'_2 \dots), \\
\text{Hermiticity}:\  & \lp t^{\al_1 \dots \al_q}_{\be_1 \dots \be_q}(\{\mbf{R}',\mbf{R}\}) \rp^* = t_{\al_1 \dots \al_q}^{\be_1 \dots \be_q}(\{\mbf{R},\mbf{R}'\}) \ . 
\eea
By imposing translation invariance, we find
\bea
\label{eq:trans}
t^{\al_1 \dots \al_q}_{\be_1 \dots \be_q}(\{\mbf{R}',\mbf{R}\}) &= t^{\al_1 \dots \al_q}_{\be_1 \dots \be_q}(\{\mbf{R}'+\mbf{a}_i,\mbf{R}+\mbf{a}_i\}) \ .
\eea
Continuing to impose that $g \in G$ is a symmetry of $H_q$ for general elements of the space group, we require
\bea
g^\dag H_q g &= \sum_{\{\mbf{R},\mbf{R}'\}} t^{\al_1 \dots \al_q}_{\be_1 \dots \be_q}(\{\mbf{R}',\mbf{R}\}) g^\dag c^\dag_{\mbf{R}'_1,\al_1} g \dots g^\dag c^\dag_{\mbf{R}_q',\al_q} g g^\dag c_{\mbf{R}_1}^{\be_1} g\dots  g^\dag c_{\mbf{R}_q}^{\be_q} g \\
%&= \sum_{\{\mbf{R},\mbf{R}'\}} ([D^\dag]_{\al_1}^{\al'_1} \dots )t^{\al_1 \dots \al_q}_{\be_1 \dots \be_q}(\{\mbf{r}',\mbf{r}\})  (D^{\be_1}_{\be'_1} \dots)  c^\dag_{g \mbf{r}',\al_1'} \dots c^\dag_{g\mbf{r}_q',\al'_q} c_{g\mbf{r}}^{\be'_1} \dots c_{g\mbf{r}_q}^{\be'_q} \\
&= \sum_{\{\mbf{R},\mbf{R}'\}} (D^\dag[g]_{\al_1}^{\al'_1} \dots  D^\dag[g]_{\al_q}^{\al'_q}  )t^{\al_1 \dots \al_q}_{\be_1 \dots \be_q}(\{g^{-1}(\mbf{R}'+ \pmb{\delta}_{\al'}) - \pmb{\delta}_{\al},g^{-1}(\mbf{R} + \pmb{\delta}_{\be'}) - \pmb{\delta}_{\be} \})  (D[g]^{\be_1}_{\be'_1} \dots)  c^\dag_{\mbf{R}_1',\al_1'} \dots c^\dag_{\mbf{R}_q',\al'_q} c_{\mbf{R}_1}^{\be'_1} \dots c_{\mbf{R}_q}^{\be'_q} \\
\eea
The $\al_1,\dots, \al_q$ and $\be_1,\dots, \be_q$ indices are fully anti-symmetric because they contract with the fully anti-symmetric $t_{\be_1,\dots,\be_q}^{\al_1,\dots,\al_q}$ tensor, and the  $\al'_1,\dots, \al'_q$ and $\be'_1,\dots, \be'_q$ indices are fully anti-symmetric because they contract with the electron operators. Now using $g H_q g^\dag = H_q$, we obtain
\bea
( D^\dag[g]_{\al_1}^{\al'_1} \dots D^\dag[g]_{\al_q}^{\al'_q} ) t^{\al_1 \dots \al_q}_{\be_1 \dots \be_q}(\{\mbf{R}'+ \pmb{\delta}_{\al'}) - \pmb{\delta}_{\al},g^{-1}(\mbf{R} + \pmb{\delta}_{\be'}) - \pmb{\delta}_{\be} \}\}(D[g]^{\be_1}_{\be'_1} \dots D[g]^{\be_q}_{\be'_q} ) &= t^{\al'_1 \dots \al'_q}_{\be'_1 \dots \be'_q}(\{\mbf{R}',\mbf{R}\})
\eea
so $t$ transforms in the representation $\bigwedge_{i = 1}^q D[g] =D[g] \wedge \dots  \wedge D[g]$. Here $\wedge$ denotes the exterior (anti-symmetric) product which anti-symmetrizes $D^\dag[g]_{\al_1}^{\al'_1} \dots D^\dag[g]_{\al_q}^{\al'_q}$ in its top and bottom indices. Time reversal behaves similarly. Repeating the calculation, we find
%\bea
%\mathcal{T}^{-1} H^{(q)} \mathcal{T} &= \sum \bar{t}^{\al_1 \dots \al_q}_{\be_1 \dots \be_q}(\{\mbf{r}',\mbf{r}\}) \mathcal{T}^{-1} c^\dag_{\mbf{R}',\al_1} \mathcal{T} \dots \mathcal{T}^{-1} c^\dag_{\mbf{R}_q',\al_q} \mathcal{T} \mathcal{T}^{-1} c_{\mbf{R}}^{\be_1} \mathcal{T}\dots  \mathcal{T}^{-1} c_{\mbf{R}_q}^{\be_q} \mathcal{T} \\
%&= \sum ([D^\dag]_{\al_1}^{\al'_1} \dots )\bar{t}^{\al_1 \dots \al_q}_{\be_1 \dots \be_q}(\{\mbf{r}',\mbf{r}\})  (D^{\be_1}_{\be'_1} \dots)  c^\dag_{\mbf{r}',\al_1'} \dots c^\dag_{\mbf{r}_q',\al'_q} c_{\mbf{r}}^{\be'_1} \dots c_{\mbf{r}_q}^{\be'_q} \\
%\eea
%with $\mathcal{T}^{-1} c^\dag_{\mbf{R},\al} \mathcal{T} = D^\be_\al c^\dag_{\mbf{R}, \be}$. For $\mathcal{T}$ to be a symmetry of the Hamiltonian, it must be that
\bea
( D^\dag[\mathcal{T}]_{\al_1}^{\al'_1} \dots D^\dag[\mathcal{T}]_{\al_q}^{\al'_q} ) {t^*}^{\al_1 \dots \al_q}_{\be_1 \dots \be_q}(\{\mbf{R}',\mbf{R}\})(D[\mathcal{T}]^{\be_1}_{\be'_1} \dots D[\mathcal{T}]^{\be_q}_{\be'_q} ) &= t^{\al'_1 \dots \al'_q}_{\be'_1 \dots \be'_q}(\{\mbf{R}',\mbf{R}\}) \ .
\eea
Finally, we remark that our expression for $H$ in \Eq{eq:HamHq} is particle-number preserving. By construction, $H$ commutes with the total $U(1)$ charge operator 
\bea
U(\theta) &= e^{i \th \hat{N}}, \quad \hat{N} =  \sum_{\mbf{R}} c^\dag_{\mbf{R},\al} c_{\mbf{R},\al} \
\eea
for all $\th$, and we have set the electron charge to 1. $U(\th)$ generates the $U(1)$ global symmetry and allows us to fix the total $U(1)$ particle number to $N$. The filling in the thermodynamic limit is $\nu = N/A = N_{occ}/N_{orb}$ where $A$ is the number of unit cells, $N_{orb}$ is the number of orbitals per unit cell, and $N_{occ}$ is the number of electrons per unit cell.

\begin{table}[H]
\centering
~~~~~
\begin{tabular}{c|cc}
PG $2$ & 1 &  $2$ \\
\hline
$A$ & 1 & 1  \\
$B$ & 1 & $-1$  \\
\hline
${}^2\ovl{E}$ & 1 & $-i$ \\
${}^1\ovl{E}$ & 1 & $i$
\end{tabular}
~~~~~
\begin{tabular}{c|c|cc}
\multicolumn{2}{c|}{PG $m$} & 1 & $m$ \\
\hline
$A'$ & $A'$ & 1 & 1 \\
$A''$ & $A''$ & 1 & $-1$ \\
\hline
$^2E_{1/2}$ & ${}^2\ovl{E}$ & 1 & $-i$ \\
$^1E_{1/2}$ & ${}^1\ovl{E}$ & 1 & $i$
\end{tabular}
~~~~~
\begin{tabular}{c|c|cccc}
\multicolumn{2}{c|}{PG $2mm$} & 1 & $2$ & $m_{100}$ & $m_{010}$\\
\hline
$A_1$ & $A_1$ & 1 & 1 & 1 & 1\\
$A_2$ & $A_2$ & 1 & 1 & $-1$ & $-1$\\
$B_1$ & $B_1$ & 1 & $-1$ & $-1$ & 1\\
$B_2$ & $B_2$ & 1 & $-1$ & 1 & $-1$\\
\hline
$E_{1/2}$ & $\ovl{E}$ & 2 & 0 & 0 & 0
\end{tabular}
~~~~~
\begin{tabular}{c|c|cccc}
\multicolumn{2}{c|}{PG $4$} & 1 & $4^+$ & $2$ & $4^-$\\
\hline
$A$ & $A$ & 1 & 1 & 1 & 1\\
$B$ & $B$ & 1 & $-1$ & 1 & $-1$\\
$^1E$ & $^1E$ & 1 & $-i$ & $-1$ & $i$\\
$^2E$ & $^2E$ & 1 &  $i$ & $-1$ & $-i$\\
\hline
$^1E_{1/2}$ & $^2\ovl{E}_1$ & 1 & $e^{-i\frac{\pi}4}$ & $-i$ & $e^{i\frac{\pi}4}$\\
$^2E_{1/2}$ & $^1\ovl{E}_1$ & 1 & $e^{i\frac{\pi}4}$ & $i$ & $e^{-i\frac{\pi}4}$\\
$^1E_{3/2}$ & $^2\ovl{E}_2$ & 1 & $e^{i\frac{3\pi}4}$ & $-i$ & $e^{-i\frac{3\pi}4}$\\
$^2E_{3/2}$ & $^1\ovl{E}_2$ & 1 & $e^{-i\frac{3\pi}4}$ & $i$ & $e^{i\frac{3\pi}4}$
\end{tabular}
\vspace{0.5cm}

\begin{tabular}{c|c|ccccc}
\multicolumn{2}{c|}{PG $4mm$} & 1 & $\{4^+,4^-\}$ & $2$ & $\{m_{010},m_{100}\}$ & $\{m_{110},m_{1-10}\}$\\
\hline
$A_1$ & $A_1$ & 1 & 1 & 1 & 1 & 1\\
$A_2$ & $A_2$ & 1 & 1 & 1 &$-1$ &$-1$\\
$B_1$ & $B_1$ & 1 &$-1$ & 1 & 1 & $-1$\\
$B_2$ & $B_2$ & 1 &$-1$ & 1 &$-1$ & 1\\
$E$ & $E$ & 2 & 0 &$-2$ & 0 & 0 \\ 
\hline
$E_{1/2}$ & $\ovl{E}_1$ & 2 & $\sqrt2$ & 0 & 0 & 0\\
$E_{3/2}$ & $\ovl{E}_2$ & 2 & $-\sqrt2$ &0 & 0 & 0
\end{tabular}
~~~~~
\begin{tabular}{c|c|ccc}
\multicolumn{2}{c|}{PG 3} & 1 & $3^+$ & $3^-$ \\
\hline
$A$ & $A$ & 1 & 1 & 1 \\
$^2E$ & $^2E$ & 1 & $e^{i\frac{2\pi}3}$ & $e^{-i\frac{2\pi}3}$ \\
$^1E$ & $^1E$ & 1 & $e^{-i\frac{2\pi}3}$ & $e^{i\frac{2\pi}3}$ \\
\hline
$A_{3/2}$ & $\ovl{E}$ & 1 & $-1$ & $-1$ \\
$^2E_{1/2}$ & ${}^1\ovl{E}$ & 1 & $e^{-i\frac{\pi}3}$ & $e^{i\frac{\pi}3}$ \\
$^1E_{1/2}$ & ${}^2\ovl{E}$ & 1 & $e^{i\frac{\pi}3}$ & $e^{-i\frac{\pi}3}$
\end{tabular}
\vspace{0.5cm}

\begin{tabular}{c|c|ccc}
\multicolumn{2}{c|}{PG $3m$} & 1 & $\{3^+,3^-\}$ & $\{m_{120},m_{210},m_{1-10}\}$ \\
\hline
$A_1$ & $A_1$ & 1 & 1 & 1 \\
$A_2$ & $A_2$ & 1 & 1 &$-1$ \\
$E$   & $E$   & 2 &$-1$ & 0 \\
\hline
$E_{1/2}$ & $\ovl{E}_1$ & 2 & 1 & 0 \\
$^1E_{3/2}$ & ${}^1\ovl{E}$ & 1 &$-1$ & $i$ \\
$^2E_{3/2}$ & ${}^2\ovl{E}$ & 1 &$-1$ & $-i$
\end{tabular}
~~~~~
\begin{tabular}{c|c|cccccc}
\multicolumn{2}{c|}{PG 6} & 1 & $6^+$ & $3^+$ & $2$ & $3^-$ & $6^-$ \\
\hline
$A$ & $A$ & 1 & 1 & 1 & 1 & 1 & 1\\
$B$ & $B$ & 1 &$-1$ & 1 &$-1$ & 1 &$-1$\\
$^1E_1$ & $^1E_2$ & 1 & $e^{-i\frac{\pi}3}$ & $e^{-i\frac{2\pi}3}$ &$-1$ & $e^{i\frac{2\pi}3}$ & $e^{i\frac{\pi}3}$\\
$^2E_1$ & $^2E_2$ & 1 & $e^{i\frac{\pi}3}$ & $e^{i\frac{2\pi}3}$ &$-1$ & $e^{-i\frac{2\pi}3}$ & $e^{-i\frac{\pi}3}$\\
$^1E_2$ & $^1E_1$ & 1 & $e^{i\frac{2\pi}3}$ & $e^{i\frac{4\pi}3}$ & 1 & $e^{i\frac{2\pi}3}$ & $e^{i\frac{4\pi}3}$\\
$^2E_2$ & $^2E_1$ & 1 & $e^{-i\frac{2\pi}3}$ & $e^{-i\frac{4\pi}3}$ & 1 & $e^{-i\frac{2\pi}3}$ & $e^{-i\frac{4\pi}3}$\\
\hline
$^1E_{1/2}$ & ${}^2\ovl{E}_3$ & 1 & $e^{i\frac{\pi}6}$ & $e^{i\frac{\pi}3}$ & $i$ & $e^{-i\frac{\pi}3}$ & $e^{-i\frac{\pi}6}$\\
$^2E_{1/2}$ & ${}^1\ovl{E}_3$ & 1 & $e^{-i\frac{\pi}6}$ & $e^{-i\frac{\pi}3}$ & $-i$ & $e^{i\frac{\pi}3}$ & $e^{i\frac{\pi}6}$\\
$^1E_{3/2}$ & ${}^2\ovl{E}_1$ & 1 & $-i$ & $-1$ & $i$ & $-1$ & $i$\\
$^2E_{3/2}$ & ${}^1\ovl{E}_1$ & 1 & $i$ & $-1$ & $-i$ & $-1$ & $-i$\\
$^1E_{5/2}$ & ${}^2\ovl{E}_2$ & 1 & -$e^{i\frac{\pi}6}$ & $e^{i\frac{\pi}3}$ & $-i$ & $e^{-i\frac{\pi}3}$ & -$e^{-i\frac{\pi}6}$ \\
$^2E_{5/2}$ & ${}^1\ovl{E}_2$ & 1 & -$e^{-i\frac{\pi}6}$ & $e^{-i\frac{\pi}3}$ & $i$ & $e^{i\frac{\pi}3}$ & -$e^{i\frac{\pi}6}$
\end{tabular}
\vspace{0.5cm}

\begin{tabular}{c|c|cccccc}
\multicolumn{2}{c|}{PG $6mm$} & 1 & $\{6^+,6^-\}$ & $\{3^+,3^-\}$ & $2$ & $\{m_{100}, m_{010}, m_{110}\}$ & $\{m_{120},m_{210},m_{1-10}\}$ \\
\hline
$A_1$ & $A_1$ & 1 & 1 & 1 & 1 & 1 & 1\\
$A_2$ & $A_2$ & 1 & 1 & 1 & 1 &$-1$ &$-1$\\
$B_1$ & $B_1$ & 1 &$-1$ & 1 &$-1$ &$-1$ & 1\\
$B_2$ & $B_2$ & 1 &$-1$ & 1 &$-1$ & 1 &$-1$\\
$E_1$ & $E_1$ & 2 & 1 &$-1$ &$-2$ & 0 & 0\\
$E_2$ & $E_2$ & 2 &$-1$ &$-1$ & 2 & 0 & 0\\
\hline
$E_{1/2}$ & $\ovl{E}_1$ & 2 & $\sqrt3$ & 1 & 0 & 0 & 0 \\
$E_{3/2}$ & $\ovl{E}_3$ & 2 & 0 &$-2$ & 0 & 0 & 0 \\
$E_{5/2}$ & $\ovl{E}_2$ & 2 & $-\sqrt3$ & 1 & 0 & 0 & 0 \\
\end{tabular}
\caption{The character tables of 2D magnetic PGs. The irrep names are shown in the first column using the Altmann-Herzig notation and second columns using the Bilbao Crystallographic notation. However, in $p6$, the notation differs from the BCS but is consistent with Ref. \cite{rsis}. In each table, the linear irreps above the second horizontal line are the single-valued irreps (no-SOC), and the projective irreps below the second horizontal line are the double-valued irreps (SOC).
\label{tab:2D-char}
}
\end{table}

\begin{table}[H]
\centering
~~~~~
\begin{tabular}{c|cc}
PG $21'$ & 1 &  $2$ \\
\hline
$A$ & 1 & 1  \\
$B$ & 1 & $-1$ \\
\hline
${}^1\ovl{E}{}^2\ovl{E}$ & 2 & $0$ \\
\end{tabular}
~~~~~
\begin{tabular}{c|c|cc}
\multicolumn{2}{c|}{PG $m1'$} & 1 & $m$ \\
\hline
$A'$ & $A'$ & 1 & 1 \\
$A''$ & $A''$ & 1 & $-1$ \\
\hline
$^2E_{1/2} {}^1E_{1/2} $ & ${}^2\ovl{E} {}^1\ovl{E}$ & 2 & $0$ \\
\end{tabular}
~~~~~
\begin{tabular}{c|c|cccc}
\multicolumn{2}{c|}{PG $2mm1'$} & 1 & $2$ & $m_{100}$ & $m_{010}$\\
\hline
$A_1$ & $A_1$ & 1 & 1 & 1 & 1\\
$A_2$ & $A_2$ & 1 & 1 & $-1$ & $-1$\\
$B_1$ & $B_1$ & 1 & $-1$ & $-1$ & 1\\
$B_2$ & $B_2$ & 1 & $-1$ & 1 & $-1$\\
\hline
$E_{1/2}$ & $\ovl{E}$ & 2 & 0 & 0 & 0
\end{tabular}
~~~~~
\begin{tabular}{c|c|cccc}
\multicolumn{2}{c|}{PG $41'$} & 1 & $4^+$ & $2$ & $4^-$\\
\hline
$A$ & $A$ & 1 & 1 & 1 & 1\\
$B$ & $B$ & 1 & $-1$ & 1 & $-1$\\
${}^1E{}^2E$ & ${}^1E{}^2E$ & 2 & $0$ & $-2$ & $0$\\
\hline
${}^1E_{1/2} {}^2E_{1/2}$ & ${}^1\ovl{E}_1{}^2\ovl{E}_1$ & 2 & $\sqrt{2}$ & $0$ & $\sqrt{2}$\\
${}^1E_{3/2} {}^2E_{3/2}$ & ${}^1\ovl{E}_2 {}^2\ovl{E}_2$ & 2 & $- \sqrt{2}$ & $0$ & $-\sqrt{2}$\\
\end{tabular}
\vspace{0.5cm}
~~~~~
\begin{tabular}{c|c|ccccc}
\multicolumn{2}{c|}{PG $4mm1'$} & 1 & $\{4^+,4^-\}$ & $2$ & $\{m_{010},m_{100}\}$ & $\{m_{110},m_{1-10}\}$\\
\hline
$A_1$ & $A_1$ & 1 & 1 & 1 & 1 & 1\\
$A_2$ & $A_2$ & 1 & 1 & 1 &$-1$ &$-1$\\
$B_1$ & $B_1$ & 1 &$-1$ & 1 & 1 & $-1$\\
$B_2$ & $B_2$ & 1 &$-1$ & 1 &$-1$ & 1\\
$E$ & $E$ & 2 & 0 &$-2$ & 0 & 0 \\ 
\hline
$E_{1/2}$ & $\ovl{E}_1$ & 2 & $\sqrt2$ & 0 & 0 & 0\\
$E_{3/2}$ & $\ovl{E}_2$ & 2 & $-\sqrt2$ &0 & 0 & 0
\end{tabular}
~~~~~
\begin{tabular}{c|c|ccc}
\multicolumn{2}{c|}{PG $31'$} & 1 & $3^+$ & $3^-$ \\
\hline
$A$ & $A$ & 1 & 1 & 1 \\
${}^1E{}^2E$ & ${}^1E{}^2E$ & 2 & $-1$ & $-1$ \\
\hline
$A_{3/2}A_{3/2}$ & $\ovl{EE}$ & 2 & $-2$ & $-2$ \\
${}^2E_{1/2}{}^1E_{1/2}$ & ${}^1\ovl{E}{}^2\ovl{E}$ & 2 & $1$ & $1$ \\
\end{tabular}
\vspace{0.5cm}
~~~~~
\begin{tabular}{c|c|ccc}
\multicolumn{2}{c|}{PG $3m1'$} & 1 & $\{3^+,3^-\}$ & $\{m_{120},m_{210},m_{1-10}\}$ \\
\hline
$A_1$ & $A_1$ & 1 & 1 & 1 \\
$A_2$ & $A_2$ & 1 & 1 &$-1$ \\
$E$   & $E$   & 2 &$-1$ & 0 \\
\hline
$E_{1/2}$ & $\ovl{E}_1$ & 2 & 1 & 0 \\
${}^1E_{3/2}{}^2E_{3/2}$ & ${}^1\ovl{E}{}^2\ovl{E}$ & 2 &-2 & $0$ \\
\end{tabular}
~~~~~
\begin{tabular}{c|c|cccccc}
\multicolumn{2}{c|}{PG $61'$} & 1 & $6^+$ & $3^+$ & $2$ & $3^-$ & $6^-$ \\
\hline
$A$ & $A$ & 1 & 1 & 1 & 1 & 1 & 1\\
$B$ & $B$ & 1 &$-1$ & 1 &$-1$ & 1 &$-1$\\
${}^1E_1{}^2E_1$ & ${}^1E_2{}^2E_2$ & 2 & $1$ & $-1$ &$-2$ & $-1$ & $1$\\
${}^1E_2{}^2E_2$ & ${}^1E_1{}^2E_1$ & 2 & $-1$ & $-1$ & 2 & $-1$ & $-1$\\
\hline
${}^1E_{1/2}{}^2E_{1/2}$ & ${}^1\ovl{E}_3{}^2\ovl{E}_3$ & 2 & $\sqrt{3}$ & $1$ & $0$ & $1$ & $\sqrt{3}$\\
${}^1E_{3/2}{}^2E_{3/2}$ & ${}^1\ovl{E}_1{}^2\ovl{E}_1$ &2 & $0$ & $-2$ & $0$ & $-2$ & $0$\\
${}^1E_{5/2}{}^2E_{5/2}$ & ${}^1\ovl{E}_2{}^2\ovl{E}_2$ & 2 & $-\sqrt{3}$ & $1$ & $0$ & $1$ & $-\sqrt{3}$ \\
\end{tabular}
\vspace{0.5cm}
~~~~~
\begin{tabular}{c|c|cccccc}
\multicolumn{2}{c|}{PG $6mm1'$} & 1 & $\{6^+,6^-\}$ & $\{3^+,3^-\}$ & $2$ & $\{m_{100}, m_{010}, m_{110}\}$ & $\{m_{120},m_{210},m_{1-10}\}$ \\
\hline
$A_1$ & $A_1$ & 1 & 1 & 1 & 1 & 1 & 1\\
$A_2$ & $A_2$ & 1 & 1 & 1 & 1 &$-1$ &$-1$\\
$B_1$ & $B_1$ & 1 &$-1$ & 1 &$-1$ &$-1$ & 1\\
$B_2$ & $B_2$ & 1 &$-1$ & 1 &$-1$ & 1 &$-1$\\
$E_1$ & $E_1$ & 2 & 1 &$-1$ &$-2$ & 0 & 0\\
$E_2$ & $E_2$ & 2 &$-1$ &$-1$ & 2 & 0 & 0\\
\hline
$E_{1/2}$ & $\ovl{E}_1$ & 2 & $\sqrt3$ & 1 & 0 & 0 & 0 \\
$E_{3/2}$ & $\ovl{E}_3$ & 2 & 0 &$-2$ & 0 & 0 & 0 \\
$E_{5/2}$ & $\ovl{E}_2$ & 2 & $-\sqrt3$ & 1 & 0 & 0 & 0 \\
\end{tabular}
\caption{The character tables of 2D PGs with time-reversal (sometimes called the gray groups). Here and throughout $1'$ denotes $\mathcal{T}$ as a group element. The irrep names are shown in the first column using the Altmann-Herzig notation and second columns using the Bilbao Crystallographic notation. In each table, the irreps above the second horizontal line are the single-valued irreps (no-SOC), and the irreps below the second horizontal line are the double-valued irreps (SOC). The traces of the anti-unitary operators are not shown because they are not invariant under unitary transforms.
\label{tab:2D-charTRS}
}
\end{table}

\section{Many-Body Atomic Limits and Construction of RSIs}
\label{app:RSIs}

In this Appendix, we construct many-body local Real Space Invariants (RSIs). First \App{app:offsite} gives a simple example of how single-particle RSIs breaks down with interactions. \App{app:RSIconstruct} then defines the many-body atomic limit states and gives explicit expressions for the many-body local RSIs of rotation groups. \App{eq:generalRSIs} generalizes these results to all 2D point groups with mirrors, time-reversal, and spin-orbit coupling. Lastly, \App{app:weakly interacting} evaluates the many-body local RSIs on product states to give expressions in terms of irrep multiplicities and single-particle RSIs.

\subsection{1D Many-Body RSI example}
\label{app:offsite} 

Ref. \cite{rsis} obtained the single-particle Real Space Invariants (RSIs) as adiabatic invariants in non-interacting Hamiltonians. States with different single-particle RSIs cannot be connected without a gap closing in non-interacting Hamiltonians. We give a minimal example of the breakdown of these single-particle RSIs when interactions are added by showing that two states with different single-particle RSIs can be brought into superposition by interaction terms. 

Let us consider a 1D system with inversion ($C_2$) symmetry. We propose a three site system with open boundaries. The site $R = 0$ has two $s$ orbitals denoted $s,s'$, and the sites at $R=1,-1$ have two $p$ orbitals each denoted $p,p'$. Inversion $\mathcal{I}$ takes $R\to -R$. We set the non-interacting Hamiltonian to be
\bea
H_0 = - \frac{t}{2} (c^\dag_{0,s}c_{0,s}+c^\dag_{0,s'}c_{0,s'}) + \frac{t}{2} (c^\dag_{-1,p}c_{-1,p}+c^\dag_{-1,p'}c_{-1,p'}+c^\dag_{1,p}c_{1,p}+c^\dag_{1,p'}c_{1,p'}), \quad t > 0
\eea
so the $s$ orbitals are at lower energy than the $p$ orbitals. The groundstate of $H_0$ at filling $2$ is given by $\ket{GS} = c^\dag_{0,s} c^\dag_{0,s'} \ket{0}$. We add a pair-hopping interaction term in the form 
\bea
H_{int} &= \frac{1}{\sqrt{2}}U (c^\dag_{1,p} c^\dag_{1,p'} c_{0,s} c_{0,s'} + c^\dag_{-1,p} c^\dag_{-1,p'} c_{0,s} c_{0,s'} )  + h.c.
\eea
which respects inversion symmetry, using $\mathcal{I} c^\dag_{r, \al} \mathcal{I}^\dag = \pm c^\dag_{-r, \al}$ for $\al = s/p$. Intuitively, $H_{int}$ connects the states $c^\dag_{0,s} c^\dag_{0,s'} \ket{0}$ and $\frac{1}{\sqrt{2}} (c^\dag_{1,p} c^\dag_{1,p'} + c^\dag_{-1,p} c^\dag_{-1,p'}) \ket{0}$. This would be impossible at the single-particle level. Alternatively, we can diagonalize $H_0 + H_{int}$ to obtain explicit energies and states. There are 6 orbitals in the model, so filling 2 gives a $15$-dimensional Hilbert space. Inversion symmetry splits the Hilbert space into $7$ even parity states and 8 odd parity states. A basis for the even parity states is
\bea
\label{eq:orderedbasis}
\ket{i} &= c^\dag_{0,s}c^\dag_{0,s'}\ket{0}, \ w^\dag_{0,s}w^\dag_{0,s'}\ket{0}, \ w^\dag_{0,p}w^\dag_{0,p'}\ket{0}, \ c^\dag_{0,s} w^\dag_{0,s}\ket{0}, \ c^\dag_{0,s'} w^\dag_{0,s}\ket{0}, \ c^\dag_{0,s} w^\dag_{0,s'}\ket{0}, \ c^\dag_{0,s'} w^\dag_{0,s'}\ket{0}
\eea
where we introduced the Wannier functions
\bea
\label{eq:wannierinvp}
w^\dag_{0,p} &= \frac{1}{\sqrt{2}} (c^\dag_{1,p} + c^\dag_{-1,p}), \qquad w^\dag_{0,s} = \frac{1}{\sqrt{2}} (c^\dag_{1,p} - c^\dag_{-1,p}) \\
w^\dag_{0,p'} &= \frac{1}{\sqrt{2}} (c^\dag_{1,p'} + c^\dag_{-1,p'}), \quad \ \ w^\dag_{0,s'} = \frac{1}{\sqrt{2}} (c^\dag_{1,p'} - c^\dag_{-1,p'}) \\ \\
\eea
which obey, for instance, $\mathcal{I} w^\dag_{0,s}\mathcal{I} = w^\dag_{0,s}$ and $\mathcal{I} w^\dag_{0,p}\mathcal{I} = -w^\dag_{0,p}$. The Wannier states are centered at $R=0$, and hence behave as $s/p$ orbitals at the $R=0$ site under inversion. The states in \Eq{eq:wannierinvp} are simply the induced representations of $G_0 = \{1, \mathcal{I}\}$ formed by the $p$ orbitals at $R = \pm1$. One can also check that this Wannier basis is orthogonal, e.g. $\{w^\dag_{0,p}, w^\dag_{0,s} \} = 0$. 

In the order of \Eq{eq:orderedbasis}, the Hamiltonian can be written (in the even parity sector) as 
\bea
\braket{i|H_0+H_{int}|j} &= \bpm
 -t & U/\sqrt{2} & U/\sqrt{2} &  &  &  &  \\
U/\sqrt{2} & t & 0 &  &  &  & \\
U/\sqrt{2} & 0 & t &  &  &  &  \\
  &  &  & 0 &  &  &  \\
  &  &  &  & 0 &  &  \\
  & &  &  &  & 0 &  \\
  & &  &  &  &  & 0 \\
\epm_{ij} 
\eea
whose spectrum is $\pm \sqrt{U^2+t^2}, t,0,0,0,0,$. The groundstate is at energy $-\sqrt{U^2+t^2}$ and remains gapped for all $U$. At $U=0$, the groundstate is  $c^\dag_{0,s} c^\dag_{0,s'} \ket{0}$, but interactions yield the entangled groundstate wavefunction
\bea
\ket{GS(U)} \propto \sqrt{2} (\sqrt{U^2+t^2}-t) c^\dag_{0,s} c^\dag_{0,s'} \ket{0} + U \lp c^\dag_{1,p} c^\dag_{1,p'} + c^\dag_{-1,p} c^\dag_{-1,p'}  \rp \ket{0}
\eea
which we left normalized, and used the simple identity
\bea
w^\dag_{0,s} w^\dag_{0,s'} + w^\dag_{0,p} w^\dag_{0,p'} &= \frac{1}{\sqrt{2}} (c^\dag_{1,p} - c^\dag_{-1,p}) \frac{1}{\sqrt{2}} (c^\dag_{1,p'} - c^\dag_{-1,p'}) + \frac{1}{\sqrt{2}} (c^\dag_{1,p} + c^\dag_{-1,p}) \frac{1}{\sqrt{2}} (c^\dag_{1,p'} + c^\dag_{-1,p'})  \\
&= c^\dag_{1,p}c^\dag_{1,p'} + c^\dag_{-1,p}c^\dag_{-1,p'} \ . 
\eea

We now show that $\ket{GS(U)}$ has a well-defined many-body RSI. The two-particle product states $c^\dag_{0,s}c^\dag_{0,s'}$, $w^\dag_{0,s} w^\dag_{0,s'}$, and $w^\dag_{0,p} w^\dag_{0,p'}$ each have a definite single-particle RSI (they are non-interacting Wannier product states), but their single-particle RSIs are different. Because $w^\dag_{0,s} w^\dag_{0,s'}$ and $w^\dag_{0,p} w^\dag_{0,p'}$ have the same \emph{many-body} RSI, they can be brought into superposition, necessarily creating entanglement. To be explicit, the single-particle RSI $\delta = m(s) - m(p)$ and many-body RSI $\Delta = \delta \mod 4$ of the states are
\bea
c^\dag_{0,s} c^\dag_{0,s'}, w^\dag_{0,s} w^\dag_{0,s'}:& \qquad \delta = 2, \quad \Delta = 2 \mod 4 , \\
w^\dag_{0,p} w^\dag_{0,p'}:& \qquad \delta = -2, \quad \Delta = 2 \mod 4 \ . \\
\eea
The fact that the two states have different single-particle RSIs means they are adiabatically distinct without interactions. But, because they have the same many-body RSIs, these two states can be adiabatically connected by interactions. As such, the correlated state $\ket{GS(U)}$ has a well-defined many-body RSI given by $\Delta = 2 \mod 4$. In the following section \App{app:RSIconstruct}, we give a rigorous argument for the definition of many-body RSIs in 2D PGs. 

\subsection{Defining Many-Body RSIs in the Atomic Limit}
\label{app:RSIconstruct}

We will now construct many-body local RSIs at a fixed Wyckoff position $\mbf{x}$ in 2D many-body atomic states protected by the symmetries of a point group $G_{\mbf{x}}$. Our construction proceeds as follows. First we define a class of many-body atomic \emph{limit} states with zero correlation length. We then show that, after imposing a cutoff to create open boundary conditions, there is a natural set of quantum numbers protected by $G_{\mbf{x}}$ which is invariant as the cutoff is changed in a symmetry-preserving fashion. We identity these invariants as many-body local RSIs and we show that they remain adiabatically well-defined beyond the zero correlation length limit. Since the cutoff can be taken to infinity, the many-body local RSIs remain well-defined as the open boundary conditions approach infinite boundary conditions, where we can simultaneously define many-body local RSIs at each Wyckoff position. 

Recall that the non-interacting atomic limit is defined by taking the lattice constant to infinity so that all hoppings in the Hamiltonian go to zero (but orbitals on the same site may still be connected by single-particle terms or interactions). The groundstate of such a system is a product state of occupied local orbitals. Our generalization of atomic limits to interacting Hamiltonians is analogous. We consider a general lattice with $N_{orb}$ electron orbitals per unit cell located at positions $\mbf{r}_\al \in \{\mbf{r}\}, \ \al = 1, \dots, N_{orb}$.  Denote the number of orbitals at position $\mbf{r}$ to be $n_\mbf{r}$. An interacting Hamiltonian in the atomic limit only contains terms that are totally local in $\mbf{r}$, e.g. onsite non-interacting potentials $H_{\mbf{R},\mbf{r}} \sim c^\dag_{\mbf{R},\al} c_{\mbf{R},\be}$ or onsite interactions $H_{\mbf{R},\mbf{r}} \sim c^\dag_{\mbf{R},\al} c^\dag_{\mbf{R},\al'} c_{\mbf{R},\be} c_{\mbf{R},\be'}$ and higher-body contact terms for $\mbf{r}_\al = \mbf{r}_{\al'}= \mbf{r}_\be = \mbf{r}_{\be'}$. Such a Hamiltonian can be written
\bea
H_{AL} &= \bigoplus_\mbf{R} \bigoplus_{\mbf{r} \in \{\mbf{r}\}} H_{\mbf{R},\mbf{r}} = \bigoplus_\mbf{R} \bigoplus_{\mbf{r} \in \{\mbf{r}\}} T^\dag_\mbf{R} H_{\mbf{r}} T_\mbf{R}
\eea
where we imposed translation invariance in the second equality. Each $H_\mbf{r}$ acts on a  strictly local Hilbert space whose Fock space is $2^{n_\mbf{r}}$ dimensional. Because $H_\mbf{r}$ is strictly local, $[H_\mbf{r}, T_\mbf{R}^\dag H_{\mbf{r}'}T_\mbf{R}] = 0$ for all $\mbf{r},\mbf{r}',\mbf{R}$. Thus $H_{AL}$ is composed of commuting terms (different sites are decoupled) and its ground state at fixed density $\nu = N_{occ}/N_{orb}, \ N_{occ} \in \mathbb{N}$ can be written
\bea
\label{eq:strictMBAL}
\ket{GS} = \prod_{\mbf{R},\mbf{r}\in \{\mbf{r}\}} T^\dag_\mbf{R} \mathcal{O}_\mbf{r} T_\mbf{R} \ \ket{0}, \qquad  [\hat{N}, \prod_{\mbf{r}\in \{\mbf{r}\}} \mathcal{O}_\mbf{r}] = N_{occ}  \prod_{\mbf{r}\in \{\mbf{r}\}} \mathcal{O}_\mbf{r} \\
\eea
where each $\mathcal{O}^\dag_\mbf{r}$ creates a groundstate of $H_\mbf{r}$ and $\hat{N}$ is the number operator. One can think of $H_\mbf{r}$ as being a quantum dot Hamiltonian, and $\mathcal{O}^\dag_\mbf{r} \ket{0}$ as the groundstate of the quantum dot. 

To move away from this strict atomic limit, hopping terms and/or interactions coupling different sites can be added to $H_{AL}$. As long as the many-body gap does not close, we say that the ground state is a many-body atomic phase since, by construction, it is adiabatically connected to an atomic limit. Note that an obstructed atomic limit\cite{rsis} is not a trivial atomic limit as defined here: there is an obstruction to deforming it into a zero correlation length state. For example, the dimerized limit of the SSH chain requires strong inter-site hoppings, and cannot be connected to the trivial atomic limit. Indeed, a trivial atomic limit is spatially decoupled and so has no corner or edge states on open boundary conditions.

We now define many-body local RSIs on the many-body atomic limit states \Eq{eq:strictMBAL}. Recall that non-interacting local RSIs\cite{rsis} are adiabatic invariants (they do not change value unless a single-particle gap closes) protected by point group symmetries, and are defined in single-particle states at fixed filling. We now explicitly show the existence of symmetry-protected adiabatic invariants which are the quantum numbers of certain symmetry operators in \emph{many-body} atomic limit groundstates (\Eq{eq:strictMBAL}). We call these quantum numbers many-body local RSIs. 

\begin{figure}[H]
 \centering
\begin{overpic}[height=0.4\textwidth,tics=10]{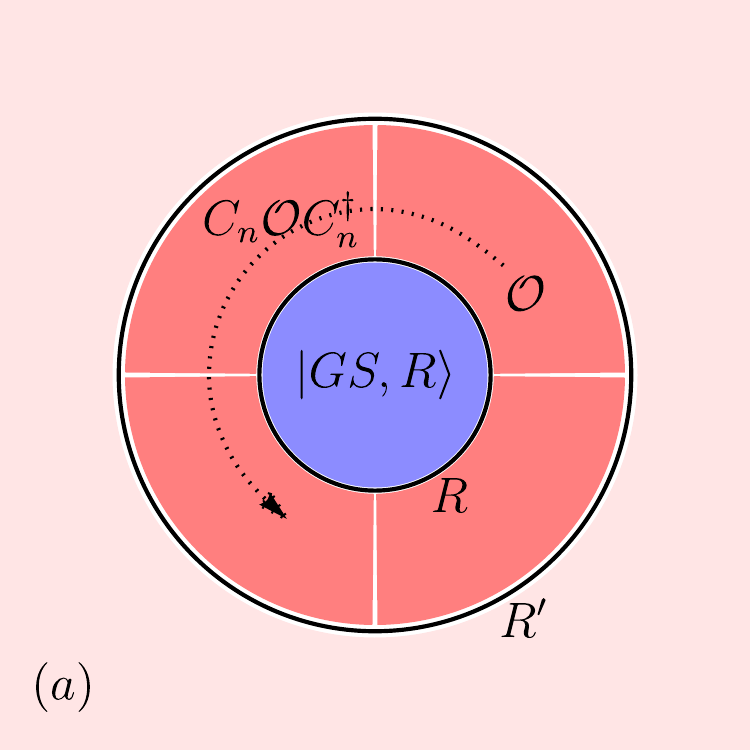}
\end{overpic} \qquad 
\begin{overpic}[height=0.4\textwidth,tics=10]{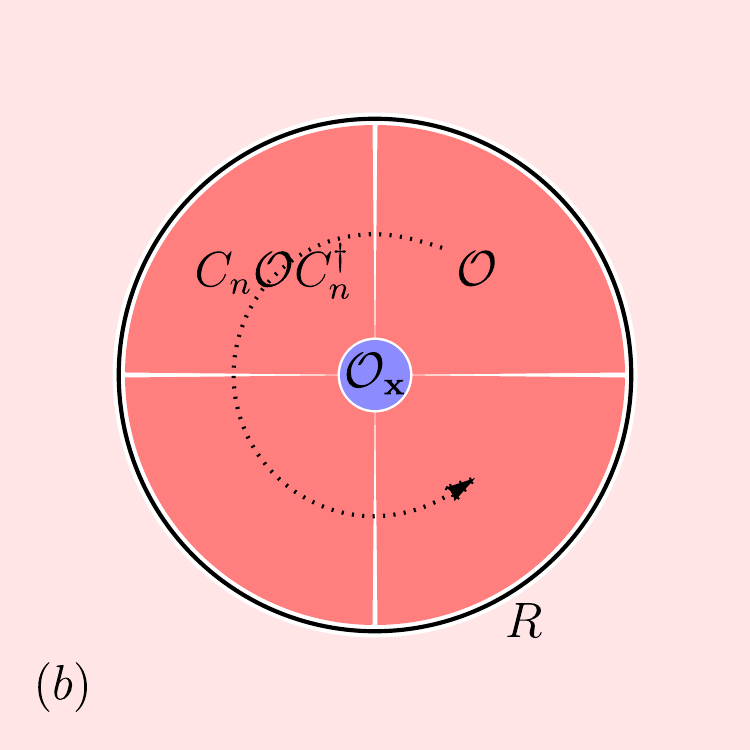}
\end{overpic} 
\caption{(a) We depict the groundstate $\ket{GS,R}$ on OBCs and the additional symmetry-related operators which are included upon expanding the cutoff to $R'$ (see \Eq{eq:groundstateansteaz}). The many-body local RSIs are invariant under the expansion. (b) Given a fixed cutoff $R$, all operators inside of $R$ but not at the $C_n$-invariant point $\mbf{x}$ are symmetry-related and so do not contribute to the many-body local RSI (see \Eq{eq:GSALOx}). Only the operator $O_\mbf{x}$ at $\mbf{x}$ transforms locally under $\mbf{x}$. Its quantum numbers determine the many-body local RSI. }
\label{fig:Ocalapp}
\end{figure}

A key part of our construction is to define the many-body local RSIs on open boundary conditions which break the space group $G$ to $G_\mbf{x}$. This serves to show that they are local (since they only depend on the groundstate within an arbitrary range of $\mbf{x}$ set by a cutoff) and are protected only by the point group symmetries of $G_\mbf{x}$. However, it is then crucial to show that the many-body local RSIs do not depend on the choice of cutoff. This ensures that the many-body local RSIs are well-defined invariants of the thermodynamic groundstate \Eq{eq:strictMBAL} since the spatial cutoff can be sent to infinity. They then serve to define invariants on infinite boundary conditions such that the gap is the thermodynamic gap. Define the open boundary conditions by
\bea
\label{eq:Hrcutoffx}
H_{AL, R, \mbf{x}} &= \bigoplus_{|\mbf{R}+\mbf{r}-\mbf{x}| < R} H_{\mbf{R},\mbf{r}} 
\eea
for a Wyckoff position $\mbf{x}$ and cutoff $R$ (see \Fig{fig:Ocalapp}a). The choice of a circular cutoff is just for convenience: the following argument hold for any cutoff that preserves $G_\mbf{x}$. At cutoff $R$, the total electron number $N_R$ is given by $[N, \mathcal{O}_{GS,R}] = N_R  \mathcal{O}_{GS,R}$ where  $ \mathcal{O}_{GS,R} = \prod_{|\mbf{r}-\mbf{x}| < R} \mathcal{O}_{\mbf{r}}$. As the cutoff $R$ is taken to infinity, the filling approaches the thermodynamic filling $\nu = N_{occ}/N_{orb}$. 

For each cutoff $R$, denote the ground state of $H_{AL, R, \mbf{x}}$ by $\ket{GS,R}$ (we omit $\mbf{x}$ henceforth because it is fixed throughout the calculation). It is important that $\ket{GS,R}$ is non-degenerate and gapped for all $R$ at the exact filling. It is natural to require these two properties because a many-body atomic phase has no edge states so the ground state is unique and is an insulator, so there is a many-body gap. Note that non-degeneracy excludes single-particle obstructed atomic or fragile topological phase (to be addressed in \App{app:fragiledef}), and spontaneously broken symmetries like a charge density wave. We now construct many-body local RSIs of $H_{AL,R,\mbf{x}}$ at a Wyckoff position $\mbf{x}$ protected by the point group symmetries of $G_\mbf{x}$. We will then use an adiabaticity argument to show that the many-body local RSIs are still well-defined in a many-body atomic phase once weak hoppings and off-site interactions are introduced. Because $\ket{GS,R}$ is non-degenerate (it is a strict atomic limit state), it transforms in a 1D irrep of $G_\mbf{x}$. Thus $\ket{GS,R}$ has the quantum numbers
\bea
\hat{N} \ket{GS,R} &= N \ket{GS,R}, \\
g \ket{GS,R} &= e^{i \la[g]} \ket{GS,R}, \qquad g \in G_\mbf{x} \ . \\
\eea
However, the quantum numbers $N, e^{i \la[g]}$ depend on the cutoff --- as we now show. To begin, we consider the spinless rotation groups $G_\mbf{x} = \{1, C_n, \dots, C_n^{n-1}\}$ with $C_n^n = +1$. (Shortly, we will study spinful electrons which have a rotation operator obeying $C_n^n = (-1)^{\hat{N}}$.) Extending the cutoff $R \to R'$ yields a new ground state which can be written
\bea
\label{eq:groundstateansteaz}
\ket{GS,R'} =  \prod_{R < |\mbf{r}| < R'} \mathcal{O}_\mbf{r} \  \ket{GS,R} = \prod_{i=0}^{n-1} C_n^i \mathcal{O} C_n^{\dag i} \  \ket{GS,R}
\eea
where, up to an irrelevant overall phase, $\mathcal{O} = \prod_{\mbf{r} \in \mathcal{D}} \mathcal{O}_\mbf{r}$ is a product of $\mathcal{O}_\mbf{r}$ in the region $\mathcal{D}$ which obeys $\{R < |\mbf{r}| < R'\} = \{\mathcal{D}, C_n \mathcal{D}, \dots, C_n^{-1} \mathcal{D}\}$ and $\mathcal{D} \cap C_n^i \mathcal{D} = \emptyset$. To be concrete, a possible choice is $\mathcal{D} = \{ \mbf{r} | R < r < R', 0 \leq \th < 2\pi/n\}$. \Eq{eq:groundstateansteaz} holds because in the strict atomic limit, $C_n\mathcal{O} C_n^\dag$ has zero overlap with $\mathcal{O}$. We call $N_\mathcal{O}$ the charge (number of particles) of $\mathcal{O}$. The quantum numbers of the state with cutoff $R'$ are
\bea
\label{eq:inductionargument}
\hat{N} \ket{GS,R'} &= (N + n N_\mathcal{O}) \ket{GS,R'} \ .\\ 
C_n \ket{GS,R'} &= \lp \prod_{i=1}^{n-1} C_n^i \mathcal{O} C_n^{\dag i} \rp C_n^n \mathcal{O} C_n^{n \dag} \ C_n\ket{GS,R} \\
&= e^{i \la[C_n]} \prod_{i=1}^{n-1} C_n^i \mathcal{O} C_n^{\dag i} \ \mathcal{O} \ \ket{GS,R} \\
&= e^{i \la[C_n]} (-1)^{(n-1) N_{\mathcal{O}}^2} \ \ket{GS,R'} \\
&= e^{i \la[C_n]} \begin{cases}
 (-1)^{N_{\mathcal{O}}}, & n \text{ even} \\
1 , & n \text{ odd} \\
\end{cases} \ \ket{GS,R'}
\eea
where in the last line we used that fact that $\mathcal{O} \, C_n^i \mathcal{O} C_n^{\dag i} = (-1)^{N_\mathcal{O}^2} C_n^i \mathcal{O} C_n^{\dag i} \, \mathcal{O}$ since $C_n\mathcal{O} C_n^\dag$ and $\mathcal{O}$ are supported on non-overlapping regions and each contain $N_\mathcal{O}$ fermionic operators. We see that both the $\hat{N}$ and $C_n$ eigenvalue (if $n$ is even) change under the expansion of the cutoff. The $(-1)^{N_\mathcal{O}}$ factor is simply from the fermionic parity of the operator. In a system of bosons, no $(-1)^{N_\mathcal{O}}$ factor appears, and the RSI groups (not considered here) are altered.  

However, there are two quantities invariant under expansion of the cutoff for arbitrary $N_\mathcal{O}$:
\bea
\label{eq:cutoffinvrsis}
e^{i \frac{2\pi}{n} \hat{N}} \ket{GS,R} &= e^{i \frac{2\pi}{n} \hat{N}} \ket{GS,R'} \\
e^{i \frac{\pi}{n} \hat{N}} C_n \ket{GS,R} &= e^{i \frac{\pi}{n} \hat{N}} C_n \ket{GS,R'} \quad \text{($n$ even)}\\
C_n \ket{GS,R} &= C_n \ket{GS,R'} \quad\quad\quad \, \text{($n$ odd)}\\
\eea
corresponding to the total charge mod $n$ and the many-body angular momentum of the $e^{i \frac{\pi}{n} \hat{N}} C_n$ \cite{2012PhRvB..86k5112F,2019PhRvX...9c1003L}. (Note that $e^{i \frac{\pi}{n} \hat{N}} C_n$ is effectively a ``spinful" rotation since $(e^{i \frac{\pi}{n} \hat{N}} C_n)^n = (-1)^{\hat{N}}$ and $C_n$ is a spin-less operator since $C_n^n = +1$.) Because the eigenvalues of $e^{i \frac{2\pi}{n} \hat{N}}, e^{i \frac{\pi}{n} \hat{N}} C_n$ (or $C_n$ if $n$ is odd) are independent of the cutoff, they are good quantum numbers for all choices of the cutoff. Then we can take $R \to \infty$ at constant density, which is the thermodynamic limit on infinite boundary conditions. Secondly, the eigenvalues of $e^{i \frac{2\pi}{n} \hat{N}} $ and $C_n$ are quantized, so they cannot change if the ground state is perturbed away from the strict atomic limit without closing the gap. Thus small hoppings and off-site interactions can be added without changing the quantum numbers in \Eq{eq:cutoffinvrsis}. Hence we have shown that the eigenvalues of the operators in \Eq{eq:cutoffinvrsis} are local, symmetry-protected quantum numbers invariant under the adiabatic expansion of the cutoff: they are many-body local RSIs. We remark that the many-body local RSIs are not typical invariants of $\ket{GS}$ in the thermodynamic limit. Usually, one computes the quantum numbers of a state with the Hilbert space fixed, whereas we demonstrated the existence of quantum numbers which are invariants upon enlarging the Hilbert space (while preserving density and symmetry). It would be desirable to have equivalent expressions for the many-body local RSIs which can be computed on a fixed Hilbert space without requiring OBCs. This is left for future work.

We now determine the group structure formed by the many-body local RSIs (\Eq{eq:cutoffinvrsis}). For even $n$, we have $(e^{i \frac{\pi}{n} \hat{N}} C_n)^{2n} = +1$, so the eigenvalues of $e^{i \frac{\pi}{n} \hat{N}} C_n$ are $\mathds{Z}_{2n}$-classified. The eigenvalues of $e^{i \frac{2\pi}{n} \hat{N}}$ are $\mathds{Z}_n$-classified, but they are not all independent from the eigenvalues of $e^{i \frac{\pi}{n} \hat{N}} C_n$ because $(e^{i \frac{\pi}{n} \hat{N}} C_n)^{n} = e^{i \pi \hat{N}} = (e^{i \frac{2\pi}{n} \hat{N}})^{n/2}$. Hence $e^{i \frac{2\pi}{n} \hat{N}}$ only provides $n/2$ more independent quantum numbers, and we obtain a $\mathds{Z}_{2n} \times \mathds{Z}_{n/2}$ classification for $n$ even. To get a set of $\mathds{Z}_{2n} \times \mathds{Z}_{n/2}$ independent quantum numbers, we define the eigenvalues of the symmetry operators in \Eq{eq:cutoffinvrsis} as
\bea
\label{eq:RSIcneven}
e^{i \frac{\pi}{n} \hat{N}} C_n \ket{GS} &= e^{i\frac{\pi}{n} \Delta_1} \ket{GS}, \qquad \Delta_1 \in \mathds{Z}_{2n} , \qquad \text{$n$ even} \\
e^{-i \frac{2\pi}{n} \hat{N}} (e^{i \frac{\pi}{n} \hat{N}} C_n)^2 \ket{GS} &= C_n^2\ket{GS} = e^{i\frac{2\pi}{n/2} \Delta_2} \ket{GS}, \qquad \Delta_2 \in \mathds{Z}_{n/2}  \\
\eea
since $(C_n^2)^{n/2} = +1$. We will refer to $\Delta_1$ and $\Delta_2$ as the many-body local RSIs. Note that \Eq{eq:inductionargument} can be used to show directly that $\Delta_2$ is invariant under the cutoff, since $C_n^2 \ket{GS,R'} = e^{i 2 \la[C_n]}(-1)^{2 N_\mathcal{O}} \ket{GS,R'} = e^{i 2 \la[C_n]}\ket{GS,R'}$ with the $N_\mathcal{O}$-dependent phase canceling.

For $n$ odd (only $n=3$ is relevant for the crystalline point groups), the $\mathds{Z}_n \times \mathds{Z}_n$ group structure is obvious from \Eq{eq:cutoffinvrsis} with $e^{i \frac{2\pi}{n}\hat{N}}$ and $C_n$ providing independent $\mathds{Z}_n$ quantum numbers. We define the eigenvalues as
\bea
\label{eq:RSIcnodd}
e^{i \frac{2\pi}{n} \hat{N}} \ket{GS} &= e^{i\frac{2\pi}{n} \Delta_1} \ket{GS}, \qquad \Delta_1 \in \mathds{Z}_{n}, \qquad \text{$n$ odd} \\
C_n \ket{GS} &= e^{i\frac{2\pi}{n} \Delta_2} \ket{GS}, \qquad \Delta_2 \in \mathds{Z}_{n} \\
\eea
We will refer to $\Delta_1$ and $\Delta_2$ as the many-body local RSIs. 

Having obtained our key results, we offer a brief alternative perspective from the strict atomic limit. At any cutoff $R$, we can write the wavefunction of the groundstate of the OBC Hamiltonian \Eq{eq:Hrcutoffx} respecting the rotation group $G_\mbf{x}$ as
\bea
\label{eq:GSALOx}
\ket{GS,R} &= \mathcal{O}_\mbf{x} \prod_{i=1}^n \mathcal{O}_i \ket{0}, \qquad \mathcal{O}_i = C^i_n \mathcal{O} C_n^{\dag i}
\eea
%[
in analogy to \Eq{eq:groundstateansteaz}, where $\mathcal{O} = \prod_{\mbf{r} \in \mathcal{W}} \mathcal{O}_\mbf{r}$ and $\mathcal{W}$ is the wedge $\{\mbf{r}| 0 < |\mbf{r}-\mbf{x}| < R, \text{arg}(\mbf{r}-\mbf{x}) \in (0,2\pi/n]\}$ (see \Fig{fig:Ocalapp}b). %) 
Note that it is important $\mathcal{W}$ not contain $\mbf{x}$, since $C_n \mathcal{W}$ must not overlap $\mathcal{W}$. For this reason, the operator $\mathcal{O}_\mbf{x}$ is separated out in \Eq{eq:GSALOx}. Since the many-body local RSIs are constructed so that the operator $\prod_{i=1}^n \mathcal{O}_i$, which depends on the cutoff, does not contribute, we observe that the many-body local RSIs of $\ket{GS,R}$ are simply the many-body local RSIs of $\mathcal{O}_\mbf{x} \ket{0}$. This statement relies on the strict atomic limit where $\mathcal{O}_\mbf{x}$ can be defined since all operators are onsite, but \Eq{eq:GSALOx} provides a useful intuitive picture. 

As we have shown, the many-body local RSIs in \Eqs{eq:RSIcneven}{eq:RSIcnodd} remain invariant as the cutoff is taken to infinity. In this infinite limit, it is natural to consider the many-body local RSIs of each Wyckoff position in the unit cell. As an example in the space group $p2$, there are four Wyckoff positions $\mbf{x} = (0,0), (1/2, 0), (0,1/2), (1/2, 1/2)$ with $G_\mbf{x} = 2$ (in each unit cell). The point groups are generated by the symmetries $C_2, T_1 C_2, T_2 C_2, T_1 T_2 C_2$ respectively where $T_i$ is the translation operator along the $i$th lattice vector. Choosing finite boundary conditions respecting the symmetries of a particular point group will necessarily break the symmetries of the other point groups, and thus our construction of the many-body local RSIs does not immediately imply they can be simultaneously defined. 

While we do not provide a rigorous general argument, we claim that the many-body local RSIs are well-defined (for interacting atomic phases) in the infinite limit at each Wyckoff position in the unit cell. Consider the following heuristic argument. On infinite boundary conditions, the many-body local RSIs at each Wyckoff positions are well-defined in the strict atomic limit because the correlation length is zero and each Wyckoff position is decoupled. Hence their many-body local RSIs can be defined at each Wyckoff position. Adding weak hoppings and off-site interactions will make the correlation length nonzero, but the quantization of the RSI eigenvalues means they remain invariant throughout this process. Of course, the formal difficulty is that the eigenvalue of $\hat{N}$, the number of particles, is not well-defined in the infinite limit. We extend all the results of this section to general 2D point groups in \App{eq:generalRSIs}. We will give explicit symmetry operators whose quantum numbers (eigenvalues) are many-body local RSIs.

Lastly, we prove that the many-body local RSIs at two symmetry-related positions are identical. This statement holds on infinite boundary conditions (where the full wallpaper group is intact) according to the argument in the preceding paragraph, such that the many-body local RSIs on infinite boundary conditions are understood as the infinite-cutoff limit on OBCs. For instance, 1a sites in different unit cells of a crystal are related by translations and have the same many-body local RSIs. (In concrete terms, we can compute the many-body local RSI at the same Wyckoff position in two different unit cells by choosing two different OBCs which respect the symmetries of each site individually, though it is impossible to respect both at once. The resulting many-body RSIs will be the same since they are related by translations on infinite boundary conditions.) Alternatively, the two sites in the 2c $=\{(1/2,0),(0,1/2)\}$ position in wallpaper group $p4$ are related by $C_4$ and have the same many-body local RSIs protected by $T_1 C_2$ and $T_2 C_2$ respectively. As such, we can define $\Delta_{2c,1} = \Delta_{(1/2,0),1} = \Delta_{(0,1/2),1} \in \mathds{Z}_4$. In other words, we can refer to the Wyckoff position 2c rather than either of its individual sites. 

We now prove the general case. Consider a wallpaper group with two sites $\mbf{x},\mbf{x}'$ related by a symmetry: $\mbf{x}' = g' \mbf{x}, g' \in G$. The site symmetry groups $G_\mbf{x}, G_{\mbf{x}'}$ are isomorphic and related by conjugation: $G_\mbf{x} \cong G_\mbf{x}'$ with $G_{\mbf{x}'} = \{h' = g' h g'^\dag| h \in G_\mbf{x}, g' \notin G_\mbf{x}\}$. For instance, $\mbf{x}= (1/2,0), \mbf{x}' = (0,1/2)$ in $G = p4$ are the two sites of the 2c position with $g' = C_4$ discussed above, or in $G=p2$, we could consider $\mbf{x}= (0,0), \mbf{x}' = (1,0)$, which are 1a sites related by a translation $g' = T_1$. In the atomic limit where $\mathcal{O}_\mbf{r}$ is the creation operator of the strictly local Hamiltonian $H_\mbf{r}$, we now prove that $\mathcal{O}_\mbf{x}$ and $\mathcal{O}_\mbf{x}'$ have the same quantum numbers. To do so, define the transformation of $\mathcal{O}_{\mbf{x}}$ by $h \mathcal{O}_\mbf{x} h^\dag = e^{i O[h]} \mathcal{O}_\mbf{x}$. Then note that
\bea
\label{eq:conjugate}
h' \mathcal{O}_{\mbf{x}'} h'^\dag &= g' h g'^\dag \mathcal{O}_{\mbf{x}'} g' h^\dag g'^\dag = e^{i O[g']} g' h \mathcal{O}_{\mbf{x}} h^\dag g'^\dag= e^{i O[h]}  e^{i O[g']} g' \mathcal{O}_{\mbf{x}} g'^\dag = e^{i O[h]} \mathcal{O}_{\mbf{x}'} 
\eea
using $g'^\dag \mathcal{O}_{\mbf{x}'} g' = e^{i O[g']}\mathcal{O}_{\mbf{x}}$, with $e^{i O[g']}$ being an irrelevant phase factor. We saw earlier in \Eq{eq:GSALOx} that the quantum numbers of $\mathcal{O}_\mbf{x}$ completely determined the many-body local RSIs of the groundstate $\ket{GS,R,\mbf{x}} = \mathcal{O}_{\mbf{x}} \prod_{j=1}^n \mathcal{O}_j \ket{0} $. Because the quantum numbers of $\mathcal{O}_{\mbf{x}}$ and $\mathcal{O}_{\mbf{x}'}$ are the same, the RSIs of $\ket{GS,R,\mbf{x}} $ and $\ket{GS,R,\mbf{x}'} $ must be the same. This holds for all adiabatically connected states, completing the proof. 

\subsection{Extension to General Point Groups}
\label{eq:generalRSIs}

In \App{app:RSIconstruct}, we gave a detailed derivation of the many-body local RSIs in the spinless point groups generated by $C_n$. To complete our RSI classification for other point groups with mirrors, time-reversal, and spin-orbit coupling (SOC), we develop the general theory of many-body RSIs. As before, we begin by working in zero correlation length atomic limit states, and then extend to adiabatically connected atomic states. Physically, expanding the cutoff and including more orbitals at a position $\mbf{x}' \neq \mbf{x}$ adds $C_n$-related copies of the orbitals to the Hilbert space. Like in the non-interacting case \cite{rsis}, these orbitals transforms as an irrep of the subgroup $G_{\mbf{x}'} \in G_{\mbf{x}}$. For instance with $C_n$ and a mirror $M_x$, the addition of orbitals in the strict atomic limit described by $\mathcal{O} = \prod_{\mbf{r} \in \mathcal{D}'} \mathcal{O}_\mbf{r}$ where $\mathcal{D}' = \{R < |\mbf{r}| < R', -\pi/n < \th < \pi/n\}$ is mapped to a distinct region by $C_n$ but is preserved under the mirror $M$ taking $y \to -y$. Thus $\mathcal{O}$ transforms under an irrep of the reflection group $M$. In the rotation group cases we have considered so far with $G_\mbf{x} = n$, the subgroup is trivial  ($G_{\mbf{x}'} = 1$), but in general there can be nontrivial subgroups of $G_\mbf{x}$. There is a further requirement on $\mathcal{O}$: because we are in the strict atomic limit with a non-degenerate ground state, $\mathcal{O}$ must transform in a 1D irrep. Explicitly, for $G_\mbf{x}  = \{g_1 H, \dots, g_{d} H\}$ in a coset construction with $G_{\mbf{x}'} = H$  and $d = |H|/|G|$, we have
\bea
h \mathcal{O} h^\dag  = e^{i O[h]} \mathcal{O}, \qquad \forall h \in H \subset G_{\mbf{x}} \\
\eea
and the operators $g_i \mathcal{O} g_i^\dag$ are supported on different orbitals for all $i = 1, \dots, d$ in the strict atomic limit. The representation $e^{i O[h]}, h \in H$ is a 1D irrep of $H$. For instance, if $\mathcal{O}$ is a product of orbital creation operators with definite parity under $M$, then $O[h=M]$ is simply the total many-body parity. Under rotations, $C_n \mathcal{O} C_n^\dag$ is a new operator supported on different orbitals in the region $C_n \mathcal{D}$.  

Now expanding the cutoff to introduce new operators at $g_1 \mbf{x}', \dots, g_{d} \mbf{x}'$ (we always take $g_1 = \mathbb{1}$) yields the ground state
\bea
\label{eq:GSinduced}
\ket{GS,R'} &= \prod_{i=1}^{d} g_i \mathcal{O} g_i^\dag \ \ket{GS,R} \ . \\
\eea

Importantly, we must find quantum numbers which are invariant under \emph{all} possible expansions of the cutoff, which can include different orbitals transforming under any of the possible subgroups. To do so systematically, we recall some basic facts from the theory of induced representations. Because $\mathcal{O}$ transforms in a representation of $H$ (in this case, a 1D irrep as required by non-degeneracy), it forms an induced representation of $G_\mbf{x}$. To be explicit, this representation is the following. Define $\mathcal{O}_i = g_i \mathcal{O} g_i^\dag, i = 1,\dots, d$ and the induced representation $R_{ij}[g]$ by
\bea
\label{eq:Rdef}
g \mathcal{O}_i g^\dag = \sum_{j=1}^d R_{ij}[g] \mathcal{O}_j , \qquad \forall g \in G_\mbf{x} \ . \
\eea
For the spinless symmetries considered here where $C_n^n = +1$, $R[g]$ is a spinless representation of $G_\mbf{x}$. Since $\mathcal{O}_i$ are distinct operators for each $i$ in the strict atomic limit ($\mathcal{O}_i,\mathcal{O}_j$ are supported on non-overlapping sites for $i \neq j$), $R_{ij}[g]$ is a complex permutation matrix \cite{rsis} in this basis: there is only one nonzero entry per row. As an example, consider $G_\mbf{x} = 4mm$ which has $G_{\mbf{x}'} = m$ as a nontrivial subgroup. Expand the cutoff to include the operator $\mathcal{O}$ supported in a finite region $\mathcal{D}'$ centered around the point $(x',0) \neq \mbf{q}$. Note that $M \in G_{\mbf{x}}$ is also a symmetry of $G_{(x,0)}$, where $M$ is a reflection acting as $M \hat{y} = - \hat{y}$, and denote the irrep of $\mathcal{O}$ by $M \mathcal{O} M^\dag = \pm \mathcal{O}$. Since $(x,0) \neq \mbf{x}$, $C_4$ ensures the groundstate also contains the operators $C_4^i \mathcal{O} C_4^{i \dag}, i = 1,2,3$ at position $(0,x),(-x,0), (0,-x)$ respectively. From \Eq{eq:Rdef}, the induced representation is
\bea
\label{eq:eeqc4M}
R[C_4] = \bpm 0 & & & 1 \\ 1 &0 & &  \\ & 1 &0 & \\ & & 1 &0 \\ \epm, \quad R[M] = \pm  \bpm 1 & & &  \\  &0 & &1  \\ &  &1 & \\ & 1&  &0 \\ \epm \ .
\eea
We now need to compute the representation of $g$ on the groundstate \Eq{eq:GSinduced}. To do so, we recall that $g \ket{GS,R} = e^{i \la[g]} \ket{GS,R}$ and $\mathcal{O}_i \mathcal{O}_j = (-1)^{N_{\mathcal{O}}^2 }\mathcal{O}_j \mathcal{O}_i = (-1)^{N_{\mathcal{O}}}\mathcal{O}_j \mathcal{O}_i $ in which case, upon enlarging the cutoff from $R'$ to $R$ gives
\bea
\label{eq:overallsign}
g \ket{GS,R'} &= g \prod_{i=1}^{d} \mathcal{O}_i \ g^\dag g \ket{GS,R}\\
&= e^{i \la[g]} \prod_{i=1}^{d} \lp \sum_j R_{ij}[g]\mathcal{O}_j \rp \  \ket{GS,R}\\
&= e^{i \la[g]} \lp \det\!{}_{(-1)^{N_\mathcal{O}}} R[g] \rp \, \prod_{i=1}^{d} \mathcal{O}_i \ \ket{GS,R}\\
&= e^{i \la[g]} \lp \det\!{}_{(-1)^{N_\mathcal{O}}} R[g] \rp \,  \ket{GS,R'} \\
\eea
where $\det{}_{\pm}$ is the permanent for $+$ and determinant for $-$. Explicitly for a $d\times d$ matrix $A$, 
\bea
\label{eq:detpm}
\det\!{}_{\pm} \, A &= \sum_\sigma (\pm 1)^{\#_\sigma} A_{1, \sigma_1}\dots A_{d, \sigma_d}\\
\eea
where $\sigma$ are the $d!$ permutations of the numbers $1,\dots, d$ and $\#_\sigma$ is the number of interchanges in the permutation. Because $D[g]$ is a complex permutation matrix with only one nonzero element per row, only one permutation $\sigma'$ in the sum \Eq{eq:detpm} over $\sigma$ is nonzero. Thus
\bea
\det\!{}_{\pm} \, R[g] &=(\pm1 )^{\#_{\sigma'}} R[g]_{1, \sigma_1'}\dots R[g]_{d, \sigma_d'}\\
\eea
so that only the overall sign in \Eq{eq:overallsign} depends on the parity of $N_\mathcal{O}$ (through the signature $(-1)^{\#_{\sigma'}}$ of $g$). 

We now give an example of this procedure in the point groups generated by $C_4$ and $M$. Consider expanding the cutoff to include the new operators $\mathcal{O}_i = C_4^i \mathcal{O} C_4^{\dag i}$ where $\mathcal{O} = \prod_{\mbf{r} \in \mathcal{D}'} \mathcal{O}_\mbf{r}$. There are two possible nontrivial subgroups that $\mathcal{O}$ could transform under: $G_{\mbf{x}'} = \{1,M\}$ if $\mathcal{D}'$ is centered around e.g. $(x,0)$ or $G_{\mbf{x}'} = \{1,C_4 M\}$ if $\mathcal{D}'$ is centered around e.g. $(x,x)$. Here $x\neq 0$ so $\mathcal{D}'$ cannot be invariant under $C_4$, and thus we do not consider the subgroups with $C_4$, as in the single-particle case \cite{rsis}. For both cases, the number of particles increases as $N \to N  + 4 N_\mathcal{O}$ where $N_\mathcal{O}$ is the number of particles in $\mathcal{O}$. Note that $N_\mathcal{O}$ can be even or odd, for instance if $\mathcal{D}$ encloses a single site with an even or odd number of orbitals. We start by considering the subgroup $\{1,M\}$. With \Eq{eq:overallsign}, we find that
\bea
\label{eq:2c4M}
C_4 \ket{GS,R'} &= e^{i \la[C_4]} (-1)^{N_\mathcal{O}} \,  \ket{GS,R'} \\
\eea
since the signature $(-1)^{\#_{\sigma'}} = -1$ of $C_4$ is negative. This obviously matches \Eq{eq:inductionargument}, recalling that $e^{i\la[C_4]}, e^{i\la[M]}$ are the eigenvalues of $C_4$ and $M$ on $\ket{GS,R}$ respectively. We now need the representation of $M$. We find
\bea
\label{eq:1M}
M \ket{GS,R'} &= e^{i \la[M]} (-1)^{N_\mathcal{O}} \,  \ket{GS,R'}, \qquad H  = \{1,M\} \\
\eea
since $R[M]$ in \Eq{eq:eeqc4M} also has odd signature. Now we consider the subgroup $\{1, C_4 M\}$. The representation matrices are now
\bea
\label{eq:eeqM}
R[C_4] = \bpm 0 & & & 1 \\ 1 &0 & &  \\ & 1 &0 & \\ & & 1 &0 \\ \epm, \quad R[M] = \pm  \bpm & & & 1  \\  &  & 1 &  \\ &  1&  & \\ 1 & &  & \\ \epm 
\eea
and notably $R[M]$ now has even signature. So in this case 
\bea
\label{eq:1c4M}
M \ket{GS,R'} &= e^{i \la[M]}  \,  \ket{GS,R'} , \qquad H  = \{1,C_4M\} \ .
\eea
Let us scrutinize \Eqs{eq:2c4M}{eq:1c4M}. On the original groundstate, the mirror eigenvalue is $M \ket{GS,R} = e^{i \la[M]} \ket{GS,R}$. Expanding the cutoff to obtain $\ket{GS,R'}$ can include operators transforming in one of two possible nontrivial subgroups. In both cases, the mirror eigenvalue is well-defined and is given by  \Eqs{eq:2c4M}{eq:1c4M}, but crucially the mirror eigenvalue \emph{differs} between the two cases by a factor of $(-1)^{N_{\mathcal{O}}}$. In order to obtain a many-body local RSI, it is essential to have a symmetry operator whose eigenvalues are invariant under any (symmetric) choice of cutoff, such as $e^{i \frac{\pi}{4} \hat{N}} C_n$. We see from \Eqs{eq:2c4M}{eq:1c4M} that, because $N_\mathcal{O}$ is unrestricted, we do not obtain another cutoff-independent quantum number from $M$. 

In fact, this is the same as in the non-interacting RSIs. Ref. \cite{rsis} showed that the non-interacting RSIs of $4mm$ can all be obtained by reducing the non-interacting RSIs of $4$ in the presence of mirrors, but that mirrors did not introduce any new non-interacting RSIs. We check that for all even $n$, inducing from the $\{1,M\}$ and $\{1,C_n M\}$ subgroups gives factors of  $e^{i \la[M]}$ and $ e^{i \la[M]} (-1)^{N_\mathcal{O}} $ as in the $4mm$ example, and thus there are no RSIs protected by $M$. For $n$ odd, we find that inducing the odd and even mirror irreps from $\{1,M\}$ brings factors of $\pm e^{i \la[M]}(-1)^{N_\mathcal{O}}$ which also prevents $M$ from protecting a new RSI. 

Lastly, we must consider how $M$ restricts the possible eigenvalues of $C_n$, thereby reducing the $C_n$-protected many-body local RSIs. First we consider the case of $n$ even. Since $M C_n M^\dag = C_n^\dag$ and $M \ket{GS} = e^{i \la[M]} \ket{GS}$ on a given groundstate, the only allowed $C_n$ eigenvalues are real: $\pm 1$. Then \Eq{eq:RSIcneven} shows that $\Delta_2 = 0$, reducing the local RSI group from $\mathds{Z}_{2n} \times \mathds{Z}_{n/2} \to \mathds{Z}_{2n}$. If $n$ is odd, then $M C_n M^\dag = C_n^\dag$ enforces the $C_n$ eigenvalue to be real, and thus $+1$ is the only possibility since $-1$ cannot be a $C_n$ eigenvalue for $n$ odd. \Eq{eq:RSIcnodd} then shows $\Delta_2 = 0$, and the many-body local RSI group is reduced from $\mathds{Z}_n \times \mathds{Z}_n \to \mathds{Z}_n$. 

We now consider spinless time-reversal. This case is simpler since the only subgroup is $1'$ which with $\mathcal{T}^2 = +1$ has only the trivial irrep, so the induction is identical to the $C_n$ case alone. Thus to derive the many-body local RSI, we only have to consider the constraints $\mathcal{T}$ imposes on the $C_n$ eigenvalues. This is also very simple since, because $\mathcal{T}^2 = +1$, we can choose the overall phase of the groundstate such that $\mathcal{T} \ket{GS,R} = \ket{GS,R}$. Then $e^{i \la[C_n]}$ must be real because $[C_n,\mathcal{T}] = 0$. This is the same condition enforced by $M$, so following the same argument we find $\Delta_2 = 0$. 

We remark that in all cases, the many-body local RSIs of all 2D point groups can obtained by reduction from their rotation subgroups. 

\Tab{tab:RSInoSOC} summarizes the full classification obtained from adding mirrors and time-reversal to the rotation groups. We also must remark on the case of a point group $G_\mbf{x} = m$ with only mirrors. This case is somewhat degenerate because $M$ is a quasi-1D symmetry and its Wyckoff positions are extended lines, not points. When expanding the cutoff in 2D, it is possible to include an arbitrary number of each irrep on the mirror axis, and thus our prescription does not yield a many-body local RSI protected by $M$ or by $e^{i \pi \hat{N}}$. In 1D there is no such issue because expanding the cutoff always induces orbitals from off the mirror plane.

\begin{table}[H]
\centering
\begin{tabular}{| c | c | c |}
\hline
PG & RSI Operators  on $\ket{GS}$  & Classification \\ 
\hline
$m$ & & $\Delta_1 \in \mathds{Z}_1$ \\
\hline
2 & $e^{i \frac{\pi}{2} \hat{N}} C_2 = e^{ i \frac{\pi}{2} \Delta_1}$ & $\Delta_1 \in \mathds{Z}_4$ \\
\hline
$2mm$& $e^{i \frac{\pi}{2} \hat{N}} C_2 = e^{ i \frac{\pi}{2} \Delta_1}$ & $\Delta_1 \in \mathds{Z}_4$ \\
\hline
3 & $e^{i \frac{2\pi}{3} \hat{N}} = e^{ i \frac{2\pi}{3} \Delta_1}, \ C_3 = e^{ i \frac{2\pi}{3} \Delta_2} $ & $\Delta_1 \in \mathds{Z}_3,\Delta_2 \in \mathds{Z}_3$ \\
\hline
$3m$ & $e^{i \frac{2\pi}{3} \hat{N}} = e^{ i \frac{2\pi}{3} \Delta_1}$ & $\Delta_1 \in \mathds{Z}_3$ \\
\hline
4 & $e^{i \frac{\pi}{4} \hat{N}} C_4 = e^{ i  \frac{\pi}{4} \Delta_1}, \ C_4^2 = e^{ i \pi \Delta_2} $ & $\Delta_1 \in \mathds{Z}_8, \Delta_2 \in \mathds{Z}_2$ \\
\hline
$4mm$ & $e^{i \frac{\pi}{4} \hat{N}} C_4 = e^{ i  \frac{\pi}{4} \Delta_1}$ & $\Delta_1 \in \mathds{Z}_8$ \\
\hline
6 & $e^{i \frac{\pi}{6} \hat{N}} C_6 = e^{ i  \frac{\pi}{6} \Delta_1}, \ C_6^2 = e^{ i \frac{2\pi}{3} \Delta_2} $ & $\Delta_1 \in \mathds{Z}_{12}, \Delta_2 \in \mathds{Z}_3$ \\
\hline
$6mm$ & $e^{i \frac{\pi}{6} \hat{N}} C_6 = e^{ i  \frac{\pi}{6} \Delta_1}$ & $\Delta_1 \in \mathds{Z}_{12}$ \\
\hline
\end{tabular}
\quad
\begin{tabular}{| c | c | c |}
\hline
PG & RSI Operators  on $\ket{GS}$ & Classification \\ 
\hline
$m$ & & $\Delta_1 \in \mathds{Z}_1$ \\
\hline
$21'$ & $e^{i \frac{\pi}{2} \hat{N}} C_2 = e^{ i \frac{\pi}{2} \Delta_1}$ & $\Delta_1 \in \mathds{Z}_4$ \\
\hline
$2mm1'$& $e^{i \frac{\pi}{2} \hat{N}} C_2 = e^{ i \frac{\pi}{2} \Delta_1}$ & $\Delta_1 \in \mathds{Z}_4$ \\
\hline
$31'$ & $e^{i \frac{2\pi}{3} \hat{N}} = e^{ i \frac{2\pi}{3} \Delta_1}$ & $\Delta_1 \in \mathds{Z}_3$ \\
\hline
$3m1'$ & $e^{i \frac{2\pi}{3} \hat{N}} = e^{ i \frac{2\pi}{3} \Delta_1}$ & $\Delta_1 \in \mathds{Z}_3$ \\
\hline
$41'$ & $e^{i \frac{\pi}{4} \hat{N}} C_4 = e^{ i  \frac{\pi}{4} \Delta_1}$ & $\Delta_1 \in \mathds{Z}_8$ \\
\hline
$4mm1'$ & $e^{i \frac{\pi}{4} \hat{N}} C_4 = e^{ i  \frac{\pi}{4} \Delta_1}$ & $\Delta_1 \in \mathds{Z}_8$ \\
\hline
$61'$ & $e^{i \frac{\pi}{6} \hat{N}} C_6 = e^{ i  \frac{\pi}{6} \Delta_1}$ & $\Delta_1 \in \mathds{Z}_{12}$ \\
\hline
$6mm1'$ & $e^{i \frac{\pi}{6} \hat{N}} C_6 = e^{ i  \frac{\pi}{6} \Delta_1}$ & $\Delta_1 \in \mathds{Z}_{12}$ \\
\hline
\end{tabular}
\caption{Many-Body Local RSIs without SOC. 
\label{tab:RSInoSOC}
}
\end{table}

We now consider spin-orbit coupling (SOC) where $C_n^n= M^2=\mathcal{T}^2 = (-1)^{\hat{N}}$. Without $M$ and $\mathcal{T}$, the SOC case can be mapped immediately to the spinless (no SOC) case via $C_n = e^{i \frac{\pi}{n} \hat{N}} \tilde{C}_n$ where $\tilde{C}_n^n = +1$ is a spinless operator. Then \Eq{eq:cutoffinvrsis} yields the same classification for \Eqs{eq:RSIcneven}{eq:RSIcnodd}. This mapping would take $\tilde{C}_n^2 \to e^{-i \frac{2\pi}{n} \hat{N}} C_n^2$ for even $n$, but we find it more convenient to use the operator $e^{+i \frac{2\pi}{n} \hat{N}} C_n^2$ in order to obtain simple expressions matching the conventions of Ref. \cite{rsis}. This is clearly equivalent because the difference is a multiple of $e^{i \frac{2\pi}{n} \hat{N}}$ which is also a symmetry operator invariant under expansions of the cutoff, and $(e^{+i \frac{2\pi}{n} \hat{N}} C_n^2)^{n/2} = e^{i 2\pi \hat{N}} = +1$. Explicitly, for $n$ even we have (dropping the cutoff dependence $R$ for brevity)
\bea
\label{eq:RSIcnevenSOC}
 C_n \ket{GS} &= e^{i\frac{\pi}{n} \Delta_1} \ket{GS}, \qquad \Delta_1 \in \mathds{Z}_{2n} \\
e^{i \frac{2\pi}{n} \hat{N}} C_n^2 \ket{GS} &= e^{i\frac{4\pi}{n} \Delta_2} \ket{GS}, \qquad \Delta_2 \in \mathds{Z}_{n/2}  \\
\eea
which is to be compared with \Eq{eq:RSIcneven}, and for $n$ odd we have
\bea
\label{eq:RSIcnoddSOC}
e^{i \frac{2\pi}{n} \hat{N}} \ket{GS} &= e^{i\frac{2\pi}{n} \Delta_1} \ket{GS}, \qquad \Delta_1 \in \mathds{Z}_{n} \\
e^{-i \frac{\pi}{n} \hat{N}} C_n \ket{GS} &= e^{i\frac{2\pi}{n} \Delta_2} \ket{GS}, \qquad \Delta_2 \in \mathds{Z}_{n}  \\
\eea
to be compared with \Eq{eq:RSIcnodd}. The addition of mirrors and time-reversal with SOC differs from the case without SOC, as is also true in the single-particle case \cite{rsis}. We first consider adding only mirror symmetry $M$. Again we must consider the mirror subgroups as discussed in \Eq{eq:1M}. Following identical steps, we check that $M$ does not protect an additional RSI (exactly like in the spinless groups) and merely reduces the $C_n$-protected RSIs. We derive these reductions now. 

Because $M C_n M^\dag = C_n^\dag$, mirrors enforce $C_n \ket{GS,R} = \pm \ket{GS,R}$ which gives a $\mathds{Z}_2$ quantum number. Then for even $n$, we find $(-1)^{\hat{N}} = C_n^{n} = (\pm 1)^{n} = +1$ on $\ket{GS,R}$, so the number of particles is even if the groundstate is non-degenerate (as we have assumed throughout). Thus $e^{ i \frac{2\pi}{n} \hat{N}}$, which computes the number of particles mod $n$, only provides a $\mathds{Z}_{n/2}$ quantum number. The many-body local RSIs are the eigenvalues of
\bea
\label{eq:RSIcnevenSOC2}
C_n \ket{GS} &= e^{i \pi \Delta_1} \ket{GS}, \qquad \Delta_1 \in \mathds{Z}_{2} \\
e^{i \frac{2\pi}{n} \hat{N}} C_n^2 \ket{GS} = e^{i \frac{2\pi}{n} \hat{N}} \ket{GS} &= e^{i\frac{2\pi}{n/2} \Delta_2} \ket{GS}, \qquad \Delta_2 \in \mathds{Z}_{n/2}  \\
\eea
 and the many-body local RSI classification is $\mathds{Z}_2 \times \mathds{Z}_{n/2}$. 
 
We now consider the case where $n$ is odd. Then because the $C_n$ eigenvalue must be real (other $\mathcal{T}$ would enforce a double-degeneracy in contraction to our assumption of a non-degenerate groundstate), it follows that $(e^{-i \frac{\pi}{n} \hat{N}} C_n)^2 = e^{-i \frac{2\pi}{n} \hat{N}} C_n^2 = e^{-i \frac{2\pi}{n} \hat{N}}$ which reduces the independent RSIs, so we could pick the eigenvalues of $e^{-i \frac{\pi}{n} \hat{N}} C_n$ as the RSIs since $e^{i \frac{2\pi}{n} \hat{N}}$ is determined from them. However, it will be simpler to pick a convention when the eigenvalues of $e^{i \frac{2\pi}{n} \hat{N}}$ are the many-body local RSIs. This is an equivalent choice because 
\bea
e^{-i \frac{\pi}{n} \hat{N}} C_n = \lp (e^{-i \frac{\pi}{n} \hat{N}} C_n)^2 \rp^{(n+1)/2} =  \lp  e^{-i \frac{2\pi}{n} \hat{N}} \rp^{(n+1)/2}
\eea
using $(e^{-i \frac{\pi}{n} \hat{N}} C_n)^n = +1$ and that fact that $n$ is odd so $(n+1)/2$ is an integer. Thus $e^{-i \frac{\pi}{n} \hat{N}} C_n$ is entirely determined by $e^{i \frac{2\pi}{n} \hat{N}}$ and we can choose the many-body local RSIs to be 
 \bea
 \label{eq:RSIoddsocmirror}
 e^{i \frac{2\pi}{n}\hat{N}} \ket{GS} = e^{i \frac{2\pi}{n} \Delta_1} \ket{GS}, \qquad \Delta_1 \in \mathds{Z}_n \ . 
 \eea

We now consider the addition of spinful $\mathcal{T}$ (without mirrors) obeying $\mathcal{T}^2 = (-1)^{\hat{N}}$. Because $\mathcal{O}$ transforms in a 1D irrep since we require the groundstate to be non-degenerate, $N_{\mathcal{O}} = 2 p_\mathcal{O}$ must be even where $p_\mathcal{O}$ is the number of Kramers pairs.  Hence $\mathcal{T}^2 = +1$ on the (non-degenerate) groundstate. If $N_\mathcal{O}$ were odd, Kramer's theorem would forbid a non-degenerate state. We now use  \Eq{eq:overallsign} to determine the $C_n$ eigenvalue recalling $N_{\mathcal{O}}$ is even:
\bea
C_n \ket{GS,R'} = e^{i \la[C_n]} \det{}_{+} R[C_n] \ket{GS,R'} =  e^{i \la[C_n]} \ket{GS,R'} \\
\eea
where $g \mathcal{O}_i g^\dag = \sum_{j} R_{ij}[g] \mathcal{O}_j$ and $R[g]$ is an \emph{spinless} representation since $\mathcal{O}$ has an even number of particles and hence $C_n^n \mathcal{O}_i C_n^{n \dag} = + \mathcal{O}_i$. Then because $R[C_n]$ is a permutation matrix, $\det_+ R[C_n] = +1$. Thus the $C_n$ eigenvalue of the groundstate is a good many-body local RSI, but $\mathcal{T}$ restricts this eigenvalue to be real. If $n$ is even, $C_n = \pm1$ is allowed since $(-1)^{\hat{N}}  \ket{GS,R}  = C_n^n \ket{GS,R} = (\pm1)^n \ket{GS} = + \ket{GS}$. If $n$ is odd, then necessarily $C_n \ket{GS} = + \ket{GS}$  since the number of particles in $\ket{GS}$ is even due to $\mathcal{T}$ and the requirement of single-degeneracy. We define the many-body local RSIs as
\bea
C_n \ket{GS,R} &= e^{i \pi \Delta_1} \ket{GS,R}, \qquad e^{i \frac{2\pi}{n} \frac{\hat{N}}{2}} \ket{GS} = e^{i \frac{2\pi}{n} \Delta_2} \ket{GS}, \qquad \Delta_1,\Delta_2 \in \mathds{Z}_2, \mathds{Z}_n \qquad (n \text{ even}) \\
e^{i \frac{2\pi}{n} \frac{\hat{N}}{2}} \ket{GS} &= e^{i \frac{2\pi}{n} \Delta_1} \ket{GS}, \qquad \Delta_1 \in \mathds{Z}_n \qquad (n \text{ odd}) \\
\eea
noting that $\hat{N}/2$ counts the number of Kramer's pairs which is defined mod $n$. We check that adding mirrors gives the identical classification. The results are summarized in \Tab{tab:RSIswithsoc}. 

\begin{table}[H]
\centering
\begin{tabular}{| c | c | c |}
\hline
PG & RSI Operators  on $\ket{GS}$  & Classification \\ 
\hline
$m$ &  & $\Delta_1 \in \mathds{Z}_1$ \\
\hline
2 & $C_2 = e^{ i \frac{\pi}{2} \Delta_1}$ & $\Delta_1 \in \mathds{Z}_4$ \\
\hline
$2mm$& $C_2 = e^{ i \pi \Delta_1}$ & $\Delta_1 \in \mathds{Z}_2$ \\
\hline
3 & $\!e^{i \frac{2\pi}{3} \hat{N}} = e^{ i \frac{2\pi}{3} \Delta_1}, e^{-\frac{i\pi \hat{N}}{3} }C_3 = e^{ i \frac{2\pi}{3} \Delta_2} \!$ & $\Delta_1 \in \mathds{Z}_3,\Delta_2 \in \mathds{Z}_3$ \\
\hline
$3m$ & $e^{i \frac{2\pi}{3} \hat{N}} = e^{ i \frac{2\pi}{3} \Delta_1}$ & $\Delta_1 \in \mathds{Z}_3$ \\
\hline
4 & $C_4 = e^{ i \frac{\pi}{4} \Delta_1}, \ e^{i \frac{2\pi}{4} \hat{N}} C_4^2 = e^{ i \pi \Delta_2} $ & $\Delta_1 \in \mathds{Z}_8, \Delta_2 \in \mathds{Z}_2$ \\
\hline
$4mm$ & $C_4 = e^{ i \pi \Delta_1}, e^{i \frac{2\pi}{4} \hat{N}} = e^{i \pi \Delta_2}$ & $\Delta_1 \in \mathds{Z}_2, \Delta_2 \in \mathds{Z}_2$ \\
\hline
6 & $C_6 = e^{ i  \frac{\pi}{6} \Delta_1}, \ e^{i \frac{2\pi}{6} \hat{N}} C_6^2 = e^{ i \frac{2\pi}{3} \Delta_2} $ & $\Delta_1 \in \mathds{Z}_{12}, \Delta_2 \in \mathds{Z}_3$ \\
\hline
$6mm$ & $C_6 = e^{i \pi \Delta_1}, e^{i \frac{2\pi}{6} \hat{N}} = e^{i \frac{2\pi}{3} \Delta_2} $ & $\Delta_1 \in \mathds{Z}_{2},\Delta_2 \in \mathds{Z}_{3}$ \\
\hline
\end{tabular}
\begin{tabular}{| c | c | c |}
\hline
PG & RSI Operators on $\ket{GS}$ & Classification \\ 
\hline
$m1'$ &  & $\Delta_1 \in \mathds{Z}_1$ \\
\hline
$21'$ & $C_2 = e^{ i \pi \Delta_1}, e^{i \pi \frac{\hat{N}}{2}} = e^{i \pi \Delta_2} $ & $\Delta_1 \in \mathds{Z}_2, \Delta_2 \in \mathds{Z}_2$ \\
\hline
$2mm1'$& $C_2 = e^{ i \pi \Delta_1}, e^{i \pi \frac{\hat{N}}{2}} = e^{i \pi \Delta_2} $ & $\Delta_1 \in \mathds{Z}_2,\Delta_2 \in \mathds{Z}_2$ \\
\hline
$31'$ & $e^{i \frac{2\pi}{3} \frac{\hat{N}}{2}} = e^{ i \frac{2\pi}{3} \Delta_1}$ & $\Delta_1 \in \mathds{Z}_3$ \\
\hline
$3m1'$ & $e^{i \frac{2\pi}{3} \frac{\hat{N}}{2}} = e^{ i \frac{2\pi}{3} \Delta_1}$ & $\Delta_1 \in \mathds{Z}_3$ \\
\hline
$41'$ & $C_4 = e^{ i \pi  \Delta_1}, e^{i \frac{\pi}{2} \frac{\hat{N}}{2}} = e^{i \frac{\pi}{2} \Delta_2} $ & $\Delta_1 \in \mathds{Z}_2, \Delta_2 \in \mathds{Z}_4$ \\
\hline
$4mm1'$ & $C_4 = e^{ i \pi  \Delta_1}, e^{i \frac{\pi}{2} \frac{\hat{N}}{2}} = e^{i \frac{\pi}{2} \Delta_2}$ & $\Delta_1 \in \mathds{Z}_2, \Delta_2 \in \mathds{Z}_4$ \\
\hline
$61'$ & $C_6 = e^{ i  \pi \Delta_1}, e^{i \frac{2\pi}{6} \frac{\hat{N}}{2}} = e^{i \frac{2\pi}{6} \Delta_2} $ & $\Delta_1 \in \mathds{Z}_{2}, \Delta_2 \in \mathds{Z}_6$ \\
\hline
$6mm1'$ & $C_6 = e^{ i  \pi \Delta_1}, e^{i \frac{2\pi}{6} \frac{\hat{N}}{2}} = e^{i \frac{2\pi}{6} \Delta_2} $ & $\Delta_1 \in \mathds{Z}_{2}, \Delta_2 \in \mathds{Z}_6$ \\
\hline
\end{tabular}
\caption{Many-Body Local RSIs with SOC.
\label{tab:RSIswithsoc}
}
\end{table}

\subsection{Weakly Interacting Limit}
\label{app:weakly interacting}

In this section, we evaluate the many-body local RSIs when acting on product states, which have been classified in terms of single-particle RSIs \cite{rsis}. Since a product state is specified by the multiplicities of each of the irreps of the point group, we will obtain expressions for the many-body local RSIs in terms of the single-particle irrep multiplicities. We then compare these expressions to the expressions for the non-interacting RSIs (which are also defined on product states) and determine the reduction of the non-interacting RSI classification due to interactions. These expressions will also give us a way to evaluate the many-body local RSIs in terms of the non-interacting RSIs in the weak coupling limit. 

We first study $G_\mbf{x} = 2$ generated by $C_2$. In a product state with $m(A)$ even irreps and $m(B)$ odd irreps (see the irrep tables in \Tab{tab:2D-char}), we find
\bea
\label{eq:c2rsisex}
e^{i \frac{\pi}{2} \Delta_1} \ket{GS} = e^{i \frac{\pi}{2} \hat{N}} C_2 \ket{GS} &= e^{i \frac{\pi}{2} (m(A) + m(B))} e^{i \pi m(B)} \ket{GS} = e^{i \frac{\pi}{2} (m(A) - m(B))} \ket{GS}  \\
\eea
so we see that the many-body local RSI is given by $\Delta_1 = m(A) - m(B) \mod 4$. 

Next we study $G_\mbf{x} = 3$ generated by $C_3$. Following \Tab{tab:2D-char}, we find that
\bea
e^{i \frac{2\pi}{3} \hat{N}} \ket{GS} &= e^{i \frac{2\pi}{3} (m(A) + m({}^1E)+ m({}^2E))}  \ket{GS}, \\
C_n \ket{GS} &= e^{i \frac{2\pi}{3} (m({}^2E)- m({}^1E))} \ket{GS} \\
\eea
so $\Delta_1 = m(A) + m({}^1E)+ m({}^2E) \mod 3$ and $\Delta_2 = m({}^2E)- m({}^1E) \mod 3$ are good many-body local RSIs giving a $\mathds{Z}_3 \times \mathds{Z}_3$ classification. Formulae for the other spinless groups follow identically. In particular, we find that all values of the many-body local RSIs in all the spinless groups can be obtained in the non-interacting limit from product states. 

However, with SOC and $M$ or $\mathcal{T}$, we find that not all possible many-body RSIs can be obtained in the single-particle limit. Recall that $M$ or $\mathcal{T}$ impose $C_n \ket{GS} = \pm \ket{GS}$. Only $+$ sign is possible for single-particle states since $M$ and $\mathcal{T}$ pair the $C_n$ eigenvalues with their complex conjugates into double irreps. For example, consider $21'$ with SOC whose single irrep is ${}^1\overline{\text{E}} {}^2\overline{\text{E}}$ with the representation $D[C_2] = i \sigma_z$. From \Tab{tab:RSIswithsoc}, the many-body local RSIs of $21'$ are $C_2 \ket{GS} = e^{i \pi \Delta_1} \ket{GS}$ and $e^{i \pi \hat{N}/2} \ket{GS} = e^{i \pi \Delta_2} \ket{GS}$. All possible non-interacting product states are built from the operators in the form $c^\dag_{+i}c^\dag_{-i}$ which carry the ${}^1\overline{\text{E}} {}^2\overline{\text{E}}$ irrep. But since $C_2 c^\dag_{+i}c^\dag_{-i} C_2^\dag = (+i)(-i) c^\dag_{+i}c^\dag_{-i}  = + c^\dag_{+i}c^\dag_{-i} $, we find that $\Delta_1 = 0 \mod 2$ on all such states. Since $[\hat{N}, c^\dag_{+i}c^\dag_{-i} ] = 2 c^\dag_{+i}c^\dag_{-i}$, we see that $\Delta_2 = m({}^1\overline{\text{E}} {}^2\overline{\text{E}}) \mod 2$ which simply counts the number of Kramers pairs. However, it is simple to write down a state with $\Delta_1 = 1 \mod 2$. For instance, we take $\ket{GS} = \frac{1}{\sqrt{2}} (c^\dag_{+i,1} c^\dag_{+i,2}+ c^\dag_{-i,1} c^\dag_{-i,2}) \ket{0}$ whose terms individually would be degenerate by Kramers theorem. We compute
\bea
C_2 \ket{GS} &= \frac{1}{\sqrt{2}} ((+i)^2 c^\dag_{+i,1} c^\dag_{+i,2}+ (-i)^2 c^\dag_{-i,1} c^\dag_{-i,2}) \ket{0} = - \ket{GS} \\
\mathcal{T} \ket{GS} &=  \frac{1}{\sqrt{2}} (c^\dag_{-i,1} c^\dag_{-i,2}+ (-1)^2 c^\dag_{+i,1} c^\dag_{+i,2}) \ket{0}= \ket{GS} \\
\eea
which shows that $\ket{GS}$ has $\Delta_1 = 1 \mod 2$ and is allowed to be non-degenerate since $\mathcal{T}$ squares to $+1$ on $\ket{GS}$. Because we showed above that $\Delta_1 = 0$ in all product states, the many-body local RSI $\Delta_1 = 1 \mod 2$ proves that $\ket{GS}$ cannot be adiabatically connected to any product state. The fact that our classification includes states with $\Delta_1 \neq 0$, which is impossible in any non-interacting atomic limit state, underscores our non-perturbative construction of the many-body local RSIs. 

\begingroup
\renewcommand\arraystretch{1.3}
\begin{longtable}{| C{.04\textwidth} | p{.25\textwidth} | p{.25\textwidth} || p{.23\textwidth} | p{.23\textwidth}|} \caption{Many-Body Local RSIs of Product States in terms of Irrep Multiplicities\label{tab:mpgrsi}} \\
\hline
\multirow{2}{*}{RSIs} 
& \multicolumn{2}{c||}{No SOC \ $ ( C_n^n = M_i^2  = +1)$}  & \multicolumn{2}{c|}{SOC  \ $ ( C_n^n = M_i^2 = (-1)^{\hat{N}})$}  \\
\cline{2-5}
 & No TRS & TRS \ $ (\mathcal{T}^2 = +1)$ & No TRS & TRS \ $ (\mathcal{T}^2 = (-1)^{\hat{N}})$\\
\hline\hline
$m$ &  & & &  \\
\hline

$2$ & $\Delta_1 = m(A) - m(B) \mod 4 $ & $\Delta_1 = m(A) - m(B) \mod 4 $ & $\Delta_1 = m({}^1\overline{\text{E}}) - m({}^2\overline{\text{E}}) \mod 4$ & $\Delta_1 = 0 \mod 2, \quad \Delta_2 = m({}^1\overline{\text{E}}{}^2\overline{\text{E}}) \mod 2$ \\
\hline

\multirow{2}{*}{$2mm$} 
 & $\Delta_1 = m(A_1) + m(A_2)  - m(B_1) - m(B_2) \mod 4$ & $\Delta_1 = m(A_1) + m(A_2)  - m(B_1) - m(B_2) \mod 4$ & $\Delta_1 = 0 \mod 2$ & $\Delta_1 = 0 \mod 2 , \qquad \Delta_2 = m(\overline{\text{E}}) \mod 2  $ \\
\hline

\multirow{4}{*}{$4$} 
 & $ \Delta_1 = m(A) - 3m(B) - m({}^1\text{E}) + 3 m({}^2\text{E})  \mod 8$ & $\Delta_1 = m(A) - 3m(B) + 2 m({}^1\text{E}{}^2\text{E})\mod 8$ & $\Delta_1 =  m({}^1\overline{\text{E}}_1) -  m({}^2\overline{\text{E}}_1) -  m({}^2\overline{\text{E}}_2) +  m({}^1\overline{\text{E}}_2)  \mod 8$  & $\Delta_1 =  0 \mod 2$  \\
 & $\Delta_2 = m({}^1\text{E})- m({}^2\text{E}) \mod 2 $ & & $ \Delta_2 = m({}^1\overline{\text{E}}_1)-m({}^1\overline{\text{E}}_2) \mod 2$  & $\Delta_2 =  m({}^1\overline{\text{E}}_1 {}^2\overline{\text{E}}_1) + m({}^1\overline{\text{E}}_2 {}^2\overline{\text{E}}_2)  \mod 4$\\
\hline

\multirow{2}{*}{$4mm$} 
 &$ \Delta_1 = m(A_1)\! +m(A_2)\! -3m(B_1) -3m(B_2) + 2m(E) \mod 8$ & $\Delta_1 = m(A_1)\! +m(A_2)\! -3m(B_1) -3m(B_2) + 2m(E) \mod 8$ &$ \Delta_1 = 0 \mod 2, \hfill \Delta_2 = m(\overline{\text{E}}_1)+m(\overline{\text{E}}_2) \!\mod 2$ & $ \Delta_1 = 0 \mod 2, \quad \Delta_2 = m(E_1) + m(E_2) \mod 4$\\
\hline

\multirow{2}{*}{$3$} 
 & $ \Delta_1 = m(A) + m({}^1E) + m({}^2E) \mod 3$ & $\Delta_1 = m(A) + 2m({}^1E{}^2E) \mod 3$ & $ \Delta_1 \! \!= m(\overline{\text{E}}) + m({}^1\overline{\text{E}})+ m({}^2\overline{\text{E}}) \mod 3$ & $\Delta_1 \! \!=  \! \!m(\overline{\text{E}} \overline{\text{E}}) + m({}^1\overline{\text{E}} {}^2\overline{\text{E}})  \mod 3$  \\
 & $ \Delta_2 = m({}^2E)-m({}^1E) \mod 3 $ & & $ \Delta_2 = m(\overline{\text{E}}) - m({}^1\overline{\text{E}}) \mod 3$ &  \\
\hline

\multirow{2}{*}{$3m$} 
 & $\Delta_1 = m(A_1) +m(A_2) + 2m(E) \mod 3$ & $ \Delta_1 = m(A_1) +m(A_2) + 2m(E) \mod 3$& $\Delta_1 = - m(\overline{\text{E}}_1) + m({}^1 \overline{\text{E}}) + m({}^2 \overline{\text{E}}) \mod 3$& $\Delta_1 = m( \overline{\text{E}}_1) + m({}^1\overline{\text{E}}{}^2 \overline{\text{E}}) \mod 3$ \\
\hline

\multirow{5}{*}{$6$} 
 & $\Delta_1 = m(A) -5m(B) - m({}^1 \text{E}_2) + 3 m({}^2 \text{E}_2) + 5 m({}^1 \text{E}_1) - 3 m({}^2 \text{E}_1) \mod 12$ & $\Delta_1 = m(A)-5m(B)+ 2 m({}^1\text{E}_2{}^2\text{E}_2) +2 m({}^1\text{E}_1{}^2\text{E}_1) \mod 12$& $ \Delta_1 = m({}^2\overline{\text{E}}_3)-m({}^1\overline{\text{E}}_3) - 3 m({}^2\overline{\text{E}}_1)+3 m({}^1\overline{\text{E}}_1) - 5 m({}^2\overline{\text{E}}_2) + 5 m({}^1\overline{\text{E}}_2)\mod 12$ &$\Delta_1 = 0\mod 2 $  \\
  & $\Delta_2 = 2 m({}^1 \text{E}_2)  + m({}^2 \text{E}_2) + 2 m({}^1 \text{E}_1) + m({}^2 \text{E}_1)\mod 3$ & & $ \Delta_2 = m({}^2\overline{\text{E}}_3)- m({}^2\overline{\text{E}}_1) - m({}^1\overline{\text{E}}_1) + m({}^2\overline{\text{E}}_2)\mod 3$ &  $\Delta_2 = m({}^1\overline{\text{E}}_1 {}^2\overline{\text{E}}_1) + m({}^1\overline{\text{E}}_2 {}^2\overline{\text{E}}_2)  +   m({}^1\overline{\text{E}}_3 {}^2\overline{\text{E}}_3) \mod 6 $\\
 \hline

 \multirow{2}{*}{$6mm$} 
 & $ \Delta_1 = m(A_1)+m(A_2) -5m(B_1)-5m(B_2) + 2 m(\text{E}_1) +2 m(\text{E}_2) \mod 12$ & $ \Delta_1 = m(A_1)+m(A_2) -5m(B_1)-5m(B_2) + 2 m(\text{E}_1) +2 m(\text{E}_2)\mod 12$ & $\Delta_1 =0 \mod 2, \quad \Delta_2 = m(\overline{\text{E}}_1) + m(\overline{\text{E}}_2) + m(\overline{\text{E}}_3) \mod 3$ & $\Delta_1 = 0 \mod 2, \quad \Delta_2 = m(\overline{\text{E}}_1) + m(\overline{\text{E}}_2) + m(\overline{\text{E}}_3) \mod 6$\\
 \hline
\end{longtable}
\endgroup

Having derived expressions for the many-body local RSIs (see \Tab{tab:mpgrsi}) in terms of the irrep multiplicities in the weak coupling limit (some of which are identically zero without interactions), we now show that they can be expressed in terms of the single-particle RSIs first derived in Ref. \cite{rsis}. This is a good consistency check of our definition of the many-body RSIs which is a priori much different than the group theoretical construction of Ref. \cite{rsis}. It will also be instructive to see how the single-particle RSIs collapse to many-body RSIs as weak interactions are turned on. In \Tab{tab:rsiredint}, we collect these reductions explicitly. If a gap is not closed, then the many-body local RSIs computed from the single-particle RSIs describe the interacting groundstate of the system. 

We now give an example of one of the calculations in \Tab{tab:rsiredint}. Let us study PG $41'$ without SOC. This group has 3 co-irreps which are presented by
\bea
A: & \qquad D_A[C_4] = 1, D_A[\mathcal{T}] = K \\
B: & \qquad D_B[C_4] = -1, D_B[\mathcal{T}] = K \\
{}^1E\,{}^2\!E: & \qquad D_{{}^1E{}^2E}[C_4] = \bpm i & 0 \\0 & -i\epm,  D_{{}^1E{}^2E}[\mathcal{T}] = \bpm 0 & 1 \\1 & 0\epm K\\
\eea
where $K$ is complex conjugation. The non-interacting RSIs are invariant under the induction/reduction of $E \uparrow 4 = A \oplus B \oplus {}^1E{}^2E$ \cite{rsis}. We find a $\mathds{Z}^2$ classification
\bea
\delta_1 &= - m(A) + m({}^1E{}^2E), \quad \delta_2 = - m(A) + m(B) \ . \\
\eea
From \Tab{tab:mpgrsi}, we see that the many-body local RSI is 
\bea
\Delta_1 = m(A) - 3m(B) + 2 m({}^1\text{E}{}^2\text{E})\mod 8
\eea
when evaluated on a product state. Now observe that
\bea
2 \delta_1 - 3 \delta_2 \mod 8 &=  - 2 m(A) + 2 m({}^1E{}^2E) + 3 m(A) -3 m(B) \mod 8 = \Delta_1 \ . \\
\eea
Identical calculations hold in the other groups.

\begingroup
\renewcommand\arraystretch{1.3}
\begin{longtable}{| C{.04\textwidth} | p{.25\textwidth} | p{.25\textwidth} || p{.23\textwidth} | p{.23\textwidth}|} 
		\caption{Many-Body Local RSI Reduction to Single-Particle RSIs in Product States  \label{tab:rsiredint}}\\ 
\hline
\multirow{2}{*}{RSIs} 
& \multicolumn{2}{c||}{No SOC \ $ ( C_n^n = M_i^2 = \mathcal{T}^2 = +1)$}  & \multicolumn{2}{c|}{SOC  \ $ ( C_n^n = M_i^2 = \mathcal{T}^2 = (-1)^{\hat{N}})$}  \\
\cline{2-5}
 & No TRS & TRS  & No TRS & TRS\\
\hline\hline
$m$ & & &   & \\
\hline

$2$ & $\Delta_1 = -\delta_1 \mod 4 $ & $\Delta_1 = -\delta_1 \mod 4 $ & $\Delta_1 = -\delta_1 \mod 4$ & $\Delta_1 = 0, \ \Delta_2 = \delta_1$ \\
\hline

\multirow{1}{*}{$2mm$} 
 & $\Delta_1 = -\delta_1 \mod 4$ & $\Delta_1 = -\delta_1 \mod 4$ & $\Delta_1 = 0$ & $\Delta_1 = 0, \ \Delta_2 = \delta_1 \mod 2  $ \\
\hline

\multirow{2}{*}{$4$} 
 & $ \Delta_1 = 3 \delta_1- 3 \delta_2 - \delta_3 \mod 8$ & $\Delta_1 =2 \delta_1 - 3 \delta_2 \mod 8$ & $\Delta_1 = 3 \delta_1- 3 \delta_2 - \delta_3 \mod 8$  & $\Delta_1 =  0$  \\
 & $\Delta_2 = \delta_1 - \delta_3 \mod 2  $ & & $ \Delta_2 =   \delta_1 - \delta_3 \mod 2$  & $\Delta_2 =  \delta_1 - 2 \delta_2  \mod 4$ \\
\hline
\multirow{1}{*}{$4mm$} 
 &$ \Delta_1 = 2 \delta_1 - 3\delta_2 \mod 8$ & $\Delta_1 = 2 \delta_1 - 3\delta_2 \mod 8$ &$\Delta_1 = 0, \ \Delta_2 = \delta_1  \mod 2$ & $\Delta_1 = 0, \ \Delta_2 = \delta_1 - 2 \delta_2 \mod 4$\\
\hline

\multirow{2}{*}{$3$} 
 & $ \Delta_1 = \delta_1 + \delta_2 \mod 3$ & $\Delta_1  = - \delta_1  \text{\,mod } 3$ & $ \Delta_1 =\delta_1 +\delta_2 \mod 3$ & $\Delta_1 =  - \delta_1 \mod 3$  \\
 & $ \Delta_2 = \delta_2 - \delta_1 \mod 3 $ & & $ \Delta_2 = \delta_2- \delta_1 \mod 3$ &  \\
\hline

\multirow{1}{*}{$3m$} 
 & $\Delta_1 = -\delta_1  \mod 3$ & $ \Delta_1 = -\delta_1 \mod 3$& $\Delta_1 = \delta_1 \mod 3$& $\Delta_1 = - \delta_1 \mod 3$ \\
\hline
\multirow{3}{*}{$6$} 
 & $\Delta_1 = - \delta_1 - 5 \delta_3 + 3 \delta_5 + 5 \delta_4 - 3 \delta_2 \mod 12$ &$\Delta_1 = 2\delta_1 + 2 \delta_2 -5 \delta_3 \mod 12$& $ \Delta_1 =  - \delta_1 - 5 \delta_3 + 3 \delta_5 + 5 \delta_4 - 3 \delta_2 \mod 12$ &$\Delta_1 = 0 $  \\
  & $\Delta_2 = 2 \delta_1 + \delta_5 + 2 \delta_4 + \delta_2 \mod 3$ & & $ \Delta_2 = 2 \delta_1 + \delta_5 + 2 \delta_4 + \delta_2 \mod 3$ & $\Delta_2 = \delta_1 + \delta_2 - 3 \delta_3 \mod 6 $  \\
 \hline
 \multirow{1}{*}{$6mm$} 
 & $ \Delta_1 = 2 \delta_1 + 2 \delta_2- 5 \delta_3 \mod 12$ & $ \Delta_1 = 2 \delta_1 + 2 \delta_2 -5 \delta_3 \mod 12$ & $\Delta_1 = 0, \Delta_2 = \delta_1 + \delta_2 \mod 3$ & $\Delta_1 = 0, \ \Delta_2 = \delta_1 + \delta_2 - 3 \delta_3 \mod 6$ \\
 \hline
\end{longtable}
\endgroup
We see that in all cases, the many-body local RSIs can be written in terms of single-particle RSIs. As we will show in \App{app:fragile}, certain single-particle fragile topology can be trivialized by adding interactions due to the reduction of single-particle RSIs to many-body local RSIs because of the mod factors in \Tab{tab:rsiredint}. Convenient expressions for the single-particle RSIs in terms of the momentum space irreps may be found in the appendices of Ref. \cite{rsis} and thus \Tab{tab:rsiredint} gives a direct map from single-particle topology to many-body fragile topology in the weak coupling limit.

\section{Fragile Topology and Constraints on Particle Number}
\label{app:fragile}

In this Appendix, we define and study many-body fragile topology using many-body local RSIs. In \App{app:constraints}, we derive constraints between the many-body RSIs and the minimum number of particles necessary to obtain the many-body local RSIs in a many-body atomic state. In many-body fragile topological phases, the many-body local RSIs can be uniquely defined at each Wyckoff position through the addition of many-body atomic states (\App{app:fragiledef}). Such phases are indicated when an inequality comparing RSIs to the $U(1)$ particle number in any many-body atomic state is violated, indicating an obstruction to adiabatic deformation into a many-body atomic state. We enumerate inequality criteria in all 2D wallpaper groups with and without SOC and time-reversal in \App{app:fragineq}. 

\subsection{Orbital Number Constraints}
\label{app:constraints}

In this section, we produce constraints between the many-body local RSIs and the total particle number per unit cell in many-body atomic states. In particular, we find that nonzero many-body local RSIs lower bound the particle number. The inequalities derived here are crucial for determining many-body fragile topological indices. We consider a fixed Wyckoff position $\mbf{x}$ and let $N$ denote the total charge of $\ket{GS,R}$, the state with OBCs respecting $G_\mbf{x}$. Although $N$ depends on $R$, the lower bound is in terms of local RSIs and hence is independent of $R$. Shrinking $R$ to include only orbitals at $\mbf{x}$ produces a lower bound for the total charge $N$ at $\mbf{x}$. 

We can systematically develop inequalities thanks to the general form of the many-body local RSI symmetry operators derived in \App{app:RSIconstruct}. To begin, we consider the spinless rotation groups which for even $n$ have the many-body local RSIs
\bea
\label{eq:rsireproduct}
e^{i \frac{\pi}{n} \hat{N}} C_n \ket{GS} = e^{i \frac{\pi}{n} \Delta_1} \ket{GS}, \qquad C^2_n \ket{GS} = e^{i \frac{2\pi}{n/2} \Delta_2} \ket{GS} 
\eea
with $\Delta_1 \in \mathds{Z}_{2n}, \Delta_2 \in \mathds{Z}_{n/2}$. Noting that $(e^{i \frac{\pi}{n} \hat{N}} C_n)^2 (C^{2}_n)^\dag = e^{i \frac{2\pi}{n} \hat{N}}$, we see immediately that $N = \Delta_1 - 2 \Delta_2 \mod n$ and thus $N \geq (\Delta_1 - 2 \Delta_2 \mod n)$ where $(a \mod n)$ is taken to be non-negative. However, this inequality is not tight and can be improved. To see that it is not tight, consider $G_\mbf{x} = 2$ in \Tab{tab:mpgrsi} where $\Delta_1 = m(A) - m(B) \mod 4$ when acting on product states ($\Delta_2 = 0$ is trivial) and the inequality is $N \geq \Delta_1 \mod 2$. The only state with zero particles is the vacuum with $\Delta_1 = 0 \mod 4$ and the only one-particle states are those with an $A$ irrep or a $B$ irrep which have $\Delta_1 = \pm 1 \mod 4$. Thus $\Delta_1 = 2 \mod 4$ can only be obtained with 2 or more particles, e.g. $m(A) = 2$ or $m(B) = 2$, even though $\Delta_1 \mod 2 = 0$.  

To tighten the bound, we prove the following. If $\Delta_1 - 2 \Delta_2 = 0 \mod n$ but $\Delta_1 \neq 0$, then $N \geq n$. The proof is by contradiction. Because $\Delta_1 - 2 \Delta_2 = N \mod n$, if $\Delta_1 - 2 \Delta_2 = 0 \mod n$ then $N$ must be a multiple of $n$. Assume for contradiction that $N = 0$, so the groundstate $\ket{0}$ is the vacuum. Then $e^{i \frac{\pi}{n} \hat{N}} C_n \ket{0} = + \ket{0}$ implying that $\Delta_1 =0$, reaching a contradiction. Thus $N$ is a nonzero multiple of $n$, so $N \geq n$. 

We now discuss the case of odd $n$ whose many-body local RSIs are $e^{\frac{2\pi i}{n} \hat{N}} \ket{GS} = e^{\frac{2\pi i}{n} \Delta_1} \ket{GS}$ and  $C_n \ket{GS} = e^{\frac{2\pi i}{n} \Delta_2} \ket{GS}$. As before, we have the simple inequality $N \geq \Delta_1 \mod n$ which is not tight at $N=n$ (for example with $n=3$, a product state with $m(A) = 1, m({}^1\text{E}) = 2$ has $\Delta_1 = 0$ from \Tab{tab:mpgrsi}). We improve bound with the same contraction argument as before: if $\Delta_1 = 0 \mod n$ but $\Delta_2 \neq 0$, then $N \mod n = 0$ but $N \neq 0$ or else $\Delta_2 = 0$. 

It is now convenient to define the function
\bea
\text{mod}_n(\Delta|\Delta') = \begin{cases}
n, & \Delta \mod n = 0, \quad \Delta' \neq 0  \\
(\Delta \mod n) \in (0,\dots, n-1), & \text{otherwise}
\end{cases} \ .
\eea
The preceding paragraphs can now be stated $N \geq \text{mod}_n(\Delta_1 + 2 \Delta_2|\Delta_1)$ for $n$ even and $N \geq \text{mod}_n(\Delta_1|\Delta_2)$ for $n$ odd in the spinless rotation groups. Since we found in \App{eq:generalRSIs} that all the many-body local RSIs in the spinless 2D point groups can be obtained from their rotation subgroups by reduction $\Delta_1 \to \Delta_1, \Delta_2 \to 0$, we then obtain the bounds $N \geq \text{mod}_n(\Delta_1|\Delta_1)$ in all the spinless groups with mirrors and/or time-reversal. 

We now consider the spinful rotation groups where the RSIs are (see \Eq{eq:RSIcnevenSOC}) $ C_n \ket{GS} = e^{i\frac{\pi}{n} \Delta_1} \ket{GS}$ and $e^{i \frac{2\pi}{n} \hat{N}} C_n^2  \ket{GS} = e^{i\frac{4\pi}{n} \Delta_2} \ket{GS}$ for even $n$. Since $e^{i \frac{2\pi}{n} \hat{N}} C_n^2 (C^\dag_n)^2 \ket{GS} = e^{i \frac{2\pi}{n} \hat{N}} \ket{GS} = e^{i \frac{2\pi}{n} (2\Delta_2 - \Delta_1)} \ket{GS}$, we see that $N =  2 \Delta_2 - \Delta_1 \mod n$, and by the same argument as above, we tighten the bound but noting that if  $2 \Delta_2 - \Delta_1 = 0 \mod n$ but $\Delta_1 \neq 0$, then $N \geq n$. The bound for groups with mirrors follows by reduction of the RSIs as shown now. Adding mirrors reduces the local RSIs (see \Eq{eq:RSIcneven}) such that $N =  2 \Delta_2 \mod n$. If $n$ is odd, \Eq{eq:RSIcnoddSOC} immediately gives $N = \Delta_1 \mod n$ and $N \geq n$ if $\Delta_1 = 0 \mod n$ and $\Delta_2 \neq 0$. With mirrors, $\Delta_2$ reduces to zero (see \Eq{eq:RSIoddsocmirror}). 

We now consider the addition of time-reversal symmetry. In the spinful groups with SOC, the many-body RSI $N/2 = \Delta_2 \mod n$ counts the number Kramers pairs, and $\Delta_2=0 \mod n, \Delta_1\neq 0$ implies $N/2 \geq n$ for $n$ even. For $n$ odd, there is only one many-body local RSI $\Delta_1$ which counts the number of Kramers pairs. 

With these results, we compile the local RSI lower bounds in \Tab{tab:boundsrsi}. 

\begingroup
\renewcommand\arraystretch{1.3}
\begin{longtable}{| C{.04\textwidth} | p{.25\textwidth} | p{.25\textwidth} || p{.23\textwidth} | p{.23\textwidth}|} 
		\caption{Local $U(1)$ Particle-number Constraints from many-body local RSIs. All mod functions are defined to be non-negative. 
		\label{tab:boundsrsi}}\\ 
\hline
\multirow{2}{*}{} 
& \multicolumn{2}{c||}{No SOC \ $ ( C_n^n = M_i^2  = +1)$}  & \multicolumn{2}{c|}{SOC  \ $ ( C_n^n = M_i^2 = -1)$}  \\
\cline{2-5}
 & No TRS & TRS \ $ (\mathcal{T}^2 = +1)$ & No TRS & TRS \ $ (\mathcal{T}^2 = -1)$\\
\hline\hline
$m$ &   & & &  \\
\hline

$2$ & $N \geq  \text{mod}_2(\Delta_1|\Delta_1)$ & $N \geq  \text{mod}_2(\Delta_1|\Delta_1)$ & $N \geq  \text{mod}_2(\Delta_1|\Delta_1)$ & $N/2 \geq \text{mod}_2(\Delta_2|\Delta_1)$ \\
\hline

\multirow{1}{*}{$2mm$} 
 & $N \geq  \text{mod}_2(\Delta_1|\Delta_1) $ & $N \geq  \text{mod}_2(\Delta_1|\Delta_1)$ & $N \geq  \text{mod}_2(0|\Delta_1)$ & $N/2 \geq \text{mod}_2(\Delta_2|\Delta_1)$ \\
\hline

\multirow{1}{*}{$4$} 
 & $N \geq  \text{mod}_4(\Delta_1- 2\Delta_2|\Delta_1)$ & $N \geq   \text{mod}_4(\Delta_1|\Delta_1)$ & $N \geq   \text{mod}_4(2\Delta_2-\Delta_1|\Delta_1)$  & $N/2 \geq \text{mod}_4(\Delta_2|\Delta_1)$  \\
\hline

\multirow{1}{*}{$4mm$} 
 &$N \geq   \text{mod}_4(\Delta_1|\Delta_1)$ & $N \geq   \text{mod}_4(\Delta_1|\Delta_1)$ &$N \geq  \text{mod}_4(2\Delta_2|\Delta_1)$ & $N/2 \geq  \text{mod}_4(\Delta_2|\Delta_1)$\\
\hline

\multirow{1}{*}{$3$} 
 & $N \geq  \text{mod}_3(\Delta_1|\Delta_2)$ & $ N \geq   \text{mod}_3(\Delta_1|\Delta_1)$ & $N \geq  \text{mod}_3(\Delta_1|\Delta_2)$ & $ N/2 \geq \text{mod}_3(\Delta_1|\Delta_1)$  \\
 \hline

\multirow{1}{*}{$3m$} 
 & $N \geq \text{mod}_3(\Delta_1|\Delta_1)$ & $N \geq  \text{mod}_3(\Delta_1|\Delta_1)$ & $N \geq  \text{mod}_3(\Delta_1|\Delta_1)$ &$N/2 \geq \text{mod}_3(\Delta_1|\Delta_1)$ \\
\hline

\multirow{1}{*}{$6$} 
 & $N \geq  \text{mod}_6(\Delta_1-2\Delta_2|\Delta_1)$ & $N \geq  \text{mod}_6(\Delta_1|\Delta_1)$& $N \geq  \text{mod}_6(2\Delta_2-\Delta_1|\Delta_1)$ & $N/2 \geq \text{mod}_6(\Delta_2|\Delta_1)$  \\
 \hline

 \multirow{1}{*}{$6mm$} 
 & $N \geq  \text{mod}_6(\Delta_1|\Delta_1)$ & $ N \geq  \text{mod}_6(\Delta_1|\Delta_1)$ & $N \geq  \text{mod}_6(2\Delta_2|\Delta_1)$ & $N/2 \geq \text{mod}_6(\Delta_2|\Delta_1)$ \\
 \hline

\end{longtable}
\endgroup

\subsection{Defining Many-Body Local RSIs in Fragile States}
\label{app:fragiledef}

We now explain how to define the  many-body local RSIs of a (many-body) fragile topological state following the method of the Main Text. First we recall that the many-body local RSIs are defined on symmetry-respecting OBCs in many-body trivial atomic limits. Crucially, such phases are non-degenerate on OBCs, while fragile phases (or obstructed atomic limit phases) \emph{are} degenerate at filling $\nu = N_{occ}/N_{orb}$. Thus we cannot \emph{directly} calculate the  many-body local RSIs of fragile states by placing them on OBCs. 

Instead, we define many-body local RSIs using the definition of a many-body fragile phase: although one cannot adiabatically deform such a state to a many-body atomic state at fixed particle number, an adiabatic deformation is possible when coupled to a many-body atomic limit (thereby increasing the particle number). This definition encompasses single-particle fragile phases \cite{rsis}. To start with, we assume PBCs so we only need to consider the non-triviality of the bulk (not the edges or corners). Explicitly, we denote the fragile state by $\ket{F} = F^\dag \ket{0}$ which is trivialized through the addition of a trivial atomic limit state $\ket{A}= A^\dag \ket{0}$. By trivialized, we mean that $A^\dag F^\dag \ket{0}$ can be adiabatically deformed to a many-body trivial atomic state $\ket{A'}$ at fixed particle number. If necessary, new orbitals may be added to the Hilbert space to define $A$. This is only required in a tight-binding model where there is a finite local Hilbert space (as opposed to a continuum model). The set of many-body local RSIs of $\ket{A}$ and $\ket{A'}$ are well-defined and are denoted by $\Delta_{A}$ and $\Delta_{A'}$ respectively. Then because the many-body local RSI groups are abelian (they are additive under stacking as discussed in the Main Text), we define the RSIs of the fragile state by $\Delta_F = \Delta_{A'} - \Delta_{A}$. Note that $\Delta_F$ is invariant under $A^\dag \to A^\dag B^\dag$ for any many-body atomic state $B^\dag$ because $\Delta_{A'B} - \Delta_{AB} = \Delta_{A'} + \Delta_{B} - \Delta_{A} - \Delta_B = \Delta_{A'} - \Delta_{A}$. It appears that $\Delta_F$ depends on the atomic state $\ket{A}$ used to trivialize $\ket{F}$ rather than being an intrinsic property of $\ket{F}$. However, we can prove that $\Delta_F$ is unique and does not depend on $A$ because the RSI groups are abelian. Consider two distinct operators  $A^\dag$, $\tilde{A}^\dag$ for which $F^\dag A^\dag\ket{0}, F^\dag \tilde{A}^\dag\ket{0}$ are both many-body atomic with corresponding RSIs $\Delta_F = \Delta_{FA}-\Delta_{A}, \Delta_{\tilde{F}} = \Delta_{F\tilde{A}}-\Delta_{\tilde{A}}$ respectively. Then $F^\dag A^\dag \tilde{A}^\dag \ket{0}$ is many-body atomic, and its many-body local RSIs $\Delta_{FAA'} - \Delta_{AA'}$ are equal to $\Delta_F$ and $\Delta_{\tilde{F}}$, since the RSIs of the fragile state are invariant under the addition of atomic states as mentioned in the prior paragraph. Hence $\Delta_F = \Delta_{\tilde{F}}$. It is worth underscoring that $A'$ must be a trivial atomic state rather than an obstructed atomic state. To understand this requirement, recall that in non-interacting Hamiltonians, the orbitals used to trivialize fragile bands may result in obstructed atomic bands. That is insufficient in our framework for computing the many-body local RSIs on OBCs because obstructed atomic bands host corner states. Momentarily, we will discuss how adding \emph{more} (appropriately chosen) atomic orbitals can resolve the obstruction and thereby obtain a trivial atomic limit.

Although the many-body local RSIs defined here do not depend on which many-body atomic operator $A$ is used to trivialize $\ket{F}$, it is necessary to find a known one in order to actually perform a computation. We now outline a strategy to find $A$ in the band theory limit where single-particle fragile states have been classified \cite{rsis,2019arXiv190503262S}. On PBCs, fragile phases can be represented in the form $(\rho \ominus \rho') \uparrow G$ where $\rho,\rho'$ represent atomic orbitals in the unit cell (see Refs. \cite{rsis} and \cite{2019arXiv190503262S} for details and the Main Text for an example). The obstruction to trivialization is represented by $\ominus \rho'$, and can be removed on PBCs by coupling orbitals $\rho'$ to the fragile band (without closing a gap). However, it is then possible that $\rho$ is an obstructed atomic state: it is consistent with the symmetries of an atomic state, but not one that appears in the tight-binding orbital basis. On OBCs, such an obstructed atomic state has a filling anomaly \cite{2018arXiv181002373W} which enforces degeneracy, preventing us from calculating the many-body local RSIs. However, we will now show that an obstructed atomic state can be connected to a trivial atomic state (at different particle number) by taking \emph{more} orbitals in the choice of $A$, so that non-degenerate OBCs are obtained upon truncating the PBC model. We explain this first on PBCs by showing how to resolve the bulk obstruction to an atomic limit. Afterwards, we give a complementary discussion on OBCs with a numerical example.

On PBCs, an obstructed atomic limit is defined by the representation $\rho_O \uparrow G$ which is compatible with atomic orbitals, but \emph{not} those which appear in the basis of the tight-binding model. This obstruction is removed simply by adding atomic orbitals in the representation $\rho_O$ to the tight-binding basis. This is intuitive because the ``obstruction" of an obstructed atomic limit state is related to the basis of atomic orbitals in the tight-binding Hamiltonian. By adding the correct orbitals to this basis, the obstruction is resolved. Explicitly, we consider an obstructed atomic insulator adiabatically tuned to the flat band limit where the Hamiltonian can be written on PBCs as
\bea
\label{eq:orig4orb}
H_{OAI} &= - t \sum_{\mbf{R}, n \in \rho_O} w^\dag_{\mbf{R},n} w_{\mbf{R},n}
\eea
where $w^\dag_{\mbf{R},n}$ are the Wannier functions carrying the representation of the $n$th orbital in $\rho_O$. In $H_{OAI}$, the obstructed valence bands are at energy $-t < 0$, and the conduction bands are at energy $0$ (thus their creation operators do not appear in the Hamiltonian). To trivialize the obstructed atomic limit, we add atomic $c^\dag_{\mbf{R},n}$ orbitals transforming in the same representation at $w^\dag_{\mbf{R},n}$. We propose the new Hamiltonian $H(\th)$ on PBCs
\bea
\label{eq:HOALIth}
H(\th) &= - t \cos \th \sum_{\mbf{R}, n \in \rho_O} w^\dag_{\mbf{R},n} w_{\mbf{R},n} + t \cos \th \sum_{\mbf{R}, n \in \rho_O} c^\dag_{\mbf{R},n} c_{\mbf{R},n} + t \sin \th \lp \sum_{\mbf{R},n\in \rho_O} w^\dag_{\mbf{R},n} c_{\mbf{R},n} + h.c. \rp \\
&= -t \sum_{\mbf{R},n \in \rho_O} \bpm w^\dag_{\mbf{R},n} & c^\dag_{\mbf{R},n} \epm (\sigma_z \cos \th + \sigma_x \sin \th ) \bpm w_{\mbf{R},n} \\ c_{\mbf{R},n} \epm
\eea
whose spectrum is $\pm t, 0$ for all $\th$. Note that the coupling $w^\dag_{\mbf{R},n} c_{\mbf{R},n}$ is symmetry-preserving. At $\th = 0$, $H(0)$ has the same spectrum as $H_{OAI}$ but with additional conduction bands of $c^\dag_{\mbf{R},n}$ orbitals at energy $+t$. Tuning $\th$ to $\pi$ does not close the gap (the spectrum is independent of $\th$), and at $\th = \pi$ where the Wannier-orbital coupling $t \sin \th$ vanishes, the Wannier states $w^\dag_{\mbf{R},n}$ are now in the conduction band at energy $+t$ and the trivial orbitals $c^\dag_{\mbf{R},n}$ are in the valence band. Hence this model tunes between the obstructed atomic state in the original Hilbert space and a trivial atomic limit in the expanded Hilbert space without a gap closing on PBCs. 

\begin{figure}[H]
 \centering
\begin{overpic}[width=0.3\textwidth,tics=10]{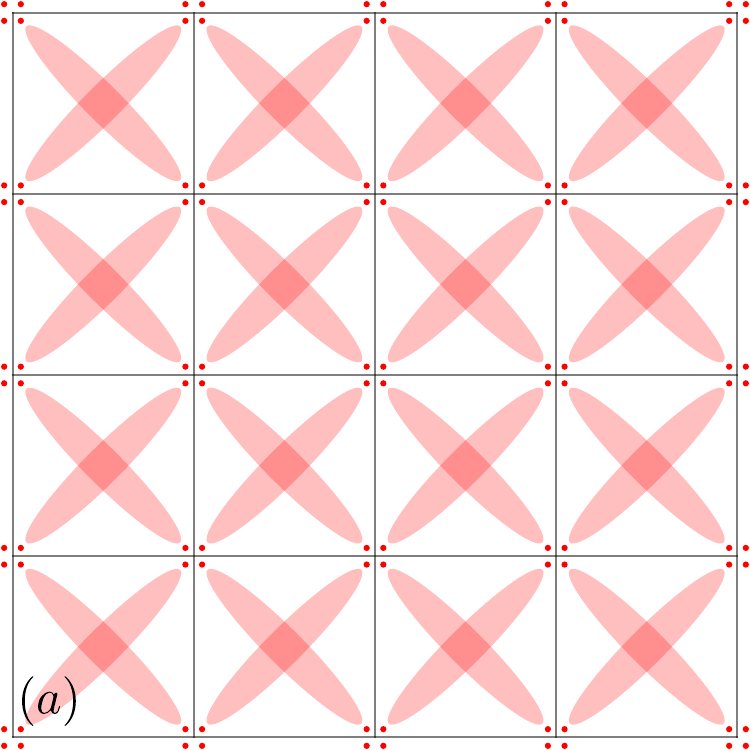}\end{overpic} \ 
\begin{overpic}[width=0.3\textwidth,tics=10]{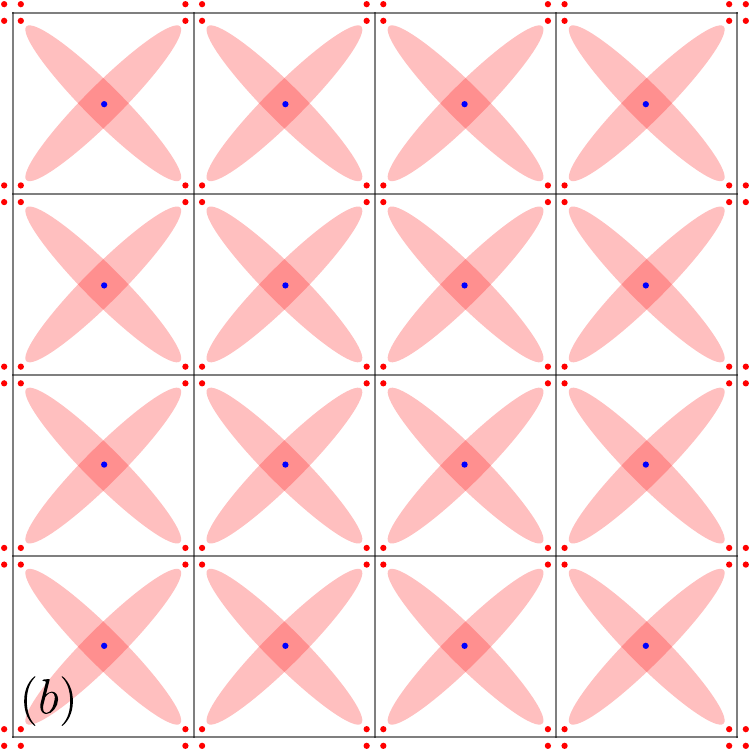}\end{overpic}  \
\begin{overpic}[width=0.3\textwidth,tics=10]{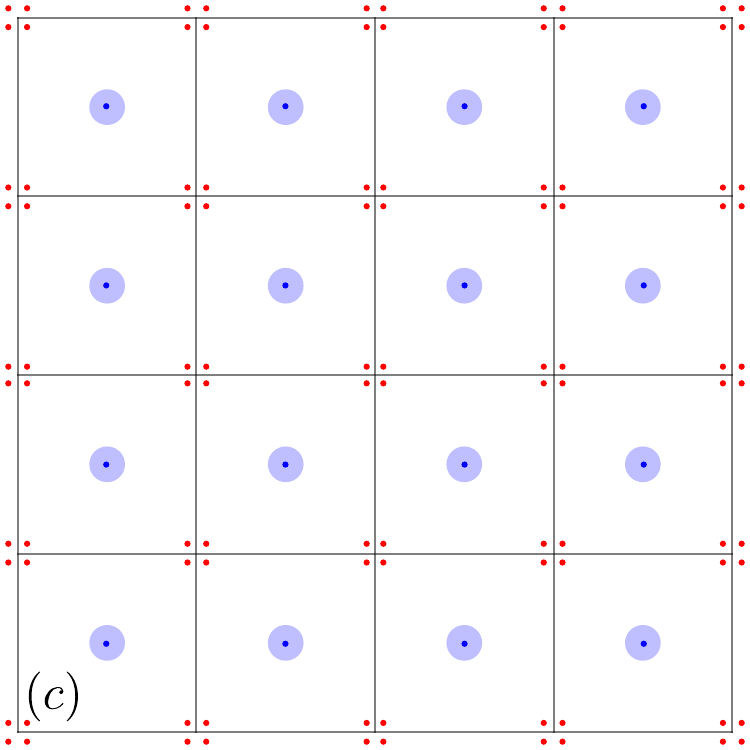}\end{overpic} 
\caption{Flat band obstructed model and auxiliary atomic orbitals. (a) We show the original 4-orbital model. The four small red points by the 1a position denote the $A,B,{}^1E{}^2E$ orbitals of the tight-binding basis. The $A_{1b}$ Wannier function is shown as a large four-pronged wavefunction to emphasize its support on the four neighboring 1a sites. On OBCs, the spatial extend of the Wannier function is responsible for corner states. (b) The auxiliary $A_{1b}$ atomic orbital is added to the tight-binding basis shown in blue. It transforms in the same way as the $A_{1b}$ Wannier state. (c) By coupling the Wannier state to the auxiliary atomic state as in \Eq{eq:HOALIth}, there is an adiabatic path on PBCs between the initial red Wannier state supported over multiple unit cells, and the final blue atomic orbitals which is delta-function supported.
}
\label{fig:p4basiswannier}
\end{figure}

We have shown how to resolve the bulk obstruction to a trivial atomic limit on PBCs by adiabatically deforming the obstructed Wannier functions into auxiliary atomic orbitals $c^\dag_{\mbf{R},n}$ via \Eq{eq:HOALIth}. The many-body local RSIs are manifestly well defined in the trivial atomic limit (upon imposing OBCs), so we have shown by construction that the many-body local RSIs of obstructed states may be defined following this procedure. However, it is instructive to track the evolution of $H(\th)$ on OBCs. We do so by using a flat band obstructed atomic limit Hamiltonian with space group $p41'$ defined in Ref \cite{2022arXiv220910559H}. The model of Ref. \cite{2022arXiv220910559H} is a four-orbital model with a basis of $A,B,{}^1E{}^2E$ orbitals at the 1a position depicted in \Fig{fig:p4basiswannier}a. The model (see \Eq{eq:hthp4}) is constructed to have a compact Wannier state in the $A_{1b}$ irrep (an $s$-orbital at the $(1/2,1/2)$ position) at energy $E=-1$. This Wannier state is obstructed because the orbital basis consists of orbitals at the 1a position (not at 1b). The three flat conduction bands are at energy $E=0$. Thus the bulk gap $E \in (-1,0)$ realizes an obstructed atomic phase in the four band model. We now add a $A_{1b}$ \emph{atomic} orbital to the Hilbert space which will trivialize the obstructed phase. Explicitly, single-particle Hamiltonian is written in the ordered basis $A_{1a},B_{1a},({}^1E{}^2E)_{1a},A_{1b}$ as
\bea
\label{eq:hthp4}
h(\th) &= - U(\mbf{k}) U(\mbf{k})^\dag \cos\th + V(\mbf{k}) V(\mbf{k})^\dag \cos\th + (U(\mbf{k}) V(\mbf{k})^\dag + V(\mbf{k}) U(\mbf{k})^\dag) \sin \th \\
U(\mbf{k}) &= \frac{1}{4} \bpm 1\\ 1\\ 1\\ 1 \\ 0\epm + \frac{1}{4} e^{- i \mbf{k} \cdot \mbf{a}_1} \bpm 1\\ -1\\ -i\\ i \\ 0 \epm + \frac{1}{4} e^{- i \mbf{k} \cdot (\mbf{a}_1 + \mbf{a}_2)}\bpm 1\\ 1\\ -1\\ -1 \\ 0 \epm + \frac{1}{4} e^{- i \mbf{k} \cdot \mbf{a}_2} \bpm 1\\ -1\\ i\\ -i \\ 0 \epm, \quad V(\mbf{k}) = e^{- i \mbf{k} \cdot (\mbf{a}_1+\mbf{a}_2)/2} \bpm 0\\ 0 \\  0 \\ 0 \\ 1 \epm  \\
\eea
so that $U(\mbf{k})$ creates a compact Wannier function at 1b $=(1/2,1/2)$ supported on the nearest four 1a sites, and $V(\mbf{k})$ creates a trivial atomic orbital 1b $=(1/2,1/2)$ (see Ref. \cite{2022arXiv220910559H} for details) shown in \Fig{fig:p4basiswannier}b. By coupling the Wannier state to the atomic state (on PBCs at fixed filling $\nu = 1/5$, noting that the addition of the $A_{1b}$ orbital means there are five orbitals in the unit cell), we trivialize the obstructed state by adiabatically connecting it to the trivial atomic phase symbolized in \Fig{fig:p4basiswannier}c. We now discuss this model numerically on OBCs to understand how the filling anomaly is resolved. 

\begin{figure}[H]
 \centering
\begin{overpic}[width=0.3\textwidth,tics=10]{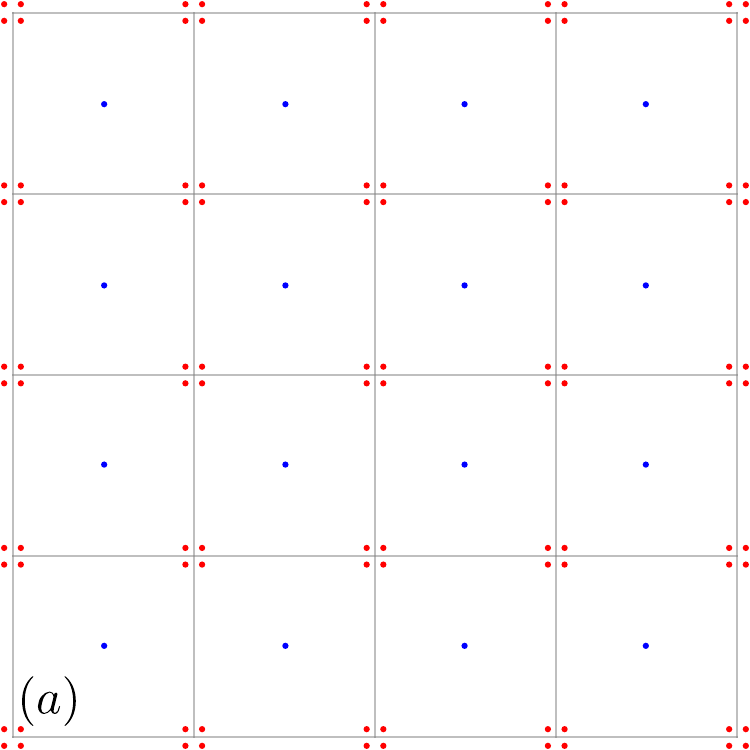}\end{overpic} \qquad 
\begin{overpic}[width=0.3\textwidth,tics=10]{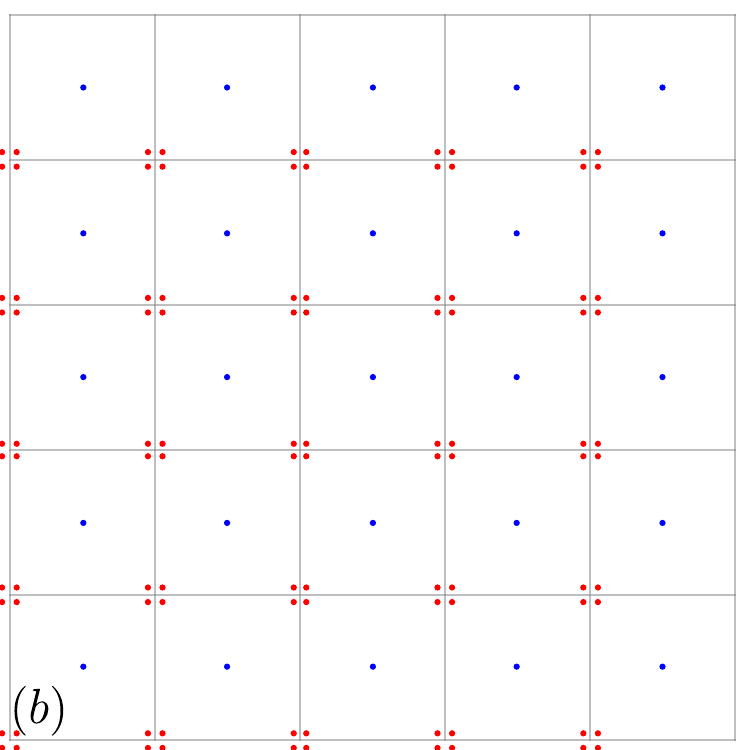}\end{overpic} \qquad
\begin{overpic}[width=0.3\textwidth,tics=10]{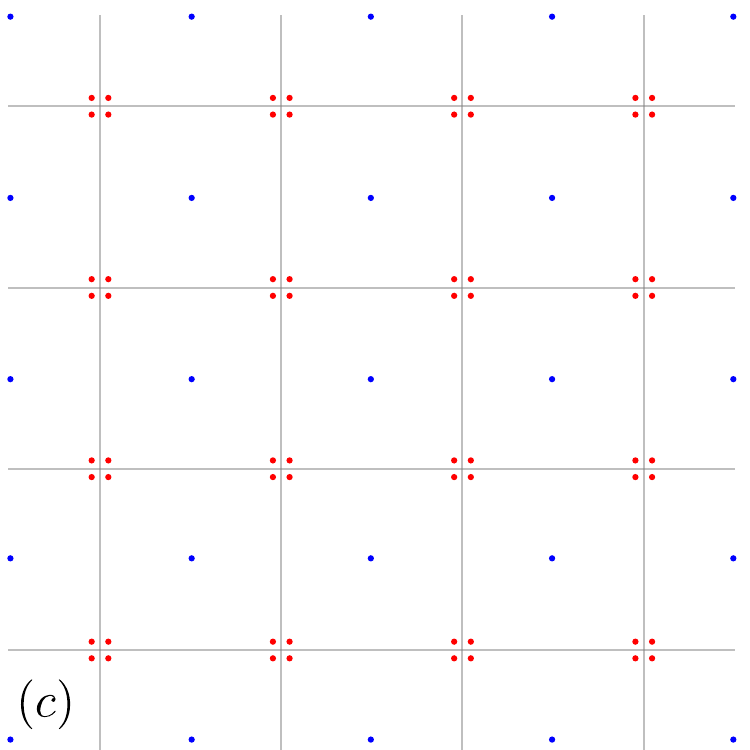}\end{overpic} 
\caption{Different OBC boundary conditions. (a) We preserve the $C_4$ symmetry of the 1a position as in \Fig{fig:OBC}a below. (b) $C_4$ symmetry is broken, but the unit cell is preserved such that there are the same number of 1a and 1b positions. (c) We preserve the $C_4$ symmetry of the 1b position as in \Fig{fig:OBC}c below. 
}
\label{fig:OBCchoices}
\end{figure}

When discussing OBCs, the precise choice of boundaries is important if there are orbitals at different Wyckoff positions in the unit cell. We depict three choices in  \Fig{fig:OBCchoices}. Both \Fig{fig:OBCchoices}a and \Fig{fig:OBCchoices}c depict $C_4$-preserving boundaries centered at the 1a and 1b positions respectively, while \Fig{fig:OBCchoices}b shows a choice of boundaries that breaks $C_4$ but preserves the unit cell. Let us discuss the geometry of these cases in detail. To preserve $C_4$ symmetry at the 1a position (as in \Fig{fig:OBCchoices}a), we center the boundary around 1a and choose a $C_4$ symmetric cutoff. Necessarily, there will be $1 + 4\mathbb{N}$ 1a sites, since the 1a site at the center transforms to itself, whereas all other 1a sites come in quartets. Similarly since the 1b position is off the center, there must be $4\mathbb{N}$ 1b positions. In \Fig{fig:OBCchoices}a, we chose $L^2 = 25 = 1 + 6\times 4$ 1a sites, and $N_{1b} \equiv (L-1)^2 = 16 = 4\times 4$ 1b sites for $L = 5$ odd. In \Fig{fig:OBCchoices}c where the 1b site is at the center, we chose $L^2 = 16 = 4\times 4$ 1a sites, and $N_{1b} \equiv (L+1)^2 = 25 = 1+ 6\times 4$ 1b sites for $L = 4$ even. When computing the many-body local RSIs, we must preserve the symmetries that protect them on OBCs, so \Fig{fig:OBCchoices}a,c are suitable while \Fig{fig:OBCchoices}b is not. We remark that the $L\times L$ OBC of \Fig{fig:OBCchoices}b is a commonly chosen boundary truncation in numerics because it preserves the unit cell, and hence the \emph{filling}, for any $L$ although it breaks the $C_4$ symmetry. 

We now discuss the OBC spectrum for the $C_4$-preserving OBCs in \Fig{fig:OBCchoices}a,c. We showed in this appendix that by adding auxiliary orbitals, we were able to connect the obstructed atomic limit to a trivial atomic limit of $A_{1b}$ orbitals on PBCs. Thus on OBCs, we always focus on the gap above occupying $N_{1b}$ states, which corresponds to the trivial atomic limit of $A_{1b}$ orbitals where we first defined many-body local RSIs (see \App{app:RSIconstruct}). We will see that at this particle number $N = N_{1b}$, there is no filling anomaly (which we discuss in detail momentarily). By tuning $\th$ in \Eq{eq:HOALIth}, we study the adiabatic connection between the trivial atomic orbitals of $H(\pi)$ and the initial obstructed Wannier states of $H(0)$. 

\begin{figure}[H]
 \centering
\begin{overpic}[width=0.45\textwidth,tics=10]{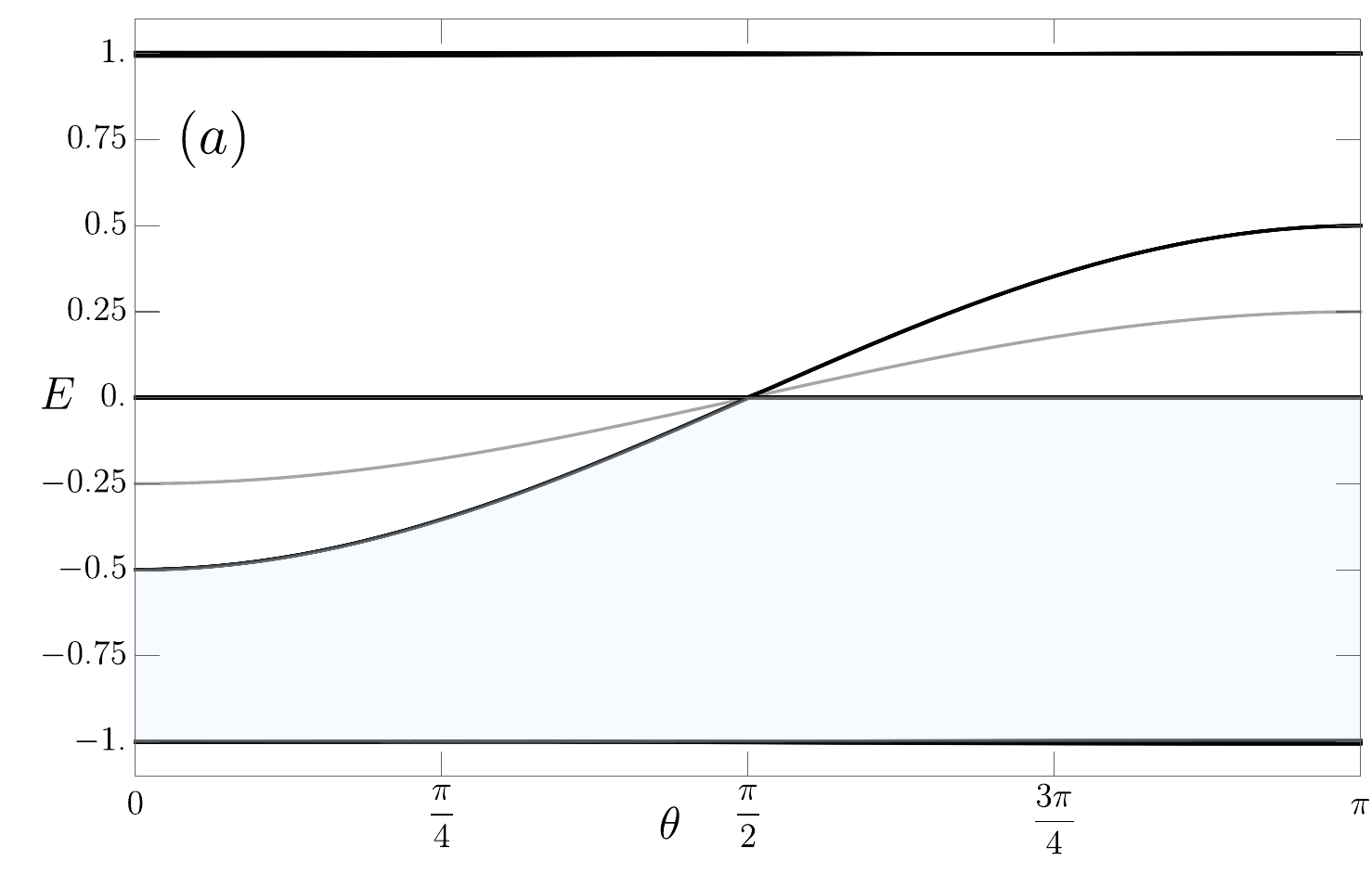}\end{overpic} 
\begin{overpic}[width=0.45\textwidth,tics=10]{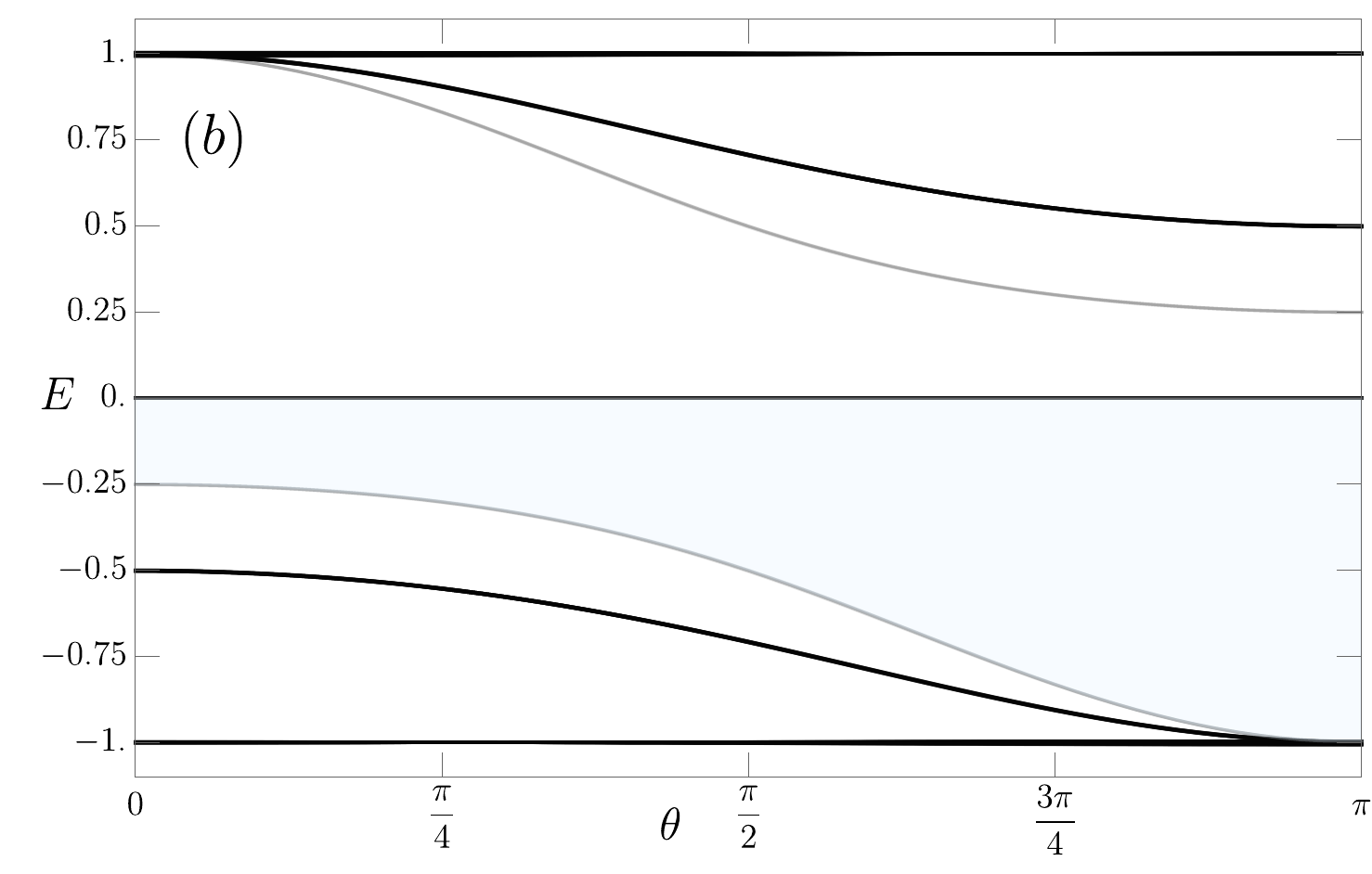}\end{overpic} \\
\caption{Adiabatic connection between an obstructed atomic phase and trivial atomic phase via the addition of trivial orbitals to the Hilbert space. Corner states originating from $E = -1/4$ appear in gray, and the edge states originating from $E = -1/2$ are dark gray. $(a)$ We fix the center to be at the 1a position and choose OBCs to have $L^2$ 1a sites (each with 4 orbitals) and $N_{1b} = (L-1)^2$ 1b sites for $L = 23$ odd. The gap above the $N_{1b} =484$ state is shaded light blue.  $(b)$ We fix the center to be at the 1b position and choose OBCs to have $L^2$ 1a sites (each with 4 orbitals) and $N_{1b} =(L+1)^2$ 1b sites for $L = 22$ even. The gap above the $N_{1b} = 529$ state is shaded light blue. 
}
\label{fig:OBC}
\end{figure}

\Fig{fig:OBC} shows the OBC spectrum of $H(\th)$ (\Eq{eq:HOALIth}) for OBCs centered at 1a and 1b, explicitly verifying the adiabatic path at constant particle number $N_b$ between the obstructed phase of the original Hilbert space (which is no longer obstructed when the $A_{1b}$ orbital is added to the Hilbert space) and the trivial phase. Let us elaborate on how this has resolved the filling anomaly of the original four orbital model in \Eq{eq:orig4orb}. One important feature of \Eq{eq:orig4orb} is that all orbitals are at the 1a position, so a $C_4$-preserving boundary condition as in \Fig{fig:OBCchoices}a will preserve the unit cell. Then at filling $\nu = 1/4$ (the bulk filling of the obstructed Wannier state) the groundstate is degenerate due to the filling anomaly. To verify this, we check in \Fig{fig:OBC}a for $H(\th = 0)$ (where the auxiliary $A_{1b}$ orbitals are decoupled and do not effect the numerics) that occupying $L^2$ states (which is $1/4$ of the $4 L^2$ orbitals at $1a$) leads to a degenerate groundstate occupying half of the four corner states at $E = -.25$. However, we showed that on these OBCs, the trivial atomic limit of $H(\th = \pi)$ occurs at occupying $N_{1b} = (L-1)^2$ states. We check that this corresponds to full filling of the $E=-1$ states and is non-degenerate with a gap above to the edge states at $E = -.5$. Of course, we could have chosen a different $C_4$-respecting boundary cutoff that also included the nearest 1b sites outside of the $L^2$ 1a sites (in \Fig{fig:OBCchoices}a, this would correspond to having $6^2 = 36$ 1b sites in 1b, enlarged from $4^2 = 16$). Then the trivial atomic limit would be obtained at occupying $(L+1)^2$ states, which we verify in \Fig{fig:OBC}a at $\th =0$ corresponds to occupying all the edge and corner states in addition to the bulk states, leading to a non-degenerate state with a gap to the bulk states at $E=0$. We see that it is the combination of symmetry-preserving boundary conditions and the number of occupying states fixed by the unit cell geometry of the auxiliary orbital that resolves the OBC filling anomaly. 

It is interesting to note that the edge states and corner states, which appear in the $E \in (-1,0)$ bulk gap at $\th = 0$ due to the obstructed Wannier state, do not contribute to the many-body local RSI. This is because they are strictly supported on the boundary and can be represented as $\prod_{i=1}^4 C_4^i \mathcal{O}_{bdy} C_4^{i \dag}$ where $ \mathcal{O}_{bdy}$ is the creation operator for all the boundary states on a single edge and corner. The many-body local RSIs are $\Delta_{1a} = (0 \mod 8,0 \mod 2), \Delta_{1b} = (1 \mod 8,0 \mod 2), \Delta_{2c} = 0 \mod 4$ from \Tab{tab:mpgrsi} and can be obtained immediately from the atomic limit state at $\th = \pi$. 

\subsection{Fragile Inequalities}
\label{app:fragineq}

We now study how the particle number constraints in \Tab{tab:boundsrsi} at each Wyckoff in the unit cell can be used to diagnose many-body fragile topology. In the non-interacting case, the classification of fragile phases has an affine monoid structure \cite{2019arXiv190503262S, rsis} because the single-particle RSIs could be $\mathds{Z}$-valued, i.e. there were infinite possible RSIs. In this (single-particle) case, the orbital number/RSI constraints analogous to those in \Tab{tab:boundsrsi} included inequality constraints and $\mathds{Z}_2$ constraints. 

In the interacting case where the many-body RSIs only take finitely many values, the fragile criteria are actually easier to derive. Our strategy is to devise inequality criteria which hold in all many-body \emph{atomic} states in terms of the local bounds RSIs (protected by the PG symmetries) and the total particle number (protected by the global $U(1)$ symmetry). If the a given set of local RSIs \emph{violate} the inequality criteria, then there is an obstruction to deformation into a many-body atomic state, proving many-body fragile topology. 

In a many-body atomic limit, the number of particles $N_{\mbf{x}}$ at each site $\mbf{x} = \mbf{R}+\mbf{r}$ is well-defined by $[\hat{N},\mathcal{O}_{\mbf{R},\mbf{r}}]  = N_{\mbf{x}} \mathcal{O}_{\mbf{R},\mbf{r}}$ using the notation in \App{app:RSIconstruct} where $\mathcal{O}_{\mbf{R},\mbf{r}}$ is the creation operator of the groundstate of $H_{\mbf{R},\mbf{r}}$. By translation symmetry, $N_{\mbf{R}+\mbf{r}} = N_{\mbf{r}}$ and we need only consider a single unit cell. At integer filling $\nu = N_{occ} / N_{orb}$, the total number of particles per unit cell $N_{occ}$ is obtained by adding up the states at all the Wyckoff positions in the unit cell:
\bea
\label{eq:ineqWannier}
N_{occ} = \sum_{\mbf{x}} m_{\mbf{x}} N_{\mbf{x}}  \geq \sum_{\text{high-symmetry } \mbf{x}} m_{\mbf{x}} N_{\mbf{x}} \qquad \text{(many-body atomic)}
\eea
where $m_{\mbf{x}}$ is the multiplicity of the Wyckoff position $\mbf{x}$ (the many-body local RSIs at each point in the Wyckoff position are equal by \Eq{eq:conjugate} because they are related by symmetries in $G$). Note that the sum in the second equality of \Eq{eq:ineqWannier} is over high-symmetry Wannier positions where the local RSIs are nontrivial, i.e. $G_\mbf{x} \neq 1$. Using \Tab{tab:boundsrsi}, $N_\mbf{x}$ can be lower bounded by the many-body RSIs at $\mbf{x}$, which remain well-defined in many-body fragile topological phases. $N_{occ}$ is also a well-defined many-body quantity: it is the $U(1)$ number density $N_{occ} = N/L^2$ where $N$ is the total number of particles in the groundstate and $L^2$ is the number of unit cells. Schematically, we write
\bea
\label{eq:ineqWannierRSIs}
N_{occ} \geq \sum_{\text{high-symmetry } \mbf{x}} m_{\mbf{x}} N_{\mbf{x}} \geq \sum_{\text{high-symmetry } \mbf{x}} m_{\mbf{x}} \, \text{bound}_{\mbf{x}} (\Delta) \\
\eea
with explicit expressions for the many-body local RSIs bound $\text{bound}_{\mbf{x}} (\Delta)$ are tabulated in \Tab{tab:boundsrsi}. \Eq{eq:ineqWannierRSIs} is obeyed in all many-body atomic states. If \Eq{eq:ineqWannierRSIs} is \emph{violated}, then the many-body local RSIs impose an obstruction to deformation into a many-body atomic state, which defines many-body fragile topology. Hence the many-body fragile topological indices are inequalities corresponding to the violation of \Eq{eq:ineqWannierRSIs}:
\bea
\label{eq:wanniercondition}
N_{occ} < \sum_{\text{high-symmetry } \mbf{x}} m_{\mbf{x}} \, \text{bound}_{\mbf{x}} (\Delta)  \qquad \text{(Many-body fragile)} \ . \\
\eea
In \Tabs{tab:fragilenoSOC}{tab:fragileSOC}, we write down the fragile criteria of \Eq{eq:wanniercondition} in all space groups.

\begin{figure}[H]
 \centering
\begin{overpic}[height=0.2\textwidth,tics=10]{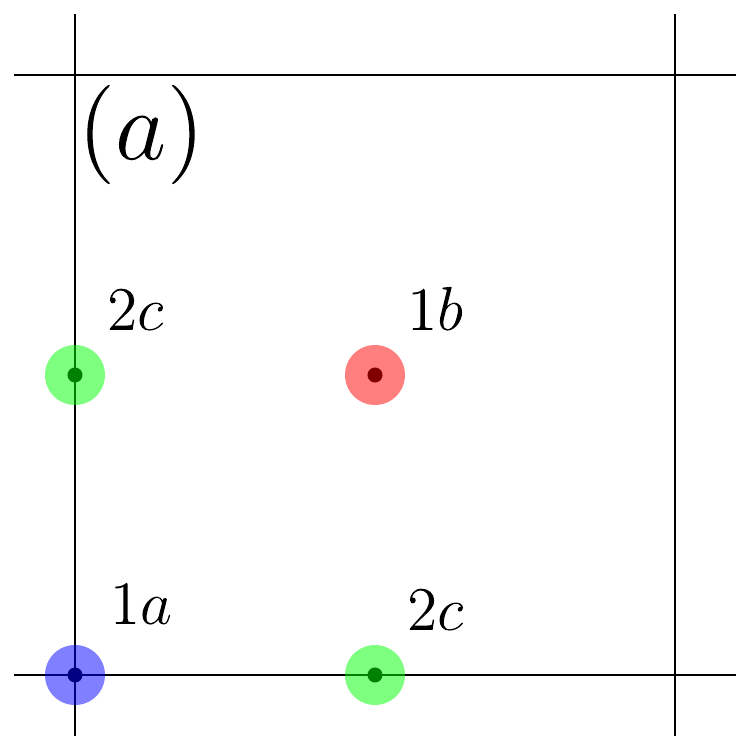}
\end{overpic} 
\begin{overpic}[height=0.2\textwidth,tics=10]{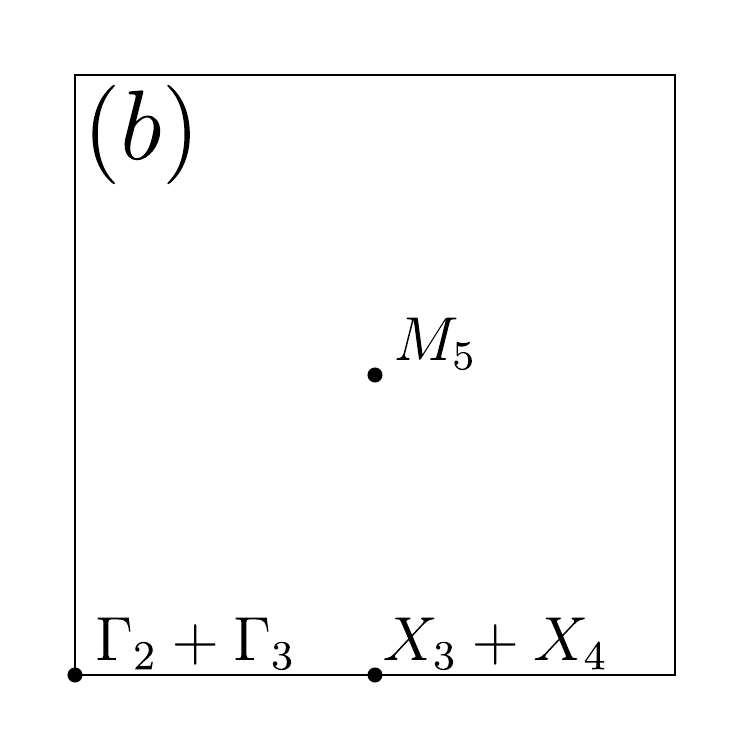}
\end{overpic} 
\caption{$(a)$ The high-symmetry Wyckoff positions in the unit cell are 1a $(4mm)$ in blue, 1b $(4mm)$ in red and 2c $(2mm)$ in green. The 1a and 1b Wyckoff positions have multiplicity 1, and 2c has multiplicity 2. $(b)$ We show the band structure of a single-particle fragile phase with is also many-body fragile. The high-symmetry points in the BZ are $\Gamma$ ($4mm$), $M$ ($4mm$), and $X$ ($2mm$). }
\label{fig:4mmfig_frag}
\end{figure}

We now give an example in the wallpaper group $p4mm$ which is generated by $C_4, M_y,$ and translations. The high-symmetry Wyckoff positions are shown in \Fig{fig:4mmfig_frag}. We do not show the mirror-symmetric lines. The 1a and 1b positions have $G_{\mbf{x}} = 4mm$, and the 2c position has $G_{\mbf{x}} = 2mm$. The many-body RSI bounds for $m_\mbf{x} N_{\mbf{x}}$ from \Tab{tab:boundsrsi} are
\bea
4mm: \qquad N_{1a} &\geq \text{mod}_4(\Delta_{1a,1}|\Delta_{1a,1}) ,\\
4mm: \qquad N_{1b} &\geq \text{mod}_4(\Delta_{1b,1}|\Delta_{1b,1}) ,\\
2mm: \qquad 2 N_{2c} &\geq 2 \text{mod}_2(\Delta_{2c,1}|\Delta_{2c,1}) \ . \\
\eea
Hence in any many-body atomic state, 
\bea
N_{occ} \geq \text{mod}_4(\Delta_{1a,1}|\Delta_{1a,1}) + \text{mod}_4(\Delta_{1b,1}|\Delta_{1b,1}) + 2 \text{mod}_2(\Delta_{2c,1}|\Delta_{2c,1}) \ .
\eea
Violating this inequality is an obstruction to the many-body atomic limit, so we deduce
\bea
\label{eq:fragcrtiex4mm}
N_{occ} < \text{mod}_4(\Delta_{1a,1}) + \text{mod}_4(\Delta_{1b,1}) + 2 \text{mod}_2(\Delta_{2c,1}) \quad \implies \text{many-body fragile topology} \ .
\eea
We now give an example of a single-particle state which is many-body fragile. Ref. \cite{rsis} contains a list of fragile roots that generate all fragile band structures in a given space group. We choose the fragile band structure $\mathcal{B} = \Gamma_2+ \Gamma_3 + M_5 + X_3 + X_4$ which has $N_{occ} =2$. The $\Gamma_2, \Gamma_3, $ and $M_5$ irreps are isomorphic to the $B_1, B_2, $ and $E$ irreps of $4mm$ respectively in momentum space (see  \Tab{tab:2D-char}), and the $X_3, X_4$ irreps are isomorphic to the $B_1, B_2$ irreps of PG $2mm$ (see \Tab{tab:2D-char}). The single-particle RSIs of Ref. \cite{rsis} are computed from $\mathcal{B}$ to be
\bea
(\delta^{1a}_1, \delta^{1a}_2) &= (1, 1), \quad (\delta^{1b}_1, \delta^{1b}_2) = (0, 1), \quad \delta^{2c}_1 = -1\\ 
\eea
and from \Tab{tab:rsiredint}, we compute the many-body RSIs by reducing the single-particle RSIs
\bea
\Delta^{1a}_1 = 7 \mod 8, \quad \Delta^{1b}_1 = 5 \mod 8, \quad \Delta^{2c}_1 = 1 \mod 4  \ .\\
\eea
We evaluate the many-body fragile criteria in \Eq{eq:fragcrtiex4mm},
\bea
N_{occ}  <  \text{mod}_4(\Delta_{1a,1}|\Delta_{1a,1}) + \text{mod}_4(\Delta_{1b,1}|\Delta_{1b,1}) + 2 \text{mod}_2(\Delta_{2c,1}|\Delta_{2c,1})  = 3 + 1 + 2(1) = 6
\eea
from which we see that the state $\mathcal{B}$ is many-body fragile since $N_{occ} = 2$. We can reproduce this calculation directly in real space checking that
\bea
\label{eq:4mmexrealspcae}
\mathcal{B} = (E^{1a} \oplus B_1^{1a} \oplus B_2^{1b} \ominus B_1^{2c}) \uparrow p4mm
\eea
using the band representations on the Bilbao Crstallographic Server (\url{https://www.cryst.ehu.es/cgi-bin/cryst/programs/bandrep.pl}). We can directly determine the many-body RSIs in real space using \Tab{tab:mpgrsi}:
\bea
\Delta^{1a}_1 = -3+2 \mod 8, \quad \Delta^{1b}_1 = -3 \mod 8, \quad \Delta^{2c} = 1 \mod 4  \ .\\
\eea
which matches the momentum space result. The intuition from the real space representation in \Eq{eq:4mmexrealspcae} is that the fragile topology results from the $\ominus B_1^{2c}$ obstruction which -- in this particular state -- cannot be removed by an interaction-enabled conversion. \App{app:modelfragile} details an exactly solvable Hamiltonian where the $\ominus$ irrep obstruction can be removed by interactions, as diagnosed by the many-body local RSIs.

\vspace{1cm}
\twocolumngrid

\begingroup
\begin{longtable}{| C{.07\textwidth} | C{.08\textwidth} | C{.32\textwidth} | } 
\caption{Fragile Criteria (No SOC)} 
\label{tab:fragilenoSOC} \\
\hline
SG & WP & $N_{occ} < \dots$  \\
\hline
$p1$ & $a(1)$ &    \\
\hline
$p1'$ & $a(1')$  &   \\
\hline
\multirow{4}{*}{$p2$} 
& $1a (2)$ & $\text{mod}_2(\Delta_{1a,1}|\Delta_{1a,1}) $ \\
\cline{2-2}
& $1b (2)$ & $+\text{mod}_2(\Delta_{1b,1}|\Delta_{1b,1})$ \\
\cline{2-2}
& $1c (2)$ & $+\text{mod}_2(\Delta_{1c,1}|\Delta_{1c,1})$ \\
\cline{2-2}
& $1d (2)$ & $+\text{mod}_2(\Delta_{1d,1}|\Delta_{1d,1})$ \\
\hline
\multirow{4}{*}{$p21'$} 
& $1a (21')$ & $\text{mod}_2(\Delta_{1a,1}|\Delta_{1a,1}) $ \\
\cline{2-2}
& $1b (21')$ & $+\text{mod}_2(\Delta_{1b,1}|\Delta_{1b,1})$ \\
\cline{2-2}
& $1c (21')$ & $+\text{mod}_2(\Delta_{1c,1}|\Delta_{1c,1})$ \\
\cline{2-2}
& $1d (21')$ & $+\text{mod}_2(\Delta_{1d,1}|\Delta_{1d,1})$ \\
\hline
\multirow{2}{*}{$pm$} 
& $1a (m)$ & \\
\cline{2-2}
& $1b (m)$ &  \\
\hline
\multirow{2}{*}{$pm1'$} 
& $1a (m1')$ & \\
\cline{2-2}
& $1b (m1')$ & \\
\hline
$pg$ &   &  \\
\hline
$pg1'$ &   &  \\
\hline
\multirow{1}{*}{$cm$} 
& $2a (m)$ &  \\
\hline
\multirow{1}{*}{$cm1'$} 
& $2a (m1')$ &  \\
\hline
\multirow{4}{*}{$p2mm$} 
& $1a(2mm)$ & $\text{mod}_2(\Delta_{1a,1}|\Delta_{1a,1}) $ \\
\cline{2-2}
& $1b(2mm)$ & $+\text{mod}_2(\Delta_{1b,1}|\Delta_{1b,1})$ \\
\cline{2-2}
& $1c(2mm)$ & $+\text{mod}_2(\Delta_{1c,1}|\Delta_{1c,1})$ \\
\cline{2-2}
& $1d(2mm)$ & $+\text{mod}_2(\Delta_{1d,1}|\Delta_{1d,1})$ \\
%\cline{2-2}
%& $e(m)$ & $+\text{mod}_2(\Delta_{1e,1}) $ \\
%\cline{2-2}
%& $f(m)$ & $+\text{mod}_2(\Delta_{1f,1})$ \\
%\cline{2-2}
%& $g(m)$ & $+\text{mod}_2(\Delta_{1g,1})$ \\
%\cline{2-2}
%& $h(m)$ & $+\text{mod}_2(\Delta_{1h,1})$ \\
\hline
\multirow{4}{*}{$p2mm1'$} 
& $1a(2mm1')$ & $\text{mod}_2(\Delta_{1a,1}|\Delta_{1a,1}) $ \\
\cline{2-2}
& $1b(2mm1')$ & $+\text{mod}_2(\Delta_{1b,1}|\Delta_{1b,1})$ \\
\cline{2-2}
& $1c(2mm1')$ & $+\text{mod}_2(\Delta_{1c,1}|\Delta_{1c,1})$ \\
\cline{2-2}
& $1d(2mm1')$ & $+\text{mod}_2(\Delta_{1d,1}|\Delta_{1d,1})$ \\
%\cline{2-2}
%& $e(m1')$ & $+\text{mod}_2(\Delta_{1e,1}) $ \\
%\cline{2-2}
%& $f(m1')$ & $+\text{mod}_2(\Delta_{1f,1})$ \\
%\cline{2-2}
%& $g(m1')$ & $+\text{mod}_2(\Delta_{1g,1})$ \\
%\cline{2-2}
%& $h(m1')$ & $+\text{mod}_2(\Delta_{1h,1})$ \\
\hline
\multirow{3}{*}{$p2mg$} 
& $2a(2)$ & $2\text{mod}_2(\Delta_{2a,1}|\Delta_{2a,1}) $ \\
\cline{2-2}
& $2b(2)$ & $+2\text{mod}_2(\Delta_{2b,1}|\Delta_{2b,1})$ \\
\cline{2-2}
& $4c(m)$ &  \\
\hline
\multirow{3}{*}{$p2mg1'$} 
& $2a(21')$ & $2\text{mod}_2(\Delta_{2a,1}|\Delta_{2a,1}) $ \\
\cline{2-2}
& $2b(21')$ & $+2\text{mod}_2(\Delta_{2b,1}|\Delta_{2b,1})$ \\
\cline{2-2}
& $4c(m1')$ &  \\
\hline
\multirow{2}{*}{$p2gg$} 
& $2a(2)$ & $2\text{mod}_2(\Delta_{2a,1}|\Delta_{2a,1}) $ \\
\cline{2-2}
& $2b(2)$ & $+2\text{mod}_2(\Delta_{2b,1}|\Delta_{2b,1})$ \\
\hline
\multirow{2}{*}{$p2gg1'$} 
& $2a(21')$ & $2\text{mod}_2(\Delta_{2a,1}|\Delta_{2a,1}) $ \\
\cline{2-2}
& $2b(21')$ & $+2\text{mod}_2(\Delta_{2b,1}|\Delta_{2b,1})$ \\
\hline
\multirow{3}{*}{$c2mm$} 
& $2a(2mm)$ & $2\text{mod}_2(\Delta_{2a,1}|\Delta_{2a,1}) $ \\
\cline{2-2}
& $2b(2mm)$ & $+2\text{mod}_2(\Delta_{2b,1}|\Delta_{2b,1})$ \\
\cline{2-2}
& $4c(2)$ & $+4\text{mod}_2(\Delta_{4c,1}|\Delta_{4c,1})$ \\
\hline
\multirow{3}{*}{$c2mm1'$} 
& $2a(2mm1')$ & $2\text{mod}_2(\Delta_{2a,1}) $ \\
\cline{2-2}
& $2b(2mm1')$ & $+2\text{mod}_2(\Delta_{2b,1})$ \\
\cline{2-2}
& $4c(21')$ & $+4\text{mod}_2(\Delta_{4c,1})$ \\
\hline
\multirow{3}{*}{$p4$} 
& $1a(4)$ & $\text{mod}_4(\Delta_{1a,1}-2\Delta_{1a,2}|\Delta_{1a,1}) $ \\
\cline{2-2}
& $1b(4)$ & $+\text{mod}_4(\Delta_{1b,1}-2\Delta_{1b,2}|\Delta_{1b,1})$ \\
\cline{2-2}
& $2c(2)$ & $+2\text{mod}_2(\Delta_{2c,1}|\Delta_{2c,1})$ \\
\hline
\multirow{3}{*}{$p41'$} 
& $1a(41')$ & $\text{mod}_4(\Delta_{1a,1}|\Delta_{1a,1}) $ \\
\cline{2-2}
& $1b(41')$ & $+\text{mod}_4(\Delta_{1b,1}|\Delta_{1b,1})$ \\
\cline{2-2}
& $2c(21')$ & $+2\text{mod}_2(\Delta_{2c,1}|\Delta_{2c,1})$ \\
\hline
\multirow{3}{*}{$p4mm$} 
& $1a(4mm)$ & $\text{mod}_4(\Delta_{1a,1}|\Delta_{1a,1}) $ \\
\cline{2-2}
& $1b(4mm)$ & $+\text{mod}_4(\Delta_{1b,1}|\Delta_{1b,1})$ \\
\cline{2-2}
& $2c(2mm)$ & $+2\text{mod}_2(\Delta_{2c,1}|\Delta_{2c,1})$ \\
\hline
\multirow{3}{*}{$p4mm1'$} 
& $1a(4mm1')$ & $\text{mod}_4(\Delta_{1a,1}|\Delta_{1a,1}) $ \\
\cline{2-2}
& $1b(4mm1')$ & $+\text{mod}_4(\Delta_{1b,1}|\Delta_{1b,1})$ \\
\cline{2-2}
& $2c(2mm1')$ & $+2\text{mod}_2(\Delta_{2c,1}|\Delta_{2c,1})$ \\
\hline
\multirow{2}{*}{$p4gm$} 
& $2a(4)$ & $2\text{mod}_4(\Delta_{2a,1}-2\Delta_{2a,2}|\Delta_{2a,1}) $ \\
\cline{2-2}
& $2b(2mm)$ & $+2\text{mod}_4(\Delta_{2b,1}|\Delta_{2b,1})$ \\
\hline
\multirow{2}{*}{$p4gm1'$} 
& $2a(41')$ & $2\text{mod}_4(\Delta_{2a,1}|\Delta_{2a,1}) $ \\
\cline{2-2}
& $2b(2mm1')$ & $+2\text{mod}_4(\Delta_{2b,1}|\Delta_{2b,1})$ \\
\hline
\multirow{3}{*}{$p3$} 
& $1a(3)$ & $\text{mod}_3(\Delta_{1a,1}|\Delta_{1a,2}) $ \\
\cline{2-2}
& $1b(3)$ & $+\text{mod}_3(\Delta_{1b,1}|\Delta_{1b,2})$ \\
\cline{2-2}
& $1c(3)$ & $+\text{mod}_3(\Delta_{1c,1}|\Delta_{1c,2})$ \\
\hline
\multirow{3}{*}{$p31'$} 
& $1a(31')$ & $\text{mod}_3(\Delta_{1a,1}|\Delta_{1a,1}) $ \\
\cline{2-2}
& $1b(31')$ & $+\text{mod}_3(\Delta_{1b,1}|\Delta_{1b,1})$ \\
\cline{2-2}
& $1c(31')$ & $+\text{mod}_3(\Delta_{1c,1}|\Delta_{1c,1})$ \\
\hline
\multirow{3}{*}{$p3m1$} 
& $1a(3m)$ & $\text{mod}_3(\Delta_{1a,1}|\Delta_{1a,1}) $ \\
\cline{2-2}
& $1b(3m)$ & $+\text{mod}_3(\Delta_{1b,1}|\Delta_{1b,1}) $ \\
\cline{2-2}
& $1c(3m)$ &$+\text{mod}_3(\Delta_{1c,1}|\Delta_{1c,1}) $ \\
\hline
\multirow{3}{*}{$p3m1'$} 
& $1a(3m1')$ & $\text{mod}_3(\Delta_{1a,1}|\Delta_{1a,1}) $ \\
\cline{2-2}
& $1b(3m1')$ & $+\text{mod}_3(\Delta_{1b,1}|\Delta_{1b,1}) $ \\
\cline{2-2}
& $1c(3m1')$ &$+\text{mod}_3(\Delta_{1c,1}|\Delta_{1c,1}) $ \\\hline
\multirow{2}{*}{$p31m$} 
& $1a(3m)$ & $\text{mod}_3(\Delta_{1a,1}|\Delta_{1a,1}) $ \\
\cline{2-2}
& $2b(3)$ & $+2\text{mod}_3(\Delta_{2b,1}|\Delta_{2b,2})$ \\
\hline
\multirow{2}{*}{$p31m1'$} 
& $1a(3m1')$ &$\text{mod}_3(\Delta_{1a,1}|\Delta_{1a,1}) $ \\
\cline{2-2}
& $2b(31')$ & $+2\text{mod}_3(\Delta_{2b,1}|\Delta_{2b,1})$ \\
\hline
\multirow{3}{*}{$p6$} 
& $1a(6)$ & $\text{mod}_6(\Delta_{1a,1}-2\Delta_{1a,2}|\Delta_{1a,1})$ \\
\cline{2-2}
& $2b(3)$ & $+2\text{mod}_3(\Delta_{2b,1}|\Delta_{2b,2})$ \\
\cline{2-2}
& $3c(2)$ & $+3\text{mod}_2(\Delta_{3c,1}|\Delta_{3c,1})$ \\
\hline
\multirow{3}{*}{$p61'$} 
& $1a(61')$ & $\text{mod}_6(\Delta_{1a,1}|\Delta_{1a,1}) $ \\
\cline{2-2}
& $2b(31')$ & $+2\text{mod}_3(\Delta_{2b,1}|\Delta_{2b,1})$ \\
\cline{2-2}
& $3c(2)$ & $+3\text{mod}_2(\Delta_{3c,1}|\Delta_{3c,1})$ \\
\hline
\multirow{3}{*}{$p6mm$} 
& $1a(6mm)$ & $\text{mod}_6(\Delta_{1a,1}|\Delta_{1a,1}) $ \\
\cline{2-2}
& $2b(3m)$ & $+2\text{mod}_3(\Delta_{2b,1}|\Delta_{2b,1})$ \\
\cline{2-2}
& $3c(2mm)$ & $+3\text{mod}_2(\Delta_{3c,1}|\Delta_{3c,1})$ \\
\hline
\multirow{3}{*}{$p6mm1'$} 
& $1a(6mm1')$ & $\text{mod}_6(\Delta_{1a,1}|\Delta_{1a,1}) $ \\
\cline{2-2}
& $2b(3m1')$ & $+2\text{mod}_3(\Delta_{2b,1}|\Delta_{2b,1})$ \\
\cline{2-2}
& $3c(2mm1')$ & $+3\text{mod}_2(\Delta_{3c,1}|\Delta_{3c,1})$ \\
\hline

\end{longtable}
\endgroup

\begingroup
\begin{longtable}{| C{.07\textwidth} | C{.08\textwidth} | C{.32\textwidth} | } 
\caption{Fragile Criteria (with SOC)} % title of Table
\label{tab:fragileSOC} \\
\hline
SG & WP & $N_{occ} < \dots$  \\
\hline
$p1$ & $a(1)$ &    \\
\hline
$p1'$ & $a(1')$  &   \\
\hline
\multirow{4}{*}{$p2$} 
& $1a (2)$ & $\text{mod}_2(\Delta_{1a,1}|\Delta_{1a,1}) $ \\
\cline{2-2}
& $1b (2)$ & $+\text{mod}_2(\Delta_{1b,1}|\Delta_{1b,1})$ \\
\cline{2-2}
& $1c (2)$ & $+\text{mod}_2(\Delta_{1c,1}|\Delta_{1c,1})$ \\
\cline{2-2}
& $1d (2)$ & $+\text{mod}_2(\Delta_{1d,1}|\Delta_{1d,1})$ \\
\hline
\multirow{4}{*}{$p21'$} 
& $1a (21')$ & $2\text{mod}_2(\Delta_{1a,2}|\Delta_{1a,1}) $ \\
\cline{2-2}
& $1b (21')$ & $+2\text{mod}_2(\Delta_{1b,2}|\Delta_{1b,1}) $ \\
\cline{2-2}
& $1c (21')$ & $+2\text{mod}_2(\Delta_{1c,2}|\Delta_{1c,1}) $ \\
\cline{2-2}
& $1d (21')$ & $+2\text{mod}_2(\Delta_{1d,2}|\Delta_{1d,1}) $ \\
\hline
\multirow{2}{*}{$pm$} 
& $1a (m)$ & \\
\cline{2-2}
& $1b (m)$ & \\
\hline
\multirow{2}{*}{$pm1'$} 
& $1a (m1')$ & \\
\cline{2-2}
& $1b (m1')$ &  \\
\hline
$pg$ &   &  \\
\hline
$pg1'$ &   &  \\
\hline
\multirow{1}{*}{$cm$} 
& $2a (m)$ &\\
\hline
\multirow{1}{*}{$cm1'$} 
& $2a (m1')$ & \\
\hline
\multirow{4}{*}{$p2mm$} 
& $1a(2mm)$ & $ \text{mod}_2(0|\Delta_{1a,1})$ \\
\cline{2-2}
& $1b(2mm)$ & $+\text{mod}_2(0|\Delta_{1b,1})$ \\
\cline{2-2}
& $1c(2mm)$ & $+\text{mod}_2(0|\Delta_{1c,1})$ \\
\cline{2-2}
& $1d(2mm)$ & $+\text{mod}_2(0|\Delta_{1d,1})$ \\
\hline
\multirow{4}{*}{$p2mm1'$} 
& $1a(2mm1')$ & $2\text{mod}_2(\Delta_{1a,2}|\Delta_{1a,1}) $ \\
\cline{2-2}
& $1b(2mm1')$ & $+2\text{mod}_2(\Delta_{1b,2}|\Delta_{1b,1}) $ \\
\cline{2-2}
& $1c(2mm1')$ & $+2\text{mod}_2(\Delta_{1c,2}|\Delta_{1c,1}) $ \\
\cline{2-2}
& $1d(2mm1')$ & $+2\text{mod}_2(\Delta_{1d,2}|\Delta_{1d,1}) $\\
\hline
\multirow{3}{*}{$p2mg$} 
& $2a(2)$ & $2\text{mod}_2(\Delta_{2a,1}|\Delta_{2a,1}) $ \\
\cline{2-2}
& $2b(2)$ & $+2\text{mod}_2(\Delta_{2b,1}|\Delta_{2b,1})$ \\
\cline{2-2}
& $4c(m)$ & \\
\hline
\multirow{3}{*}{$p2mg1'$} 
& $2a(21')$ & $4\text{mod}_2(\Delta_{2a,2}|\Delta_{2a,1}) $ \\
\cline{2-2}
& $2b(21')$ & $+4 \text{mod}_2(\Delta_{2b,2}|\Delta_{2b,1})$ \\
\cline{2-2}
& $4c(m1')$ &  \\
\hline
\multirow{2}{*}{$p2gg$} 
& $2a(2)$ & $2\text{mod}_2(\Delta_{2a,1}|\Delta_{2a,1}) $ \\
\cline{2-2}
& $2b(2)$ & $+2\text{mod}_2(\Delta_{2b,1}|\Delta_{2b,1})$ \\
\hline
\multirow{2}{*}{$p2gg1'$} 
& $2a(21')$ & $4 \text{mod}_2(\Delta_{2a,2}|\Delta_{2a,1}) $ \\
\cline{2-2}
& $2b(21')$ & $+4\text{mod}_2(\Delta_{2b,2}|\Delta_{2b,1}) $ \\
\hline
\multirow{3}{*}{$c2mm$} 
& $2a(2mm)$ & $2 \text{mod}_2(0|\Delta_{2a,1})  $ \\
\cline{2-2}
& $2b(2mm)$ & $ +2 \text{mod}_2(0|\Delta_{2b,1}) $ \\
\cline{2-2}
& $4c(2)$ & $+4\text{mod}_2(\Delta_{4c,1}|\Delta_{4c,1})$ \\
\hline
\multirow{3}{*}{$c2mm1'$} 
& $2a(2mm1')$ & $4 \text{mod}_2(\Delta_{2a,2}|\Delta_{2a,1})  $ \\
\cline{2-2}
& $2b(2mm1')$ & $+4 \text{mod}_2(\Delta_{2b,2}|\Delta_{2b,1}) $ \\
\cline{2-2}
& $4c(21')$ & $+8 \text{mod}_2(\Delta_{4c,2}|\Delta_{4c,1}) $ \\
\hline
\multirow{3}{*}{$p4$} 
& $1a(4)$ & $\text{mod}_4(2\Delta_{1a,2}-\Delta_{1a,1}|\Delta_{1a,1}) $ \\
\cline{2-2}
& $1b(4)$ & $+\text{mod}_4(2\Delta_{1b,2}-\Delta_{1b,1}|\Delta_{1b,1})$ \\
\cline{2-2}
& $2c(2)$ & $+2\text{mod}_2(\Delta_{2c,1}|\Delta_{2c,1})$ \\
\hline
\multirow{3}{*}{$p41'$} 
& $1a(41')$ & $2 \text{mod}_4(2\Delta_{1a,2}|\Delta_{1a,1}) $ \\
\cline{2-2}
& $1b(41')$ & $+2 \text{mod}_4(2\Delta_{1b,2}|\Delta_{1b,1})$ \\
\cline{2-2}
& $2c(21')$ & $+4 \text{mod}_4(2\Delta_{2c,2}|\Delta_{2c,1})$ \\
\hline
\multirow{3}{*}{$p4mm$} 
& $1a(4mm)$ & $\text{mod}_4(2\Delta_{1a,2}|\Delta_{1a,1})$ \\
\cline{2-2}
& $1b(4mm)$ & $\text{mod}_4(2\Delta_{1b,2}|\Delta_{1b,1})$ \\
\cline{2-2}
& $2c(2mm)$ & $ 2\text{mod}_4(0|\Delta_{2c,1})$ \\
\hline
\multirow{3}{*}{$p4mm1'$} 
& $1a(4mm1')$ & $\text{mod}_4(\Delta_{1a,2}|\Delta_{1a,1})$ \\
\cline{2-2}
& $1b(4mm1')$ & $+\text{mod}_4(\Delta_{1b,2}|\Delta_{1b,1})$ \\
\cline{2-2}
& $2c(2mm1')$ & $+2 \text{mod}_4(\Delta_{2c,2}|\Delta_{2c,1})$ \\
\hline
\multirow{2}{*}{$p4gm$} 
& $2a(4)$ & $2\text{mod}_4(2\Delta_{2a,2}-\Delta_{2a,2}|\Delta_{2a,2}) $ \\
\cline{2-2}
& $2b(2mm)$ & $2 \text{mod}_4(0|\Delta_{2b,2})$ \\
\hline
\multirow{2}{*}{$p4gm1'$} 
& $2a(41')$ & $4 \text{mod}_4(\Delta_{2a,2}|\Delta_{2a,1}) $ \\
\cline{2-2}
& $2b(2mm1')$ & $+4 \text{mod}_4(\Delta_{2b,2}|\Delta_{2b,1})$ \\
\hline
\multirow{3}{*}{$p3$} 
& $1a(3)$ & $\text{mod}_3(\Delta_{1a,1}|\Delta_{1a,2}) $ \\
\cline{2-2}
& $1b(3)$ & $+\text{mod}_3(\Delta_{1b,1}|\Delta_{1b,2})$ \\
\cline{2-2}
& $1c(3)$ & $+\text{mod}_3(\Delta_{1c,1}|\Delta_{1c,2})$ \\
\hline
\multirow{3}{*}{$p31'$} 
& $1a(31')$ & $2 \text{mod}_3(\Delta_{1a,1}|\Delta_{1a,1})$ \\
\cline{2-2}
& $1b(31')$ & $+2 \text{mod}_3(\Delta_{1b,1}|\Delta_{1b,1})$ \\
\cline{2-2}
& $1c(31')$ & $+2 \text{mod}_3(\Delta_{1c,1}|\Delta_{1c,1})$ \\
\hline
\multirow{3}{*}{$p3m1$} 
& $1a(3m)$ & $\text{mod}_3(\Delta_{1a,1}|\Delta_{1a,1}) $ \\
\cline{2-2}
& $1b(3m)$ & $+\text{mod}_3(\Delta_{1b,1}|\Delta_{1b,1})$ \\
\cline{2-2}
& $1c(3m)$ & $+\text{mod}_3(\Delta_{1c,1}|\Delta_{1c,1})$ \\
\hline
\multirow{3}{*}{$p3m1'$} 
& $1a(3m1')$ & $2\text{mod}_3(\Delta_{1a,1}|\Delta_{1a,1})  $ \\
\cline{2-2}
& $1b(3m1')$ & $+2\text{mod}_3(\Delta_{1b,1}|\Delta_{1b,1}) $ \\
\cline{2-2}
& $1c(3m1')$ & $+2\text{mod}_3(\Delta_{1c,1}|\Delta_{1c,1}) $ \\
\hline
\multirow{2}{*}{$p31m$} 
& $1a(3m)$ & $\text{mod}_3(\Delta_{1a,1}|\Delta_{1a,1})$ \\
\cline{2-2}
& $2b(3)$ & $+2\text{mod}_3(\Delta_{2b,1}|\Delta_{2b,2})$ \\
\hline
\multirow{2}{*}{$p31m1'$} 
& $1a(3m1')$ & $2\text{mod}_3(\Delta_{1a,1}|\Delta_{1a,1}) $ \\
\cline{2-2}
& $2b(31')$ & $+4\text{mod}_3(\Delta_{2b,1}|\Delta_{2b,1}) $ \\
\hline
\multirow{3}{*}{$p6$} 
& $1a(6)$ & $\text{mod}_6(2\Delta_{1a,2}-\Delta_{1a,1}|\Delta_{1a,1}) $ \\
\cline{2-2}
& $2b(3)$ & $+2\text{mod}_3(\Delta_{2b,1}+\Delta_{2b,2})$ \\
\cline{2-2}
& $3c(2)$ & $+3\text{mod}_2(\Delta_{3c,1})$ \\
\hline
\multirow{3}{*}{$p61'$} 
& $1a(61')$ & $2 \text{mod}_6(\Delta_{1a,2}|\Delta_{1a,1}) $ \\
\cline{2-2}
& $2b(31')$ & $+4 \text{mod}_3(2\Delta_{2b,1}|\Delta_{2b,1}) $ \\
\cline{2-2}
& $3c(2)$ & $+6 \text{mod}_2(\Delta_{3c,2}|\Delta_{3c,1}) $ \\
\hline
\multirow{3}{*}{$p6mm$} 
& $1a(6mm)$ & $ \text{mod}_6(2\Delta_{1a,2}|\Delta_{1a,1}) $ \\
\cline{2-2}
& $2b(3m)$ & $+2 \text{mod}_3(\Delta_{2b,1}|\Delta_{2b,1}) $ \\
\cline{2-2}
& $3c(2mm)$ & $+3 \text{mod}_2(0|\Delta_{3c,1})  $ \\
\hline
\multirow{3}{*}{$p6mm1'$} 
& $1a(6mm1')$ & $2 \text{mod}_6(\Delta_{1a,2}|\Delta_{1a,1}) $ \\
\cline{2-2}
& $2b(3m1')$ & $+4 \text{mod}_3(\Delta_{2b,1}|\Delta_{2b,1}) $ \\
\cline{2-2}
& $3c(2mm1')$ & $+6 \text{mod}_6(\Delta_{3c,2}|\Delta_{3c,1}) $ \\
\hline
\end{longtable}
\endgroup
\newpage 
\onecolumngrid

\section{Stable Topology and Global RSIs}
\label{app:strong}

In this Appendix, we study many-body stable topology using global many-body RSIs, which are defined using the same operators in \App{app:RSIconstruct} but evaluated on periodic boundary conditions. \App{eq:reductionstable} proves that the global many-body RSIs defined in this way vanish in all many-body atomic and many-body fragile topological states. \App{eq:Chernglobalrsi} then evaluates
 the global many-body RSIs on general product states, finding that they can be nonzero in symmetry-indicated Chern insulators. 

\subsection{Global Many-body RSIs in Many-body Atomic and Fragile States}
\label{eq:reductionstable}

In this section, we show that global many-body RSIs, defined in the Main Text, vanish in all many-body atomic states. Since the RSIs of many-body fragile states are the differences of the RSIs of many-body atomic states, this result shows that the global RSIs vanish there as all. We can state both results succinctly as: global many-body RSIs vanish if the many-body local RSIs are well-defined. We first prove our result in the spinless rotation groups, and then we discuss the addition of mirrors, time-reveral, and SOC. In the following, we fix a specific Wyckoff position $\mbf{x}$ with point group $G_\mbf{x}$. We then define global many-body RSIs at that Wyckoff position. In a given wallpaper group, there are global many-body RSIs defined at each Wyckoff position in the unit cell. For instance in $p2$, the global many-body RSI $\Delta^G_{1a,1}$ at $\mbf{x} = $1a with $G_{1a}  = \{1,C_2\}$ is computed from $e^{i \frac{\pi}{2} \hat{N}} C_2\ket{GS,PBC} = e^{i \frac{\pi}{2} \Delta^G_{1a,1}} \ket{GS,PBC} $, whereas the global many-body RSI at $\mbf{x} = $1b with $G_{1b} = \{1,T_1C_2\}$ is computed from $e^{i \frac{\pi}{2} \hat{N}} T_1 C_2\ket{GS,PBC} = e^{i \frac{\pi}{2} \Delta^G_{1b,1}} \ket{GS,PBC}$. We see that $\Delta^G_{1a,1}$ and $\Delta^G_{1b,1}$ are related by the total many-body momentum of the state. In the following, we consider a fixed Wyckoff position with $C_n$ understood to be the rotation centered at that Wyckoff position. For convenience, we choose the origin so that $\mbf{x} = (0,0)$, so the formulae we prove are for the 1a many-body global RSIs $\Delta^G_{1a,i}$, but the many-body global RSIs at other Wyckoff positions can be obtained using the total many-body momentum from translations. 

\begin{figure}[H]
 \centering
\begin{overpic}[height=0.2\textwidth,tics=10]{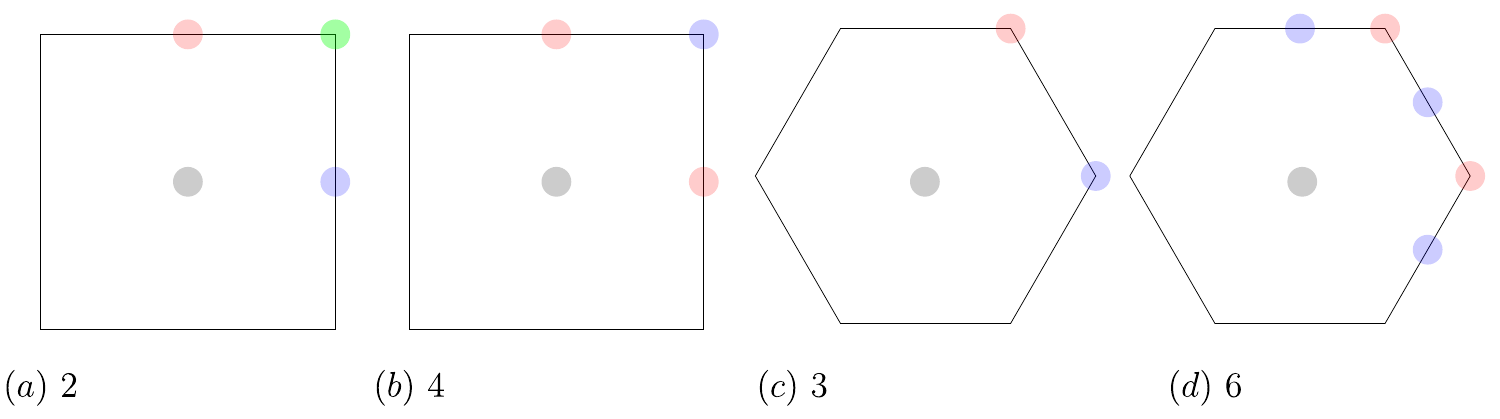}
\end{overpic} 
\begin{overpic}[height=0.2\textwidth,tics=10]{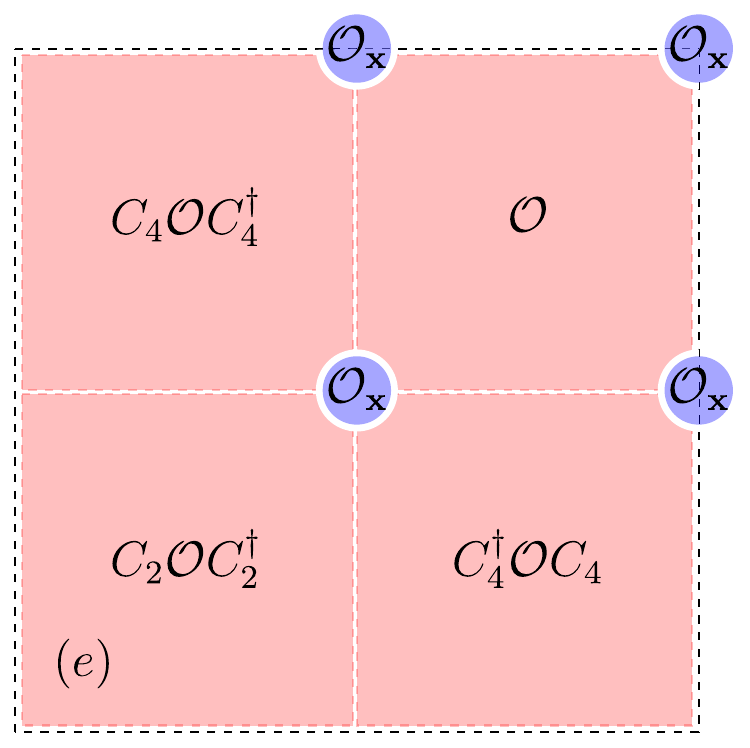}
\end{overpic} 
\caption{We show the Wyckoff positions $\mbf{x}^G$ that contribute to the global RSI at the 1a position in the PGs $2,3,4,6$. The plots here show the entire sample on periodic boundary conditions, and we assume the center of the $C_n$ operator is at the 1a position (center of the sample). Adding mirrors and TRS simply reduces the RSI expressions. Points of the same color are mapped to each other by the elementary rotation of the group. (a) There are four points symmetric under $C_2$. (b) There are two points (gray, blue) symmetric under $C_4$, and the red points are exchanged under $C_4$.  (c) There are three points symmetric under $C_3$. (d) There is one point (gray) symmetric under $C_6$, two points (red) that are exchanged under $C_6$, and three points (blue) that are permuted under $C_6$. (e) We depict the partitioning of a many-body atomic limit state on a spatial torus (PBCs) in rotation-related operators $\mathcal{O}$ and the finite $\mathcal{O}_\mbf{x}$ operators at fixed points of the rotation $\mbf{x} \in \mbf{x}^G$.}
\label{fig:PGglobal}
\end{figure}

The crucial feature of calculating the many-body global RSIs on periodic boundary conditions (PBCs) is that multiple points in the large spatial torus are invariant under the same rotation symmetry, unlike for OBCs where only a single point (the rotation center) is invariant. All other effective rotation symmetries like $T_1 C_n$ can be obtained from $C_n$ with translation operators, or equivalently by shifting the origin. Thus it is enough to consider only $C_n$. For each rotation $C_n$, we denote the set of points
\bea
\mbf{x}^G = \{\mbf{x}|C_n^m \mbf{x} = \mbf{x} \mod L_1 \mbf{a}_1, L_2 \mbf{a}_2 \ \forall \ 1\leq m < n \} \ .
\eea
\Fig{fig:PGglobal} depicts the various cases with $\mbf{x}^G$ shown by colored dots. Mathematically, for periodic boundary conditions on an $L_1 \mbf{a}_1 \times L_2 \mbf{a}_2$ lattice, a $C_n$-invariant point $\mbf{r}$ is defined by $C_n \mbf{r} = \mbf{r} \mod L_1 \mbf{a}_1, L_2 \mbf{a}_2$. We also require that $L_1,L_2$ are defined such that the high-symmetry points in the Brillouin zone exist, e.g. $L_1,L_2$ must be even with $C_2$ symmetry and multiples of three with $C_3$ symmetry.

Following identical steps as in \App{app:RSIconstruct}, we consider many-body atomic limit states on periodic boundary conditions. In the notation of \App{app:RSIconstruct}, they can be written as
\bea
\ket{GS,PBC} &= \prod_{\mbf{R},\mbf{r}} \mathcal{O}_{\mbf{R},\mbf{r}} \ket{0} = \prod_{\mbf{x} \in \mbf{x}^G} \mathcal{O}_{\mbf{x}} \ \prod_{i=1}^n C_n^i \mathcal{O} C_n^{i \dag} \ \ket{0}
\eea
where $\mathcal{O}= \prod_{\mbf{R}+\mbf{r} \in \mathcal{D}} \mathcal{O}_{\mbf{R},\mbf{r}}$ (see \App{app:RSIconstruct}) and $\mathcal{D}$ is $1/n$th of the total space excluding $\mbf{x}^G$, and is depicted as one of the shaded red squares in \Fig{fig:PGglobal}(e) for $C_n = C_4$. \App{app:RSIconstruct} showed explicitly that the RSI operators $Q_i = e^{i \frac{\pi}{n} \hat{N}} C_n, C_n^2$ for $n$ even and $Q_i = C_n, e^{i \frac{2\pi}{n} \hat{N}}$ for $n$ odd commute with $\prod_{i=1}^n C_n^i \mathcal{O} C_n^{i \dag}$ as shown in \Eq{eq:inductionargument}. In the context of \App{app:RSIconstruct}, the factor $\prod_{i=1}^n C_n^i \mathcal{O} C_n^{i \dag}$ appeared when increasing the size of the cutoff, under which the RSI operators are invariant. Analogously on PBCs, increasing $L_1, L_2$ increases the number of particle in $\prod_{i=1}^n C_n^i \mathcal{O} C_n^{i \dag}$, but crucially does not change the number of sites in $\mbf{x}^G$. Because the RSI operators commute with $\prod_{i=1}^n C_n^i \mathcal{O} C_n^{i \dag}$, we have
\bea
\label{eq:Qcharge}
Q_i \ket{GS,PBC} &= Q_i \prod_{\mbf{x} \in \mbf{x}^G} \mathcal{O}_{\mbf{x}} \ Q^\dag_i  Q_i \prod_{i=1}^n C_n^i \mathcal{O} C_n^{i \dag} Q_i^\dag \ket{0} \\
&=  \lp Q_i \prod_{\mbf{x} \in \mbf{x}^G} \mathcal{O}_{\mbf{x}}\  Q_i^\dag\rp \  \prod_{i=1}^n C_n^i \mathcal{O} C_n^{i \dag} \ket{0} \ . \\
\eea 
Thus we only need to compute the object in parentheses for each rotation group. We can perform the computation systematically using the translation operators which yield $\mathcal{O}_{\mbf{x}} = T_{\mbf{x}} \mathcal{O}_{(0,0)} T_{\mbf{x}} ^\dag$ since $\mbf{x} \in \mbf{x}^G$ are lattice vectors. Then we have
\bea
C_n \mathcal{O}_\mbf{x} C_n^\dag &= C_n T_{\mbf{x}} \mathcal{O}_{(0,0)} T_{\mbf{x}} ^\dag C_n  = T_{C_n \mbf{x}} C_n \mathcal{O}_{(0,0)} C_n^\dag T_{C_n \mbf{x}}^\dag = e^{i \la[C_n]} \mathcal{O}_{C_n\mbf{x}}
\eea
where $\la[C_n] \in \frac{2\pi}{n} \mathds{Z}_n$ is the angular momentum of $\mathcal{O}_{(0,0)}$. It follows that
\bea
\label{eq:usefulidentities}
e^{i \frac{\pi}{n} \hat{N}} \prod_{\mbf{x} \in \mbf{x}^G} \mathcal{O}_\mbf{x} \ e^{-i \frac{\pi}{n} \hat{N}}&= e^{i \frac{\pi}{n}n_G N_{\mathcal{O}} } \prod_{\mbf{x} \in \mbf{x}^G} \mathcal{O}_{\mbf{x}} \\
C_n \prod_{\mbf{x} \in \mbf{x}^G} \mathcal{O}_\mbf{x} \ C_n^\dag &= e^{i n_G \la[C_n]} \prod_{\mbf{x} \in \mbf{x}^G} \mathcal{O}_{C_n\mbf{x}} = e^{i n_G \la[C_n]} \lp \det{}_{(-1)^{N_{\mathcal{O}}}} R[C_n] \rp  \prod_{\mbf{x} \in \mbf{x}^G} \mathcal{O}_{\mbf{x}} \\
\eea
where $n_G$ is the number of sites in $\mbf{x}^G$ (the multiplicity of $\mbf{x}^G$ on PBCs), $N_{\mathcal{O}}$ is the total charge of $\mathcal{O}_{(0,0)}$, and $R[C_n]$ is the permutation matrix defined by $R_{ij}[C_n] = 1$ if $C_n \mbf{x}_i = \mbf{x}_j$ for $\mbf{x}^G = \mbf{x}_1,\dots, \mbf{x}_{n^G}$ (see \Eq{eq:detpm}). From \Fig{fig:PGglobal} which marks points in the same color if they are permuted to each other under $C_n$, we find
\bea
R[C_2] = \bpm 1 && & \\ & 1& & \\ & & 1 &  \\ &&&1 \epm, \quad R[C_3] = \bpm 1 & & \\ & 1&  \\ & & 1 \epm, \quad R[C_4] = \bpm 1 && & \\ & 1& & \\ & &  0&1  \\ &&1&0 \epm, \quad R[C_6] = \bpm 1 && & & & \\ & 0& 1& & & \\ & 1& 0& & & \\ & & &0 & 1 & \\& & & & 0 &1 \\ & & & 1&  &0 \epm
\eea
for which it is easy to check that $\det{}_\pm R[C_2] =\det{}_\pm R[C_3]  = +1$ and  $\det{}_\pm R[C_4] =\det{}_\pm R[C_6]  = \pm1$. We now compute the global RSIs $\Delta^G_1$ using \Eq{eq:usefulidentities}:
\bea
C_2:& \quad e^{i \frac{\pi}{2} \Delta_1^G}  \prod_{\mbf{x} \in \mbf{x}^G} \mathcal{O}_\mbf{x} = e^{i \frac{\pi}{2} \hat{N}} C_2  \prod_{\mbf{x} \in \mbf{x}^G} \mathcal{O}_\mbf{x} \ (e^{i \frac{\pi}{2} \hat{N}} C_2)^\dag = e^{i \frac{\pi}{2} 4 N_{\mathcal{O}} + i 4 \la[C_2]}  \prod_{\mbf{x} \in \mbf{x}^G} \mathcal{O}_\mbf{x} = +  \prod_{\mbf{x} \in \mbf{x}^G} \mathcal{O}_\mbf{x} \\
C_3:& \quad e^{i \frac{2\pi}{3} \Delta_1^G}  \prod_{\mbf{x} \in \mbf{x}^G} \mathcal{O}_\mbf{x} = C_3 \prod_{\mbf{x} \in \mbf{x}^G} \mathcal{O}_\mbf{x} \ C_3^\dag = e^{i 3 \la[C_3]}  \prod_{\mbf{x} \in \mbf{x}^G} \mathcal{O}_\mbf{x} = +  \prod_{\mbf{x} \in \mbf{x}^G} \mathcal{O}_\mbf{x} \\
C_4:& \quad e^{i \frac{\pi}{4} \Delta_1^G}  \prod_{\mbf{x} \in \mbf{x}^G} \mathcal{O}_\mbf{x} = e^{i \frac{\pi}{4} \hat{N}} C_4  \prod_{\mbf{x} \in \mbf{x}^G} \mathcal{O}_\mbf{x} \ (e^{i \frac{\pi}{4} \hat{N}} C_4)^\dag = e^{i \frac{\pi}{4} 4 N_{\mathcal{O}} + i 4 \la[C_4]} (-1)^{N_{\mathcal{O}}} \prod_{\mbf{x} \in \mbf{x}^G} \mathcal{O}_\mbf{x} = +  \prod_{\mbf{x} \in \mbf{x}^G} \mathcal{O}_\mbf{x} \\
C_6:& \quad e^{i \frac{\pi}{6} \Delta_1^G}  \prod_{\mbf{x} \in \mbf{x}^G} \mathcal{O}_\mbf{x} = e^{i \frac{\pi}{6} \hat{N}} C_6  \prod_{\mbf{x} \in \mbf{x}^G} \mathcal{O}_\mbf{x} \ (e^{i \frac{\pi}{6} \hat{N}} C_6)^\dag = e^{i \frac{\pi}{6} 6 N_{\mathcal{O}} + i 6 \la[C_6]} (-1)^{N_{\mathcal{O}}} \prod_{\mbf{x} \in \mbf{x}^G} \mathcal{O}_\mbf{x} = +  \prod_{\mbf{x} \in \mbf{x}^G} \mathcal{O}_\mbf{x} \\
\eea
showing that $\Delta_1^G = 0$ in all cases, using $n \la[C_n] = 0 \mod 2\pi$. Next we show that the $\Delta_2^G$ global RSIs also vanish:
\bea
C_3:& \quad e^{i \frac{2\pi}{3} \Delta_2^G}  \prod_{\mbf{x} \in \mbf{x}^G} \mathcal{O}_\mbf{x} = e^{i \frac{2\pi}{3} \hat{N}} \prod_{\mbf{x} \in \mbf{x}^G} \mathcal{O}_\mbf{x} \ e^{-i \frac{2\pi}{3} \hat{N}} = e^{i \frac{2\pi}{3} 3 N_{\mathcal{O}}}  \prod_{\mbf{x} \in \mbf{x}^G} \mathcal{O}_\mbf{x} = +  \prod_{\mbf{x} \in \mbf{x}^G} \mathcal{O}_\mbf{x} \\
C_4:& \quad e^{i \pi \Delta_2^G}  \prod_{\mbf{x} \in \mbf{x}^G} \mathcal{O}_\mbf{x} = C_4^2  \prod_{\mbf{x} \in \mbf{x}^G} \mathcal{O}_\mbf{x} C_4^{2 \dag} = (e^{i 4 \la[C_4]} (-1)^{N_{\mathcal{O}}})^2 \prod_{\mbf{x} \in \mbf{x}^G} \mathcal{O}_\mbf{x} = +  \prod_{\mbf{x} \in \mbf{x}^G} \mathcal{O}_\mbf{x} \\
C_6:& \quad e^{i \frac{2\pi}{3} \Delta_2^G}  \prod_{\mbf{x} \in \mbf{x}^G} \mathcal{O}_\mbf{x} = C_6^2  \prod_{\mbf{x} \in \mbf{x}^G} \mathcal{O}_\mbf{x} \ C_6^{2\dag} = (e^{i 6 \la[C_6]} (-1)^{N_{\mathcal{O}}})^2 \prod_{\mbf{x} \in \mbf{x}^G} \mathcal{O}_\mbf{x} = +  \prod_{\mbf{x} \in \mbf{x}^G} \mathcal{O}_\mbf{x} \\
\eea
showing that $\Delta_2^G = 0$ in all cases. Note that the $C_2$ case is trivial because there is no $\Delta_2^G$ invariant. 

Thus in all spinless rotation groups with PBCs preserving the high-symmetry points, we have shown that the global many-body RSI vanish in many-body atomic limits. In the spinless groups, this immediately extends to all the points groups adding mirrors and/or time-reversal because the RSIs in these cases can be obtained from the rotation subgroups by reduction (see \App{eq:generalRSIs}). Similarly, the rotation groups with SOC are isomorphic to the rotation groups without SOC and thus all the global many-body RSIs vanish there as well. With SOC, adding mirrors simply reduces the RSIs of the rotation groups, but we must consider the SOC groups with time-reversal separately. In these groups, there is always a many-body local RSI $\Delta_2 = N/2 \mod n$ which is not obtained by reduction. $\Delta_2$ is the eigenvalue of $e^{i \frac{2\pi}{n} \frac{\hat{N}}{2}}$, which intuitively counts the number of Kramers pairs (since $\mathcal{T}^2 = -1$) mod $n$. However, the global many-body RSI $e^{i \frac{2\pi}{n} \frac{\hat{N}}{2}} \ket{GS,PBC} = e^{i \frac{2\pi}{n} \Delta_2^G} \ket{GS,PBC}$ is always trivial as we now show. This follows because $\mathcal{O}_\mbf{x}$ must always create an even number of electrons since the state is non-degenerate, and $\mbf{x}^G$ always contains a multiple of $n$ points with $C_n$ symmetry. Thus $e^{i \frac{2\pi}{n} \frac{\hat{N}}{2}} \ket{GS,PBC} = +\ket{GS,PBC} $.

\subsection{Many-Body Global RSIs in the Band Theory Limit}
\label{eq:Chernglobalrsi}

We now show that nontrivial many-body global RSIs can be obtained from non-interacting Chern insulators. A similar calculation has been formed in Ref. \cite{2012PhRvB..86k5112F}, but to be self-contained we give the full details in our formalism here. First we define the momentum operators
\bea
c^\dag_{\mbf{k},\al} &= \frac{1}{\sqrt{L_1L_2}} \sum_{\mbf{R}} e^{- i \mbf{k} \cdot (\mbf{R} + \mbf{r}_\al)} c^\dag_{\mbf{R},\al}, \qquad \mbf{k} = k_1\mbf{b}_1 + k_2 \mbf{b}_2
\eea
where $\mbf{b}_i \cdot \mbf{a}_j = \delta_{ij}$, $\al = 1, \dots, N_{orb}$ is the orbital index, $\mbf{r}_\al$ is the location of the $\al$ orbital, and the sum is over lattice vectors $\mbf{R} = m \mbf{a}_1 + n \mbf{a}_2, m = 0,\dots, L_1-1, n = 0,\dots, L_2-1$. The momenta (defined mod $2\pi$) are
\bea
\label{eq:allowedk}
k_1 = 0, \frac{2\pi}{L_1}, \dots,  \frac{2\pi}{L_1}(L_1-1), \quad k_2 = 0, \frac{2\pi}{L_2}, \dots,  \frac{2\pi}{L_2}(L_2-1) .
\eea
In general, we require that $L_1,L_2 \in \mathbb{N}$ are chosen such that all high symmetry momenta appear in \Eq{eq:allowedk}, e.g. $L_1,L_2$ are even in systems with a $C_2$ symmetry. The Bloch states transform as eigenstates under translation via
\bea
\label{eq:Dperiodictrans}
T_i c^\dag_{\mbf{k},\al} T_i^\dag = e^{i \mbf{k} \cdot \mbf{a}_i} c^\dag_{\mbf{k},\al} \\
\eea
and under the (symmorphic) PG symmetries, e.g.  rotations, $g$ as
\bea
\label{eq:Dperiodicapp}
g c^\dag_{\mbf{k},\al} g^\dag &= \sum_{\be} c^\dag_{g \mbf{k},\be} D_{\be \al}[g] 
\eea
where $D[g]$ is the $N_{orb} \times N_{orb}$ representation matrix of $g$ on the orbitals. For spinless/spinfull electrons, $D[C_n^n]=D[M^2]=D[\mathcal{T}^2] = \pm1$. 

In non-interacting Hamiltonian, the eigenstate creation operators are $\gamma^\dag_{\mbf{k},n} = \sum_\al c^\dag_{\mbf{k},\al} U_{\al n}(\mbf{k})$ where $U_{\al n}(\mbf{k})$ are the elements of an $N_{orb} \times N_{orb}$ unitary matrix which rotates between the orbital index $\al$ and the band index $n$. The energies determined by the Hamiltonian are $[H, \gamma^\dag_{\mbf{k},n}] = E_n(\mbf{k})  \gamma^\dag_{\mbf{k},n}$. In an insulator, there is a gap between the $N_{occ}$ occupied bands and the $N_{orb} - N_{occ}$ unoccupied bands. We define the rectangular $N_{orb} \times N_{occ}$ matrix $[U(\mbf{k})]_{\al n},\  \al = 1,\dots, N_{orb}, n = 1, \dots, N_{occ}$ to be the eigenvectors of the occupied bands. It obeys $U^\dag(\mbf{k})U(\mbf{k}) = \mathbb{1}_{N_{occ}\times N_{occ}}$ and $U(\mbf{k})U^\dag(\mbf{k}) = P(\mbf{k})$ where $P(\mbf{k})^2 = P(\mbf{k})$ is the rank $N_{occ}$ projector \cite{2022PhRvL.128h7002H}. Symmetry enforces $E_n(g \mbf{k}) = E_n(\mbf{k})$ and
\bea
\label{eq:UgUB}
U(g \mbf{k}) = D[g] U(\mbf{k}) B^\dag_g(\mbf{k}), \qquad B_g(\mbf{k}) = U^\dag(g \mbf{k}) D[g] U(\mbf{k}) \\
\eea
where $B_g(\mbf{k})$ is called the sewing matrix \cite{andreibook} and is an $N_{occ} \times N_{occ}$ unitary matrix with nonzero elements only between states of the same energy. At high symmetry points $\mbf{K} = g \mbf{K} \mod 2\pi \mbf{b}_i$ for $g \in G_\mbf{K}$, $B_g(\mbf{K})$ is the representation matrix of the little group $G_\mbf{K}$ \cite{2022PhRvL.128h7002H}. A simple calculation gives
\bea
\label{eq:ggamma}
g \gamma^\dag_{\mbf{k},n} g^\dag &= \sum_\al g c^\dag_{\mbf{k},\al} g^\dag U_{\al n}(\mbf{k}) \\
&= \sum_\be c^\dag_{g\mbf{k},\be} [ D[g] U(\mbf{k})]_{\be n} \\
&= \sum_\be c^\dag_{g\mbf{k},\be} [ D[g] U(\mbf{k}) B^\dag_g(\mbf{k}) B_g(\mbf{k})]_{\be n} \\
&= \sum_\be c^\dag_{g\mbf{k},\be} [U(g\mbf{k})B_g(\mbf{k})]_{\be n} \\
&= \sum_m \gamma^\dag_{g\mbf{k},m} [B_g(\mbf{k})]_{m n} \\
\eea
using \Eq{eq:UgUB}. With this result, we can evaluate the action of a symmetry operators on any state. However, it will be useful to express \Eq{eq:ggamma} in a more general form:
\bea
\label{eq:SGrep}
g \gamma^\dag_{\mbf{k},n} g^\dag &= \sum_{\mbf{k}'m} \gamma^\dag_{\mbf{k}',m} [\mathcal{B}_g]_{\mbf{k}'m,\mbf{k}n}, \qquad [\mathcal{B}_g]_{\mbf{k}'m,\mbf{k}n} =  \delta_{\mbf{k}',g\mbf{k}} [B_g(\mbf{k})]_{m n} 
\eea
such that $\mathcal{B}_{\mbf{k}'m,\mbf{k}n}[g]$ is the $L_1L_2 N_{occ} \times L_1L_2 N_{occ}$ representation matrix of $g$ on \emph{all} the occupied states (it is a space group representation). 

We now consider the groundstate of a band insulator denoted
\bea
\ket{GS} &= \prod_{\mbf{k},n} \gamma^\dag_{\mbf{k},n} \, \ket{0} \\
\eea 
where the product is taken over the $L_1L_2 N_{occ} $ occupied states in the BZ. \Eq{eq:SGrep} now immediately gives
\bea
\label{eq:detBg}
g \ket{GS} &= \det \mathcal{B}_g \, \ket{GS}
\eea
because there are $L_1L_2 N_{occ}$ anti-commuting $\gamma^\dag_{\mbf{k},n} $ operators which fully anti-symmetrize the $L_1L_2 N_{occ}\times L_1L_2 N_{occ}$ matrix $\mathcal{B}_g$. It is easy to evaluate $ \det \mathcal{B}_g$ since $\mathcal{B}_g$ is block-diagonalized into representations, and the determinant of a direct sum is the product of their determinants. We now calculate the representations at all points in the BZ, which divide into high-symmetry points with nontrivial little groups (of which there are only a finite number in each space group) and non-high-symmetry points where the little group is trivial (of which there is a macroscopic number). We will see that the high-symmetry point irrep determine the Chern number mod $n$ and the non-high-symmetry points contribute overall phases which are canceled due to our formula for the global RSI. 

Let us study $G=p2$ with $C_2$ and translations. All states off the high-symmetry momenta with $\mbf{k} \neq - \mbf{k} \mod BZ$ transform in a 2D representation of $C_2$ (if there are accidental degeneracies, we allow for a sum of 2D representations), since $C_2$ exchanges the distinct states $\mbf{k},-\mbf{k}$ with an phase factor $e^{i \th}$ determined by the eigenvectors which cancels in the determinant:  
\bea
\label{eq:C2rep}
C_2 \begin{pmatrix}
\gamma^\dag_{\mbf{k},n} \\
\gamma^\dag_{-\mbf{k},n} \\
\end{pmatrix} C_2^\dag  &= \begin{pmatrix}
e^{i \th} \gamma^\dag_{-\mbf{k},n} \\
e^{-i \th} \gamma^\dag_{\mbf{k},n} \\
\end{pmatrix} = \bpm 0 & e^{i \th} \\ e^{-i \th} & 0 \epm \begin{pmatrix}
\gamma^\dag_{\mbf{k},n} \\
\gamma^\dag_{-\mbf{k},n} \\
\end{pmatrix}, \qquad \det \bpm 0 & e^{i \th} \\ e^{-i \th} & 0 \epm   = -1 \ .
\eea
From the group theory perspective, states off high-symmetry momenta have a trivial little group, and their representation can be expressed as $E \uparrow G = A \oplus B$ where $E$ is the irrep of the trivial group. Note that $\det A \oplus B = (+1)(-1) = -1$ in agreement with \Eq{eq:C2rep}. Thus we have shown that all states off the high-symmetry momenta contribute a factor of $(-1)$ per pair. The only remaining states at the four high-symmetry momenta given by
\bea
\Gamma = (0,0), \quad X = \pi \mbf{b}_1, \quad Y = \pi \mbf{b}_2, \quad M = \pi \mbf{b}_1 +\pi \mbf{b}_2 
\eea
whose irreps are denoted $\Gamma_1,\Gamma_2, \dots, M_1, M_2$ with $\rho_1 \cong A, \rho_2 \cong B$ so only the $\Gamma_2, X_2,Y_2, M_2$ irrep contribute nontrivial factors to $\det \mathcal{B}_g$. Letting $m(\rho)$ denote the multiplicity of the $\rho$ irrep in $\ket{GS}$, we now have proven
\bea
C_2 \ket{GS} = (-1)^{m(\Gamma_2)+m(X_2)+m(Y_2)+m(M_2)} (-1)^{ (L_1 L_2 -4)N_{occ}/2} \ket{GS} \\
\eea
since there are $(L_1 L_2 -4)N_{occ}/2$ pairs of states off the high-symmetry momenta. We can now evaluate the global many-body RSI $\Delta_1$ according to $e^{i \frac{\pi}{2} \hat{N}} C_2 \ket{GS} = e^{i \frac{\pi}{2} \Delta_1} \ket{GS}$:
\bea
\label{eq:C2globalrsi}
e^{i \frac{\pi}{2} \hat{N}} C_2 \ket{GS} &= (-1)^{m(\Gamma_2)+m(X_2)+m(Y_2)+m(M_2)} (-1)^{ (L_1 L_2 -4)N_{occ}/2} e^{i \frac{\pi}{2} L_1 L_2 N_{occ} }\ket{GS} \\
&= (-1)^{m(\Gamma_2)+m(X_2)+m(Y_2)+m(M_2)} e^{i \frac{\pi}{2} L_1 L_2 N_{occ}} e^{i \frac{\pi}{2} L_1 L_2 N_{occ} }\ket{GS} \\
&= (-1)^{m(\Gamma_2)+m(X_2)+m(Y_2)+m(M_2)} \ket{GS} \\
\eea
where we used that $L_1$ and $L_2$ must be even for the $X,Y,$ and $M$ points to exist, and that there are $N = N_{occ} L_1 L_2$ electrons in the groundstate. We see that the $e^{i \frac{\pi}{2} \hat{N}}$ operator which naturally appears in the definition of the many-body global RSI cancels the size-dependent phase factor. Finally, the band theory invariant $m(\Gamma_2)+m(X_2)+m(Y_2)+m(M_2) = \th_2 \mod 2$ is called a symmetry
indicator and is known to obey $\th_2 = C \mod 2$ where $C$ is the Chern number \cite{2012PhRvB..86k5112F}. Thus \Eq{eq:C2globalrsi} proves
\bea
\Delta^G_1 = 2C \mod 4 \ . 
\eea
We observe that although the $e^{i \frac{\pi}{2} \hat{N}} C_2$ operator gives $ \Delta^G_1 \in \mathds{Z}_4$, only even $\Delta^G_1 = 2C \mod 4$ is possible in band insulators.

We now return to the general case of \Eq{eq:detBg}. In the spinless groups, we only need to study the rotations $C_n$ since, as proved in \App{eq:generalRSIs}, the RSIs with mirrors and/or time-reversal are obtained by reduction from the rotation groups. To evaluate $\det \mathcal{B}_g$, we consider the possible $C_n$ representations induced from each $\mbf{k}$ point. We now enumerate the possibilities. 

In $p4$, the $\Gamma,M$ points have $G_\Gamma=G_M = 4$, the $X,Y$ points are interchanged by $C_4$ and have $G_X,G_Y = 2$, and all other points have $G_\mbf{k} =1$, the trivial group. The irrep inductions at the high-symmetry points are
\bea
\Gamma_1  \downarrow 4 &= A, \quad \Gamma_2 \downarrow 4 = B, \quad \Gamma_3  \downarrow 4 ={}^2E, \quad \Gamma_4 \downarrow 4 = {}^1E  \\
M_1 \downarrow 4 &= A, \quad M_2 \downarrow 4 = B, \quad M_3 \downarrow 4 ={}^2E, \quad M_4 \downarrow 4 = {}^1E  \\
X_1 \downarrow 4 & = A \oplus B, \qquad X_2 \downarrow 4 = {}^1E \oplus {}^2E \\
\eea
and at every non-high-symmetry point, the four $C_4$ related states induce $A \oplus B \oplus {}^1E \oplus {}^2E$ with $\det (A \oplus B \oplus {}^1E \oplus {}^2E) = (+1)(-1)(-i)(+i) = -1$. Thus we find
\bea
C_4 \ket{GS} &= e^{i \frac{2\pi}{4}( 2m(\Gamma_2)+2m(M_2) +  m(\Gamma_3)+m(M_3)-  m(\Gamma_4) - m(M_4) + 2 m(X_1) ) } (-1)^{ (L_1 L_2 - 4) N_{occ}/4} \ket{GS} \\
e^{i \frac{\pi}{4} \hat{N}} C_4 \ket{GS} &= e^{i \frac{2\pi}{4}( 2m(\Gamma_2)+2m(M_2) +  m(\Gamma_3)+m(M_3)-  m(\Gamma_4) - m(M_4) + 2 m(X_1) ) } e^{i \frac{\pi}{4} (L_1 L_2 - 4) N_{occ} +i \frac{\pi}{4} L_1 L_2 N_{occ}} \ket{GS} \\
&= e^{i \frac{2\pi}{4}( 2m(\Gamma_2)+2m(M_2) +  m(\Gamma_3)+m(M_3)-  m(\Gamma_4) - m(M_4) + 2 m(X_1) ) } e^{i \pi N_{occ}} \ket{GS} \\
%C_4^2 \ket{GS} &= e^{i \pi(m(\Gamma_3)+m(M_3) +m(\Gamma_4) + m(M_4)) } (-1)^{ (L_1 L_2 - 4) N_{occ}/2} \ket{GS} \\
\eea
where we used that $L_1 L_2 \in 4 \mathbb{N}$ for the high-symmetry points to exist. We now also use $N_{occ} = m(X_1)+ m(X_2)$ to find $e^{i \frac{2\pi}{4} 2 m(X_1)  } e^{i \pi N_{occ}} = e^{i \frac{2\pi}{4}( 2 m(X_2) )}$ yielding
\bea
e^{i \frac{\pi}{4} \hat{N}} C_4 \ket{GS}  &= e^{i \frac{2\pi}{4}( 2m(\Gamma_2)+2m(M_2) +  m(\Gamma_3)+m(M_3)-  m(\Gamma_4) - m(M_4) + 2 m(X_2) ) } \ket{GS} = e^{i \frac{2\pi}{4} C} \ket{GS} \\
\eea
where the relation between the irrep multiplicities and the Chern number was first proved in Ref. \cite{2012PhRvB..86k5112F} using the Wilson loop. Thus we obtain
\bea
\Delta^G_1 = 2 C \mod 8, \quad \Delta^G_2 = C \mod 2
\eea
where the second equality follows from $e^{i \pi \Delta_2} \ket{GS}  = C_4^2 \ket{GS} = (e^{i \frac{\pi}{4} \hat{N}} C_4)^2 \ket{GS} $ since $\hat{N}$ is a multiple of 4 on $\ket{GS}$. 

We now study $G=p6$ with three high-symmetry points where the irrep inductions are
\bea
\Gamma_1 \downarrow 6 &= A, \ \Gamma_2  \downarrow 6 = B, \ \Gamma_3  \downarrow 6 = {}^2E_1, \ \Gamma_4  \downarrow 6 = {}^2E_2, \ \Gamma_5  \downarrow 6 = {}^1E_1, \ \Gamma_5  \downarrow 6 = {}^1E_2 \\
K_1  \downarrow 6 &= A \oplus B, \ K_2 \downarrow 6 = {}^2E_1 \oplus {}^2E_2, \ K_3  \downarrow 6 = {}^1E_1 \oplus {}^1E_2 \\
M_1 \downarrow 6 &= A \oplus {}^1E_1 \oplus {}^2E_1, \ M_2 \downarrow 6 = B\oplus {}^1E_2 \oplus {}^2E_2
\eea
and at every non-high-symmetry point, the six $C_6$-related states induce $A\oplus B \oplus {}^1E_1  \oplus {}^1E_2  \oplus {}^1E_2  \oplus {}^2E_2$ with $\det (A\oplus B \oplus {}^1E_1  \oplus {}^1E_2  \oplus {}^1E_2  \oplus {}^2E_2) = -1$. We compute
\bea
C_6 \ket{GS} &= e^{i \frac{2\pi}{6}(3 m(\Gamma_2) + m(\Gamma_4) +2 m(\Gamma_5) -2 m(\Gamma_3) - m(\Gamma_6) + 3m(K_1) - m(K_2) + m(K_3) + 3 m(M_2) )} (-1)^{(L_1L_2-6) N_{occ}/6} \ket{GS} \\
e^{i \frac{\pi}{6} \hat{N}} C_6 \ket{GS} &= e^{i \frac{2\pi}{6}(3 m(\Gamma_2) + m(\Gamma_4) +2 m(\Gamma_5) -2 m(\Gamma_3) - m(\Gamma_6) + 3m(K_1) - m(K_2) + m(K_3) + 3 m(M_2) )} e^{i \frac{\pi}{6}(2L_1L_2-6) N_{occ}} \ket{GS} \\
\eea
and now use that $L_1L_2$ must be a multiple of $6$ for the $K$ and $M$ points to exist and that $N_{occ} = m(K_1)+m(K_2) + m(K_3)$. Plugging in yields
\bea
e^{i \frac{\pi}{6} \hat{N}} C_6 \ket{GS} &= e^{i \frac{2\pi}{6}(3 m(\Gamma_2) + m(\Gamma_4) +2 m(\Gamma_5) -2 m(\Gamma_3) - m(\Gamma_6) + 3m(K_1) - m(K_2) + m(K_3) + 3 m(M_2) )} e^{i \frac{2\pi}{6} 3(m(K_1)+m(K_2) + m(K_3))} \ket{GS} \\
&= e^{i \frac{2\pi}{6}(3 m(\Gamma_2) + m(\Gamma_4) +2 m(\Gamma_5) -2 m(\Gamma_3) - m(\Gamma_6) + 2m(K_2) - 2m(K_3) + 3 m(M_2) )}  \ket{GS} \\
&= e^{i \frac{2\pi}{6} C} \ket{GS} \\
\eea
using the results of Ref. \cite{2012PhRvB..86k5112F} in the last line. Thus we obtain
\bea
\Delta^G_1 = 2C \mod 12, \quad \Delta^G_2 = C \mod 3 \\
\eea
again using $(e^{i \frac{\pi}{6} \hat{N}} C_6)^2 \ket{GS} = C_6^2 \ket{GS}$ since $\ket{GS}$ contains a multiple of 6 electrons to obtain the last equality. 

Finally, we consider $G = p3$ which has an odd rotation. The irrep reductions are
\bea
\Gamma_1 \downarrow 6 &= A, \quad \Gamma_2 \downarrow 6 = {}^2E, \quad \Gamma_3 \downarrow 6 = {}^1E \\
K_1 \downarrow 6 &= A, \quad K_2 \downarrow 6 = {}^2E, \quad K_3 \downarrow 6 = {}^1E \\
K'_1 \downarrow 6 &= A, \quad K'_2 \downarrow 6 = {}^1E, \quad K'_3 \downarrow 6 = {}^2E \\
\eea
and at every non-high-symmetry point, the three $C_3$ related states induce $A\oplus {}^1E\oplus {}^2E$ which has $\det(A\oplus {}^1E\oplus {}^2E) = +1$. It is now direct to compute
\bea
C_3 \ket{GS} &= e^{i \frac{2\pi}{3} (m(\Gamma_2)-m(\Gamma_3)+m(K_2)-m(K_3) - m(K'_2)+m(K'_3))}\ket{GS} = e^{i \frac{2\pi}{3}C}\ket{GS}
\eea
which, using the results of Ref. \cite{2012PhRvB..86k5112F}, leads to
\bea
\Delta^G_1 = C \mod 3, \qquad \Delta^G_2 = 0 \mod 3
\eea
where we used that $e^{i \frac{2\pi}{3} \hat{N}} \ket{GS} = e^{i \frac{2\pi}{3} \Delta_2}\ket{GS} = +\ket{GS}$ since must contain a multiple of 3 electrons. This is because $L_1,L_2$ must be multiples of 3 for the $K$ points to be defined. 

We summarize the results of this section as follows. We have proven
\bea
\label{eq:summaryresults}
\Delta^G_1 = 2C \mod 2n, \qquad \Delta^G_2 = C \mod n/2, \qquad G_\mbf{x} = C_n, \quad \text{$n$ even} \\
\Delta^G_1 = C \mod n, \qquad \Delta^G_2 = 0 \mod n, \qquad G_\mbf{x} = C_n, \quad \text{$n=3$ odd} \ . \\
\eea
in a convention where $C_n$ is a rotation about the origin. Because mirrors and time-reversal set $C=0$ and the many-body global RSIs of general spinless point groups are reduced from those of their rotation subgroups, \Eq{eq:summaryresults} completes our classification. With SOC and time-reversal, there is a global many-body RSI $\Delta_2$ which counts the Kramers pairs mod $n$ and is not reduced from the $C_n$ subgroup, but $\Delta^G_2 = 0$ is easily proven because all irreps are $2D$ because of spinful time-reversal, and the number of points in the $BZ$ is $L_1 L_2$ which is a multiple of $n$ since we require the high-symmetry points to exist. Thus in all point groups (with and without SOC) with mirrors and/or time-reversal, all the global many-body RSIs vanish on non-interacting states. However, the global many-body RSIs in these point groups do not necessarily vanish on correlated states and therefore diagnose topology which is only possible with strong interactions. 

In wallpaper groups with multiple Wyckoff positions, the global many-body RSIs can be evaluated with \Eq{eq:summaryresults} and the total many-body momentum, which is zero in a 2D band insulator. For instance in $p4$, consider the Wyckoff positions 1a$=(0,0)$ and 1b$=(1/2,1/2)$ whose point groups are generated by $C_4$ and $T_1 C_4,$ respectively. Because $T_1$ acts trivially on the filled band groundstate, we have $(\Delta_{1a, 1}, \Delta_{1a, 2}) = (\Delta_{1b, 1}, \Delta_{1b, 2}) = (2C,C)$. At the 2c $= \{(1/2,0),(0,1/2)\}$ position, the site symmetry groups are generated by $T_1 C_2$ and $T_2 C_2$ respectively. Evaluating the global many-body RSI at 2c is easily accomplished by computing
\bea
e^{i \frac{\pi}{2} \hat{N}} T_1 C_2 \ket{GS,PBC} &= (e^{i \frac{\pi}{4} \hat{N}}  C_4)^2 T_1\ket{GS,PBC} = e^{i 2 \frac{\pi}{4} \Delta_{1a,1}^G} \ket{GS,PBC} \\
e^{i \frac{\pi}{2} \hat{N}} T_2 C_2 \ket{GS,PBC} &= (e^{i \frac{\pi}{4} \hat{N}}  C_4)^2 T_2\ket{GS,PBC} = e^{i 2 \frac{\pi}{4} \Delta_{1a,1}^G} \ket{GS,PBC} \ . \\
\eea
In a general wallpaper group, only the many-body global RSI at the 1a positions needs to be evaluated, and the rest are determined by the many-body translation operators.

\section{Details of the Exactly Solvable Hamiltonian}
\label{app:modelfragile}

In this Appendix, we present a model with single-particle fragile topology that is trivialized by the introduction of interactions. In \App{app:spham} we construct a non-interacting Hamiltonian with two phases, fragile and trivial, as we compute using single-particle RSIs and band representations. However, we show that the interacting RSIs are trivial in both phases. To illustrate this physically, in \App{app:wannierham} we add an interaction term that adiabatically connects the two single-particle phases. We can show this explicitly because the Hamiltonian is exactly solvable. 

\subsection{Construction of the Fragile Flat Band Hamiltonian}
\label{app:spham}

\begin{figure}
 \centering
\begin{overpic}[height=0.25\textwidth,tics=10]{fragmodellattice}
\end{overpic} \quad
\begin{overpic}[height=0.25\textwidth,tics=10]{fragmodelorb}
\end{overpic} \qquad
\begin{overpic}[height=0.25\textwidth,tics=10]{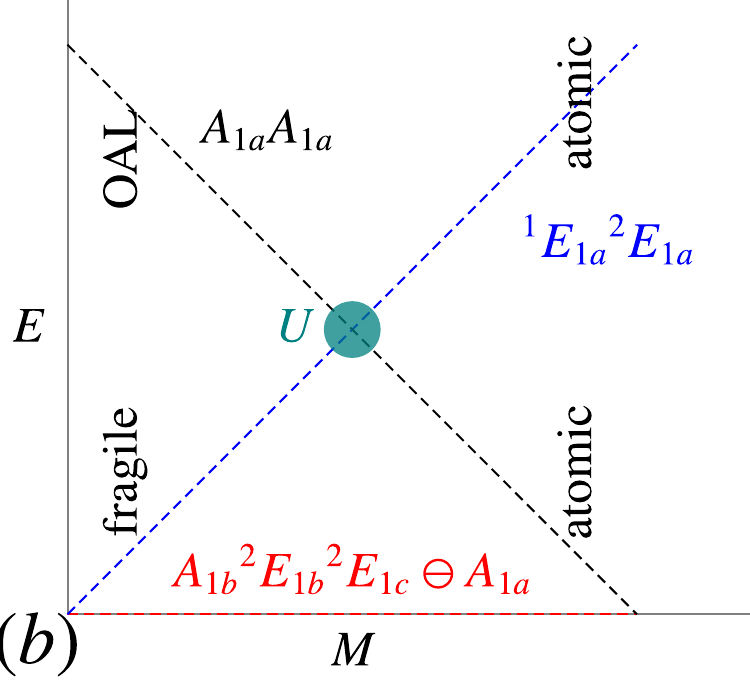}
\end{overpic} 
\caption{(a) We show the atomic orbitals in the space group $p3$ appearing on the Wyckoff positions 1a (grey), 1b (red), and 1c (blue) which are shown within a dashed unit cell. The fragile bands and OAL complement are formed from the three orbitals at 1b and 1c. The 1a orbitals are decoupled form the rest, and are only acted on by the interaction term. (b) We show the single-particle phase diagram. The fragile bands (red) are fixed at zero energy. The trivial atomic ${}^1E_{1a}{}^2E_{1a}$ bands (blue) increase in energy as a function of $M$ and the trivial $A_{1a}A_{1a}$ bands (black), one of which is the obstructed atomic Wannier state, decrease in energy. For $M < 1/2$, the ${}^1E_{1a}{}^2E_{1a}$ do not trivial the fragile topology, but for $M > 1/2$, the $A_{1a} A_{1a}$ bands remove the fragile obstruction. The band crossing is opened by the interaction $U$.}
\label{fig:modelorbs}
\end{figure}

Out starting point is a non-interacting Hamiltonian on a triangular lattice with 6 bands which has the space group $G = p3$ consisting of translations and a $C_3$ rotation. The lattice and atomic orbitals are shown in \Fig{fig:modelorbs}. We choose $\mbf{a}_1 = (1,0) / \sqrt{\frac{\sqrt{3}}{2}}, \mbf{a}_2 = C_3 \mbf{a}_1$ which satisfy $\mbf{a}_1 \times \mbf{a}_2 = 1$. The Wyckoff positions are
\bea
\text{1a } = (0,0), \quad \text{1b } = \frac{2}{3} \mbf{a}_1 + \frac{1}{3} \mbf{a}_2, \quad \text{1c } = \frac{1}{3} \mbf{a}_1 + \frac{2}{3} \mbf{a}_2
\eea
and each has PG 3. We choose our model to be spinless and denote the irreps $A,{}^1\!E,{}^2\!E$ of PG 3 by 
\bea
D_A[C_3] &= 1, \quad D_{{}^1\!E}[C_3] = \omega , \quad D_{{}^2\!E}[C_3] = \omega^*, \qquad \omega = \exp \lp i \frac{2\pi}{3} \rp \ . \\
\eea
We begin the construction of our Hamiltonian by creating the orbitals $A_{1b}, {}^2\!E_{1b},{}^2\!E_{1c}$ where $\rho_{\mbf{x}}$ denotes the $\rho$ irrep of $C_3$ at Wyckoff position $\mbf{x}$. From these orbitals, we construct the following Wannier function at $\mbf{R} = 0$:
\bea
\label{eq:defwannier}
w^\dag_{0,A_{1a}} = \frac{1}{3} \sum_{j=0}^2 C_3^j (c^\dag_{0, A_{1b}} + c^\dag_{0, {}^2\!E_{1b}} + c^\dag_{0, {}^2\!E_{1c}} ) {C_3^\dag}^j
\eea
where $C_3$ denotes a rotation about the origin. Note that
\bea
C_3 w^\dag_{0,A_{1a}} C_3^\dag &=  \frac{1}{3} \sum_{j=1}^3 C_3^j (c^\dag_{0, A_{1b}} + c^\dag_{0, {}^2\!E_{1b}} + c^\dag_{0, {}^2\!E_{1c}} ) {C_3^\dag}^j = (+1) w^\dag_{0,A_{1a}} 
\eea
so this state transforms like an $A$ irrep at the $1a$ position, justifying our notation. Secondly, we are justified in calling $w^\dag_{0,A_{1a}}$ a Wannier function because it obeys the orthogonality condition
\bea
\label{eq:wAorth}
\{w^\dag_{\mbf{R},A_{1a}}, w_{\mbf{R}',A_{1a}} \} &= \delta_{\mbf{R},\mbf{R}'}
\eea
where $w^\dag_{\mbf{R},A_{1a}}$ is formed from $w^\dag_{0,A_{1a}}$ by translation. \Eq{eq:wAorth} may be proved as follows. Clearly $\{w^\dag_{\mbf{R},A_{1a}}, w_{\mbf{R}',A_{1a}} \} = 0$ if $\mbf{R} \neq \mbf{R}'$ are \emph{not} nearest neighbors. Let us then compute $\{w^\dag_{0,A_{1a}}, w_{\mbf{a}_1,A_{1a}} \}$, which comes from contributions in the $\mbf{R} =0$ and $\mbf{R} = -\mbf{a}_2$ unit cells. We find
\bea
\{w^\dag_{0,A_{1a}}, w_{\mbf{a}_1,A_{1a}} \} &= \frac{1}{9} \Big( 
\{c^\dag_{0, A_{1b}} + c^\dag_{0, {}^2\!E_{1b}} + {\omega^*}^2 c^\dag_{-\mbf{a}_2, {}^2\!E_{1b}},c^\dag_{0, A_{1b}} + \omega c^\dag_{0, {}^2\!E_{1b}} + \omega c^\dag_{-\mbf{a}_2, {}^2\!E_{1b}} \} \Big) \\
&= \frac{1}{9}(1 + \omega + \omega^*) = 0 \ .
\eea
We can check all other anti-commutation relations with other nearest neighbors identically, and we find that they are zero by construction. Invoking translation symmetry, this proves \Eq{eq:wAorth}. 

We now develop a single-particle Hamiltonian whose bands form $w^\dag_{\mbf{R},A_{1a}}$. Performing a Fourier transform, we find the momentum-space eigenvector
\bea
U_{A_{1a}}(\mbf{k}) &= \frac{1}{3} \bpm 1 \\ 1 \\ 1\epm + \frac{1}{3} \bpm 1 \\ \omega^* \\ 0 \epm e^{-i \mbf{k} \cdot (-\mbf{a}_1)} + \frac{1}{3} \bpm 1 \\ \omega \\  \omega^* \epm e^{-i \mbf{k} \cdot (-\mbf{a}_1-\mbf{a}_2)}  + \frac{1}{3} \bpm 0 \\ 0 \\ \omega \epm e^{-i \mbf{k} \cdot (-\mbf{a}_2)} 
\eea
in the basis $(A_{1b}, {}^2\!E_{1b}, {}^2\!E_{1c})^T$ which satisfies
\bea
w^\dag_{\mbf{R},A_{1a}} &= \int \frac{d^2k}{(2\pi)^2} e^{i \mbf{R} \cdot \mbf{k}} U^\al_{A_{1a}}(\mbf{k}) c^\dag_{\mbf{k}, \al}
\eea
and the sum over $\al = A_{1b}, {}^2\!E_{1b}, {}^2\!E_{1c}$ is implied. Here $\mbf{k} = k_1 \mbf{b}_1 + k_2 \mbf{b}_2, \mbf{b}_i\cdot \mbf{a}_j = \delta_{ij}$ is a momentum in the Brillouin zone defined by $k_1,k_2 \in (-\pi,\pi)$ and $c^\dag_{\mbf{k},\al}$ is the electron creation operator at momentum $\mbf{k}$ at orbital $\al$. If we choose this Wannier band to be a \emph{conduction} band at energy $1-M>0$, then the single-particle Hamiltonian is proportional to a projection matrix:
\bea
\label{eq:handfvectors}
(1-M) h_{f}(\mbf{k}) &= (1-M) U_{A_{1a}}(\mbf{k}) U^\dag_{A_{1a}}(\mbf{k})
\eea
and thus has three perfectly flat bands. Note that $h_{f}(\mbf{k})$ projects out the fragile bands at zero energy, whose orthonormal eigenvectors $U_{f,1}(\mbf{k}), U_{f,2}(\mbf{k})$ are orthogonal to $U_{A_{1a}}(\mbf{k})$. By construction from the $w^\dag_{\mbf{R},A_{1a}}$ states, $h_{f}(\mbf{k})$ has only nearest neighbor hoppings in real space. The complement of the fragile bands are the obstructed atomic limit band. 

Using the Bilbao Crystallographic server (\url{https://www.cryst.ehu.es/cgi-bin/cryst/programs/bandrep.pl}), we compute the band representation of the unoccupied Wannier band at energy $1-M>0$ to be
\bea
 \Gamma_1 + K_1 + K'_1 = A_{1a} \uparrow G
\eea
as expected by construction. The band representation of the two occupied \emph{valence} bands is 
\bea
2 \Gamma_3 + 2 K_2 + K'_2 + K'_3 = ( A_{1b} \oplus {}^2\!E_{1b} \oplus {}^2\!E_{1c}) \uparrow G \ominus  A_{1a} \uparrow G
\eea
which is single-particle fragile. We expected the occupied bands to have fragile topology because the conduction bands is an \emph{obstructed} atomic limit (OAL), and the difference of two atomic bands can be fragile. Shortly, we show that the valence bands are fragile using their single-particle RSIs. 

First, we add three more uncoupled orbitals to the Hamiltonian to be used when we add interactions. We add $A,{}^1\!E,{}^2\!E$ orbitals at the $1a$ position, giving 6 orbitals total in the unit cell. In order to preserve the fragile topology of the occupied valence bands, we must put the $A_{1a}$ orbital in the conduction band, and we add the ${}^1\!E_{1a}, {}^2\!E_{1a}$ orbitals to the valence band. Thus there are $4$ occupied bands and $6$ orbitals, so our model is at filling $\nu = 2/3$. In summary, the single-particle Hamiltonian in the basis $A_{1b}, {}^2\!E_{1b},{}^2\!E_{1c}, A_{1a},{}^1\!E_{1a},{}^2\!E_{1a}$ reads:
\bea
h_0(\mbf{k}) &= \bpm
(1-M) h_{f}(\mbf{k}) &  & & \\
& 1-M&  & \\
&  &M & \\
&  && M\\
\epm, \qquad \text{spec } h_0(\mbf{k}) = \{E(\mbf{k})\} = \{0,0,M,M,1-M, 1-M\} 
\eea
where $h_f$ is a $3\times 3$ matrix and we emphasize that all bands are exactly flat. We show the single particle spectrum in \Fig{fig:modelorbs}b. When $M \in (0, 1/2)$, the groundstate is fragile at filling $2/3$ (we work at fixed particle number). It consists of the two zero energy flat bands, and the $M$ energy trivial ${}^1\!E_{1a}, {}^2\!E_{1a}$ bands. The band representation is
\bea
\mathcal{B}_{val} = \Gamma_2 + 3 \Gamma_3 + 3 K_2 + K_3 + 2K'_2 + 2K'_3 = ( A_{1b} \oplus {}^2\!E_{1b} \oplus {}^2\!E_{1c} \oplus {}^1\!E_{1a} \oplus {}^2\!E_{1a} ) \uparrow G \ominus  A_{1a} \uparrow G \\
\eea
which, using the momentum space tables of Ref. \cite{rsis}, gives the non-interacting RSIs
\bea
%(
\text{fragile:} \qquad \delta_{1a, 1} = 2, \delta_{1a, 2} = 2, \qquad \delta_{1b, 1} = -1, \delta_{1b, 2} = 0, \qquad \delta_{1c,  1} = 0, \delta_{1c,2} = 1,\qquad M \in [0,1/2) \ . \\
%]
\eea
Checking the inequality criteria in Ref. \cite{rsis}, for instance
\bea
N_{occ} = 4 < \delta_{1a,1} + \delta_{1a,2} - 2 \delta_{1b,1} + \delta_{1b,2} + \delta_{1c,1} + \delta_{1c,2} = 2 + 2 + 2 + 1 = 7, \\
\eea
we get a further confirmation of the fragile topology of the occupied bands. Note that these RSIs can be directly computed using the Wannier representation $A_{1b} \oplus {}^2\!E_{1b} \oplus {}^2\!E_{1c} \ominus  A_{1a} \oplus {}^1\!E_{1a} \oplus {}^2\!E_{1a}$  and the definitions
\bea
\label{eq:realspaceRSIs3}
\delta_{\mbf{x},1} = -m(A_{\mbf{x}}) + m({}^1\!E_{\mbf{x}}), \quad \delta_{\mbf{x},2} = -m(A_{\mbf{x}}) + m({}^2\!E_{\mbf{x}}), \qquad \mbf{x} = 1a, \ 1b, \ 1c \ .
\eea
%[
At $M=1/2$, there is a gap closing at $E=1/2$ where the Hamiltonian undergoes a phase transition. For $M \in (1/2, 1]$, the occupied bands are the two 0 energy fragile bands and the two $1a$ bands at energy $1-M$. Because the fragile bands and their OAL complement are all occupied, this phase is trivial. Technically, an infinitesimally small perturbation to $h_f$ which couples the OAL and fragile bands is needed to mix and trivialize them. This can be confirmed from the single-particle RSIs, which are
%)
\bea
%[
\text{trivial:} \qquad \delta_{1a, 1} = -1, \delta_{1a, 2} = -1, \qquad \delta_{1b, 1} = -1, \delta_{1b, 2} = 0, \qquad \delta_{1c,  1} = 0, \delta_{1c,2} = 1,\qquad M \in (1/2,1]  \\
%)
\eea
which can be obtained directly from the atomic limit state $A_{1a} \oplus A_{1b} \oplus {}^2\!E_{1b} \oplus{}^2\!E_{1c}$ using \Eq{eq:realspaceRSIs3}. Hence it is manifestly trivial. 

We will now show that the gap closing at $M=1/2$ which separates the single-particle phases can be circumvented when interactions are included. 

Consulting \Tab{tab:rsiredint}, we find that the interacting RSIs are the same \emph{in both phases}. Explicitly, 
\bea
(\Delta_{1a,1},\Delta_{1a,2}) = (1,0),  \quad (\Delta_{1b,1},\Delta_{1b,2}) = (2,1) ,\quad (\Delta_{1c,1},\Delta_{1c,2}) = (1,1)
\eea
Despite the nontrivial single-particle topology, the interacting RSIs are many-body trivial. This is shown using \Tab{tab:fragilenoSOC} where the fragile inequality is
\bea
\text{fragile if: } 4 = N_{occ} < \text{mod}_3(\Delta_{1a,1}|\Delta_{1a,2})+ \text{mod}_3(\Delta_{1b,1}|\Delta_{1b,2} )+ \text{mod}_3(\Delta_{1c,1}| \Delta_{1c,2} ) = 1 + 2 + 1 = 4
\eea
which is \emph{false}. Thus the interacting RSIs imply that even single-particle fragile state can be connected to a trivial state. The phase diagram is summarized in \Fig{fig:model1}(a). We now show this explicitly by introducing an interaction term that adiabatically connects this state to a trivial atomic limit. 

\subsection{Trivialization by Interactions}
\label{app:wannierham}

Before we discuss the interaction term, we will find a convenient expression for the non-interacting Hamiltonian. In momentum space, we have
\bea
H_0 &= \sum_{\mbf{k}} c^\dag_{\mbf{k},\al} h^{\al}_\be(\mbf{k}) c^\be_{\mbf{k}}, \qquad c^\dag_{\mbf{k},\al} = \frac{1}{\sqrt{\mathcal{N}}} \sum_{\mbf{R}} e^{-i \mbf{k} \cdot (\mbf{R} + \pmb{\delta}_{\al})} c^\dag_{\mbf{R},\al} \\
\eea
with $\mathcal{N}$ defined as the number of unit cells and the orbital basis is ordered $\al = A_{1b},{}^2\!E_{1b}, {}^2\!E_{1c} ,A_{1a},{}^1\!E_{1a}, {}^2\!E_{1a}$. In the single-particle eigenbasis, defined by the unitary transformation of the electron operators of the obstructed atomic phase $\gamma^\dag_{\mbf{k}} = U^\al_{A_{1a}}(\mbf{k}) c^\dag_{\mbf{k},\al}$, the Hamiltonian can be rewritten
\bea
H_0 &= \sum_{\mbf{k}} \left[ (1-M) \gamma^\dag_{\mbf{k}} \gamma_{\mbf{k}} + (1-M) c^\dag_{\mbf{k}, A_{1a}} c_{\mbf{k}, A_{1a}} + M c^\dag_{\mbf{k}, {}^1\!E_{1a}} c_{\mbf{k}, {}^1\!E_{1a}} + M c^\dag_{\mbf{k}, {}^2\!E_{1a}} c_{\mbf{k}, {}^2\!E_{1a}} \right] \ . 
\eea
Importantly, the electron operators defining the fragile band do not appear because they are at exactly zero energy. Taking $\mathcal{N} \to \infty$, we now perform another change of basis to the Wannier basis defined by
\bea
\label{eq:wbasis}
w^\dag_{\mbf{R},n} &= \int \frac{d^2k}{(2\pi)^2} e^{i \mbf{R} \cdot \mbf{k}} U^\al_n(\mbf{k}) c^\dag_{\mbf{k}, \al}, \qquad \sum_{\mbf{R}} e^{-i \mbf{R} \cdot \mbf{k}} w^\dag_{\mbf{R},n} =  U^\al_n(\mbf{k}) c^\dag_{\mbf{k}, \al}
\eea
where $U^\al_n(\mbf{k})$ are the eigenvectors of the $n$th band. For the three uncoupled 1a orbitals, the eigenvectors are trivial: $U^\al_n(\mbf{k})$, and the Wannier basis is simply the orbital basis. Because the bands are perfectly flat, we can solve the Fourier transforms and find
\bea
\label{eq:Hwannier}
H_0 &= \sum_{\mbf{R}} \left[ (1-M) w^\dag_{\mbf{R},A_{1a}} w_{\mbf{R},A_{1a}}  + (1-M) c^\dag_{\mbf{R}, A_{1a}} c_{\mbf{R}, A_{1a}} + M c^\dag_{\mbf{R}, {}^1\!E_{1a}} c_{\mbf{R}, {}^1\!E_{1a}} + M c^\dag_{\mbf{R}, {}^2\!E_{1a}} c_{\mbf{R}, {}^2\!E_{1a}} \right] \ . 
\eea
The Wannier basis has diagonalized the Hamiltonian in energy \emph{and} in the local unit cell. Recall from \Eq{eq:defwannier} that $w^\dag_{\mbf{R},A_{1a}}$ the sum electron operators on the 6 nearest neighbor sites. Hence all operators in $H_0$ are local. 

Inspired by the simple form of \Eq{eq:Hwannier}, we now add in the interaction term
\bea
H_{int} &= U \sum_{\mbf{R}} w^\dag_{\mbf{R},A_{1a}} c^\dag_{\mbf{R},A_{1a}} c_{\mbf{R}, {}^1\!E_{1a}} c_{\mbf{R}, {}^2\!E_{1a}} + h.c.\\
\eea
which introduces scattering between the ${}^1E,{}^2E$ irreps and $A,A$ irreps at 1a. Notably, the interacting involves pair-hopping because the $w^\dag_{A_{1a}}$ Wannier state is supported off the 1a position. First, we verify that $H_{int}$ is local because $w^\dag_{\mbf{R},A_{1a}}$ is local. Second, we check that $H_{int}$ preserves the symmetries of $H_0$. By construction, $H_{int}$ is invariant under translations. To show that $C_3^\dag H_{int} C_3 = H_{int}$, we recall
\bea
C^\dag_3 w^\dag_{\mbf{R},A_{1a}} C_3 &=  w^\dag_{C_3\mbf{R},A_{1a}} \\
C^\dag_3 c^\dag_{\mbf{R},A_{1a}} C_3 &=  c^\dag_{C_3 \mbf{R},A_{1a}} \\
C^\dag_3  c^\dag_{\mbf{R}, {}^1\!E_{1a}}  C_3 &= \omega  c_{C_3 \mbf{R}, {}^1\!E_{1a}}  \\
C^\dag_3  c^\dag_{\mbf{R}, {}^2\!E_{1a}}  C_3 &= \omega^*  c_{C_3 \mbf{R}, {}^2\!E_{1a}}  \\
\eea
from which we compute
\bea
C_3^\dag H_{int} C_3 &= U \sum_{\mbf{R}} w^\dag_{C_3\mbf{R},A_{1a}} c^\dag_{C_3 \mbf{R},A_{1a}} \omega \omega^* c_{C_3 \mbf{R}, {}^1\!E_{1a}} c_{C_3 \mbf{R}, {}^2\!E_{1a}} + h.c. = H_{int} \ . \\
\eea
In summary, we have shown that
\bea
H &= H_{0} + H_{int} =  \sum_{\mbf{R}} \left[ (1-M) (w^\dag_{\mbf{R},A_{1a}} w_{\mbf{R},A_{1a}}  +  c^\dag_{\mbf{R}, A_{1a}} c_{\mbf{R}, A_{1a}} ) + M (c^\dag_{\mbf{R}, {}^1\!E_{1a}} c_{\mbf{R}, {}^1\!E_{1a}} + c^\dag_{\mbf{R}, {}^2\!E_{1a}} c_{\mbf{R}, {}^2\!E_{1a}}) \right. \\ & \qquad \qquad \qquad\qquad \left.+ U w^\dag_{\mbf{R},A_{1a}} c^\dag_{\mbf{R},A_{1a}} c_{\mbf{R}, {}^1\!E_{1a}} c_{\mbf{R}, {}^2\!E_{1a}} + h.c. \right] 
\eea
is a fully local Hamiltonian with $G = p3$ \emph{and} is diagonalized in the Wannier basis, e.g. each term in the $\mbf{R}$ sum commutes with every other. $H$ can be interpreted as a ``stabilizer code'' in the Wannier basis.

\subsection{Exact Solution of $H$}
\label{app:modelsolution}

\begin{figure}
 \centering
\begin{overpic}[height=0.35\textwidth,tics=10]{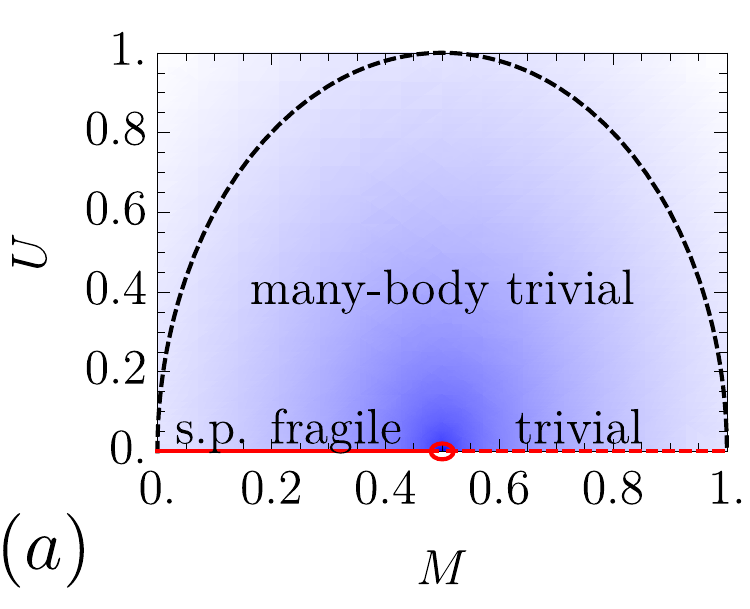}
\end{overpic}  
\begin{overpic}[height=0.37\textwidth,tics=10, trim = 0 0 0  0, clip]{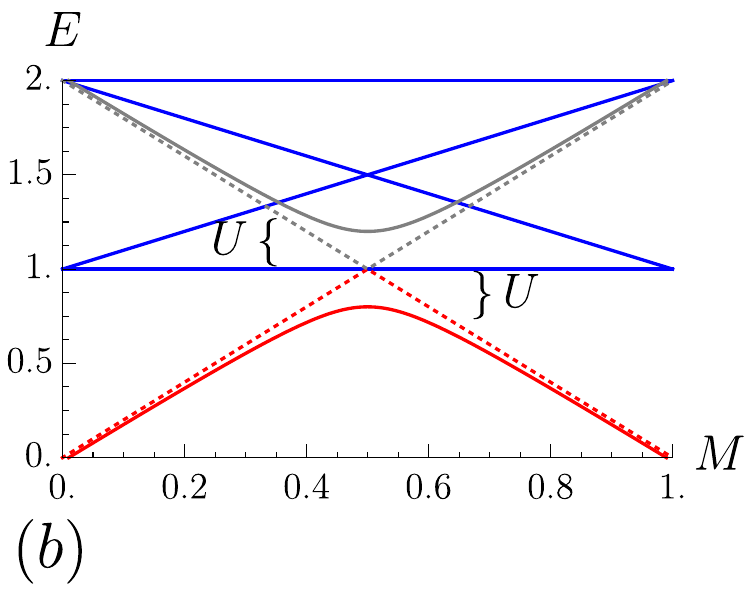}
\end{overpic}  \\
\caption{(a) We sketch the phase diagram of $H = H_0 + H_{int}$. At $U=0$, the single-particle (s.p.) fragile phase is denoted as a solid red line. A gap closing (open oval) at $M=1/2, U=0$ separates this phase from the trivial atomic limit, denoted by a dashed red line. However, if $U\neq 0$, the single-particle classifications are destroyed and the model is many-body trivial. In particular, the black dashed line \Eq{eqLMUparam} connects the two single-particle phases without a gap closing, showing that interactions trivialize the single-particle fragile topolgy. (b) $H$ is exactly solvable because it is decoupled in the Wannier basis. We show the spectrum of $H$ in a single unit cell determined in \Eq{eq:modelspectrum15}. The groundstate is shown in solid red for nonzero $U$ and dashed for $U=0$. When $U=0$, we see there is a gap closing with a higher band at $M=1/2$ as expected from the single-particle physics. }
\label{fig:model1}
\end{figure}

We now solve $H$ exactly by exploiting its diagonalization in Wannier space. In a given unit cell at $\nu = 4/6$, the Hilbert space is $\binom{6}{4}$-dimensional and is spanned by all $4$-operator products chosen from the set $w^\dag_{\mbf{R},A_{1a}}, c^\dag_{\mbf{R},A_{1a}} , c^\dag_{\mbf{R}, {}^1\!E_{1a}} , c^\dag_{\mbf{R}, {}^2\!E_{1a}}, f^\dag_{\mbf{R},1}, f^\dag_{\mbf{R},2}$. Here $f^\dag_{\mbf{R},1},f^\dag_{\mbf{R},2}$ denote the electron operators of the fragile band in the ``Wannier" basis defined by
\bea
\label{eq:basischange}
f^\dag_{\mbf{R},j} &= \int \frac{d^2k}{(2\pi)^2} e^{i \mbf{R} \cdot \mbf{k}} U^\al_{f,j}(\mbf{k}) c^\dag_{\mbf{k}, \al}, \qquad j = 1,2
\eea
where $U_{f,j}(\mbf{k})$ are the single-particle eigenvectors defined below \Eq{eq:handfvectors} and satisfy $U_{f,j}^\dag(\mbf{k}) U_{A_{1a}}(\mbf{k}) = 0$. Because of the single-particle fragile topology, there is no choice of $U_{f,j}(\mbf{k})$ where $f^\dag_{\mbf{R},j}$ are transform local under the symmetries and are exponentially decaying. For concreteness, we pick $U_{f,j}(\mbf{k})$ such that $f^\dag_{\mbf{R},j}$ \emph{are} exponentially localized \cite{PhysRevLett.98.046402}, at the expense of a local action of the symmetries on $f^\dag_{\mbf{R},j}$. Most importantly, $f^\dag_{\mbf{R},j}$ and $w^\dag_{\mbf{R},A_{1a}}$ form a complete orthonormal basis of the Hilbert space at $\mbf{R}$. Moreover, the fragile bands are at zero energy and are annihilated by $H$. Finally, we emphasize that the operators $ w^\dag_{\mbf{R},A_{1a}}, c^\dag_{\mbf{R},A_{1a}} , c^\dag_{\mbf{R}, {}^1\!E_{1a}} , c^\dag_{\mbf{R}, {}^2\!E_{1a}}, f^\dag_{\mbf{R},1}, f^\dag_{\mbf{R},2}$ are all orthonormal.  The operators $w^\dag_{\mbf{R},A_{1a}}, c^\dag_{\mbf{R},A_{1a}} , c^\dag_{\mbf{R}, {}^1\!E_{1a}} , c^\dag_{\mbf{R}, {}^2\!E_{1a}}$ are local (and compactly supported). The operators $ f^\dag_{\mbf{R},1}, f^\dag_{\mbf{R},2}$ are non-local but do not appear in the Hamiltonian:
\bea
\null [H, f^\dag_{\mbf{R},1}] = [H, f^\dag_{\mbf{R},2}] = 0 \ .
\eea
Thus $H$ is manifestly local. (It has mutually commuting local terms.)

Writing $H$ explicitly in the $15$-dimensional local Hilbert space, we find that $H$ is diagonal except for a $U$-coupling between the four-particle states $f^\dag_{\mbf{R},1} f^\dag_{\mbf{R},2} w^\dag_{\mbf{R}, A_{1a}} c^\dag_{\mbf{R}, A_{1a}}$ and $f^\dag_{\mbf{R},1} f^\dag_{\mbf{R},2} c^\dag_{\mbf{R}, {}^1\!E_{1a}} c^\dag_{\mbf{R}, {}^2\!E_{1a}}$. This is guaranteed by the orthogonality of the operators. Nonzero $U$ mixes these states, which we can think of as performing the adiabatic process $AA \to {}^1\!E{}^2\!E$ at the 1a site. This process is the one responsible for trivializing the single-particle state, e.g.
\bea
A_{1b} \oplus {}^2\!E_{1b} \oplus {}^2\!E_{1c} \ominus  A_{1a} \oplus {}^1\!E_{1a} \oplus {}^2\!E_{1a} \to A_{1b} \oplus {}^2\!E_{1b} \oplus {}^2\!E_{1c} \ominus  A_{1a} \oplus A_{1a} \oplus {}A_{1a} = A_{1b} \oplus {}^2\!E_{1b} \oplus {}^2\!E_{1c} \oplus {}A_{1a} 
\eea
where we see that the Wannier obstruction has been removed. 

We now show this explicitly. The spectrum of $H$ in the local Hilbert space is (dropping the unit cell subscript for clarity)
\bea
\label{eq:modelspectrum15}
\frac{1}{\sqrt{(E_+-2M)^2 + U^2}} \big( (E-2M) f^\dag_1 f^\dag_2 w^\dag_{A_{1a}} c^\dag_{A_{1a}} + U  f^\dag_1 f^\dag_2 c^\dag_{{}^1\!E_{1a}} c^\dag_{{}^2\!E_{1a}} \big) \ket{0}  \quad  & E_+ = 1 + \sqrt{(1-2M)^2 + U^2} \\
\big(w^\dag_{A_{1a}} c^\dag_{A_{1a}}  c^\dag_{{}^1\!E_{1a}} c^\dag_{{}^2\!E_{1a}}\big) \ket{0}: \quad & E = 2 \\
f^\dag_{1} f^\dag_{2}c^\dag_{A_{1a}} c^\dag_{{}^2\!E_{1a}} \ket{0}, f^\dag_{1} f^\dag_{2}c^\dag_{A_{1a}} c^\dag_{{}^1\!E_{1a}}\ket{0} ,f^\dag_{1} f^\dag_{2}w^\dag_{A_{1a}} c^\dag_{{}^2\!E_{1a}} \ket{0}, f^\dag_{1} f^\dag_{2}w^\dag_{A_{1a}} c^\dag_{{}^1\!E_{1a}} \ket{0}: \quad & E = 1 \\
w^\dag_{A_{1a}} c^\dag_{A_{1a}}  c^\dag_{{}^1\!E_{1a}} f^\dag_{1} \ket{0} ,w^\dag_{A_{1a}} c^\dag_{A_{1a}}  c^\dag_{{}^1\!E_{1a}} f^\dag_{2}  \ket{0} ,w^\dag_{A_{1a}} c^\dag_{A_{1a}}  c^\dag_{{}^2\!E_{1a}} f^\dag_{1}  \ket{0} ,w^\dag_{A_{1a}} c^\dag_{A_{1a}}  c^\dag_{{}^2\!E_{1a}} f^\dag_{2}  \ket{0}: \quad & E = 2-M \\
 c^\dag_{{}^1\!E_{1a}}  c^\dag_{{}^2\!E_{1a}}  f^\dag_{1} c^\dag_{A_{1a}}  \ket{0}, c^\dag_{{}^1\!E_{1a}}  c^\dag_{{}^2\!E_{1a}}  f^\dag_{2} c^\dag_{A_{1a}}  \ket{0},
  c^\dag_{{}^1\!E_{1a}}  c^\dag_{{}^2\!E_{1a}}  f^\dag_{1} w^\dag_{A_{1a}}  \ket{0}, c^\dag_{{}^1\!E_{1a}}  c^\dag_{{}^2\!E_{1a}}  f^\dag_{2} w^\dag_{A_{1a}}  \ket{0}: \quad  & E = 1+M \\
\frac{1}{\sqrt{(E_--2M)^2 + U^2}}  \big( (E_--2M) f^\dag_1 f^\dag_2 w^\dag_{A_{1a}} c^\dag_{A_{1a}} + U  f^\dag_1 f^\dag_2 c^\dag_{{}^1\!E_{1a}} c^\dag_{{}^2\!E_{1a}} \big)  \ket{0}  \quad  & E_- = 1- \sqrt{(1-2M)^2 + U^2}\ .  \\
\eea

The spectrum is plotted as a function of $M$ in \Fig{fig:model1}(b). The last state is the unique groundstate everywhere except $M=1/2, U = 0$. We can see that the interaction as coupled the two groundstates of the two single-particle phases ($ f^\dag_1 f^\dag_2 c^\dag_{{}^1\!E_{1a}} c^\dag_{{}^2\!E_{1a}} $ which is fragile and $ f^\dag_1 f^\dag_2 w^\dag_{A_{1a}} c^\dag_{A_{1a}} $ which is trivial) and has opened the gap at $M=1/2$ for nonzero $U$. In fact, the gap is given by $\sqrt{(1-2M)^2 + U^2}$ and is nonzero everywhere except $M=1/2, U=0$ where there is a six-fold degeneracy. 

For concreteness, we can choose a parameterization in the $U,M$ phase diagram that connects the two single-particle regions where $U=0$ without closing the many-body gap. We take
\bea
\label{eqLMUparam}
M = \frac{1}{2}(1- \cos \theta), \quad U = \sin \theta, \qquad \theta \in (0,\pi)
\eea
which tunes between $M= 0,U=0$ and $M=1, U=0$. The full many-body groundstate along this path is given by
\bea
\ket{GS(\th)} &= \prod_{\mbf{R}} \lp \cos \frac{\theta}{2}  f^\dag_{\mbf{R},1} f^\dag_{\mbf{R},2} c^\dag_{\mbf{R},{}^1\!E_{1a}} c^\dag_{\mbf{R},{}^2\!E_{1a}} -\sin \frac{\theta}{2} f^\dag_{\mbf{R},1} f^\dag_{\mbf{R},2} w^\dag_{\mbf{R},A_{1a}} c^\dag_{\mbf{R},A_{1a}} \rp \ket{0}
\eea
and the many-body gap $\sqrt{(1-2M)^2 + U^2}$ is equal to 1 for all $\theta$. At $\theta = \pi$, the groundstate is a Slater determinant and hence
\bea
\ket{GS(\pi)} &= \prod_{\mbf{R}} \lp - f^\dag_{\mbf{R},1} f^\dag_{\mbf{R},2} w^\dag_{\mbf{R},A_{1a}} c^\dag_{\mbf{R},A_{1a}} \rp \ket{0} \propto  \prod_{\mbf{R}} c^\dag_{\mbf{R},A_{1b}} c^\dag_{\mbf{R},{}^2\!E_{1b}} c^\dag_{\mbf{R},{}^2\!E_{1c}} c^\dag_{\mbf{R},A_{1a}}  \ket{0}
\eea
because $f^\dag_{\mbf{R},1}, f^\dag_{\mbf{R},2}, w^\dag_{\mbf{R}A_{1a}}$ form a complete basis of the $A_{1b},{}^2\!E_{1b},{}^2\!E_{1c}$ orbitals, and hence by total anti-symmetry it follows that 
\bea
\label{eq:trivfromfrag}
\prod_{\mbf{R}} f^\dag_{\mbf{R},1} f^\dag_{\mbf{R},2} w^\dag_{\mbf{R},A_{1a}} = e^{i \varphi} \prod_\mbf{R} c^\dag_{\mbf{R},A_{1b}} c^\dag_{\mbf{R},{}^2\!E_{1b}} c^\dag_{\mbf{R},{}^2\!E_{1c}} 
\eea
where $e^{i \varphi}$ is the determinant of the unitary operator that changes the orbital basis to the Wannier basis in \Eqs{eq:wbasis}{eq:basischange}. The righthand side of \Eq{eq:trivfromfrag} makes it clear that $\ket{GS(\pi)}$ is a trivial atomic limit. 

\end{document}